\documentclass[twocolumn]{aastex63}

\usepackage{lipsum}
\usepackage{xcolor}
\usepackage{graphicx}
\usepackage{enumitem}
\usepackage{amsmath}
\usepackage{makecell}
\usepackage{comment}
\usepackage{array}
\usepackage{diagbox}
\usepackage{booktabs}
\usepackage{diagbox}

\shorttitle{ALeRCE Stamp Classifier}
\shortauthors{R. Carrasco-Davis and E. Reyes}
\graphicspath{{./}{}}
\begin{document}

\title{Alert Classification for the ALeRCE Broker System: The Real-time Stamp Classifier}

\newcommand\MAS{Millennium Institute of Astrophysics (MAS), Nuncio Monse{\~{n}}or S{\'{o}}tero Sanz 100, Providencia, Santiago, Chile}
\newcommand\CMM{Center for Mathematical Modeling, Universidad de Chile, Beauchef 851, North building, 7th floor, Santiago 8320000, Chile}
\newcommand\DIE{Department of Electrical Engineering, Universidad de Chile, Av. Tupper 2007, Santiago 8320000, Chile}
\newcommand\DAS{Departamento de Astronom\'ia, Universidad de Chile, Casilla 36D, Santiago, Chile}
\newcommand\PSANDRESBELLO{Departamento de Ciencias F\'isicas, Universidad Andres Bello,
Av. Republica 230, Santiago 8370146, Chile}
\newcommand\ASTROPUC{Instituto de Astrof{\'{\i}}sica and Centro de Astroingenier{\'{\i}}a, Facultad de F{\'{i}}sica, Pontificia Universidad Cat{\'{o}}lica de Chile, Casilla 306, Santiago 22, Chile}
\newcommand\ENGADOLFO{Faculty of Engineering and Sciences, Universidad Adolfo Iba\~nez, Diagonal Las Torres 2700, Pe\~nalol\'en, Santiago, Chile}
\newcommand\UDEC{Department of Computer Science, Universidad de Concepci\'on, Edmundo Larenas 219, Concepci\'on, Chile}

\correspondingauthor{Rodrigo Carrasco-Davis}
\email{rodrigo.carrasco.davis@gmail.com \\ rodrigo.carrasco.d@ing.uchile.cl}

\author[0000-0003-4673-8791]{R. Carrasco-Davis}
\altaffiliation{These authors contributed equally to this work}
\affiliation{\MAS}
\affiliation{\DIE}

\author[0000-0003-3455-9358]{E. Reyes}
\altaffiliation{These authors contributed equally to this work}
\affiliation{\MAS}
\affiliation{\DIE}

\author[0000-0001-5306-1390]{C. Valenzuela}
\affiliation{\CMM}
\affiliation{\MAS}
\affiliation{\ENGADOLFO}
\affiliation{Data Observatory, Santiago, Chile}

\author[0000-0003-3459-2270]{F. F\"orster}
\affiliation{\CMM}
\affiliation{\MAS}
\affiliation{\DAS}

\author[0000-0001-9164-4722]{P. A. Est\'evez}
\affiliation{\MAS}
\affiliation{\DIE}

\author[0000-0003-0006-0188]{G. Pignata}
\affiliation{\PSANDRESBELLO}
\affiliation{\MAS}

\author[0000-0002-8686-8737]{F. E. Bauer}
\affiliation{\ASTROPUC}  
\affiliation{\MAS}
\affiliation{Space Science Institute, 4750 Walnut Street, Suite 205, Boulder, Colorado 80301, USA}

\author[0000-0003-3627-0216]{I. Reyes}
\affiliation{\MAS} 
\affiliation{\CMM}
\affiliation{\DIE}

\author[0000-0003-0820-4692]{P. S\'anchez-S\'aez}
\affiliation{\MAS} 
\affiliation{Inria Chile Research Center, Av. Apoquindo 2827, Las Condes, Chile}
\affiliation{\ASTROPUC}
\affiliation{\ENGADOLFO}

\author[0000-0002-2720-7218]{G. Cabrera-Vives}
\affiliation{\UDEC}
\affiliation{\MAS}

\author[0000-0003-4723-9660]{S. Eyheramendy}
\affiliation{\ENGADOLFO}
\affiliation{\MAS}

\author[0000-0001-6003-8877]{M. Catelan}
\affiliation{\ASTROPUC}
\affiliation{\MAS}

\author{J. Arredondo}
\affiliation{\MAS}

\author{E. Castillo-Navarrete}
\affiliation{\CMM}
\affiliation{\MAS}

\author{D. Rodr\'iguez-Mancini}
\affiliation{\MAS} 
\affiliation{\UDEC}

\author[0000-0002-1292-2374]{D. Ruz-Mieres}
\affiliation{\CMM}
\affiliation{\MAS}
\affiliation{Institute of Astronomy, University of Cambridge, Madingley Road, Cambridge CB3 0HA, UK}

\author[0000-0002-7003-5087]{A. Moya}
\affiliation{\MAS}
\affiliation{\CMM} 

\author{L. Sabatini-Gacit\'ua}
\affiliation{\MAS}
\affiliation{\CMM} 

\author{C. Sep\'ulveda-Cobo}
\affiliation{\MAS}
\affiliation{\CMM} 

\author{A. A. Mahabal}
\affiliation{Cahill Center for Astrophysics, California Institute of Technology, 1200 E. California Boulevard, Pasadena, CA 91125, USA} 
\affiliation{Center for Data Driven Discovery, California Institute of Technology, Pasadena, CA 91125, USA}

\author{J. Silva-Farf\'an}
\affiliation{\DAS}

\author{E. Camacho-Iñiguez}
\affiliation{\ASTROPUC}

\author{L. Galbany}
\affiliation{Departamento de F\'isica Te\'orica y del Cosmos, Universidad de Granada, E-18071 Granada, Spain}

%%%%%%%%%%%%%%%%%%%%%%%%%%%%%%%%%%%%%%%%%%%%%%%%%%%%%%%%%%%%%%
%%%%%%%%%%%%%%%%%% PAPER BEGINS HERE %%%%%%%%%%%%%%%%%%%%%%
%%%%%%%%%%%%%%%%%%%%%%%%%%%%%%%%%%%%%%%%%%%%%%%%%%%%%%%%%%%%%%

\begin{abstract}
We present a real-time stamp classifier of astronomical events for the ALeRCE (Automatic Learning for the Rapid Classification of Events) broker. The classifier is based on a convolutional neural network, trained on alerts ingested from the Zwicky Transient Facility (ZTF). Using only the \textit{science, reference} and \textit{difference} images of the first detection as inputs, along with the metadata of the alert as features, the classifier is able to correctly classify alerts from active galactic nuclei, supernovae (SNe), variable stars, asteroids and bogus classes, with high accuracy ($\sim$94\%) in a balanced test set. In order to find and analyze SN candidates selected by our classifier from the ZTF alert stream, we designed and deployed a visualization tool called SN Hunter, where relevant information about each possible SN is displayed for the experts to choose among candidates to report to the Transient Name Server database. From June 26th 2019 to February 28th 2021, we have reported 6846 SN candidates to date (11.8 candidates per day on average), of which 971 have been confirmed spectroscopically. Our ability to report objects using only a single detection means that 70\% of the reported SNe occurred within one day after the first detection. ALeRCE has only reported candidates not otherwise detected or selected by other groups, therefore adding new early transients to the bulk of objects available for early follow-up. Our work represents an important milestone toward rapid alert classifications with the next generation of large etendue telescopes, such as the Vera C. Rubin Observatory.

\end{abstract}


\section{Introduction}\label{sec:intro}

The amount of data generated by modern survey telescopes cannot be directly handled by humans. Therefore, automatic data analysis methods are necessary to fully exploit their scientific return. A particularly challenging problem is the real-time classification of transient events. 
Nevertheless, the possibility to generate a quick probabilistic evaluation of which type of transient has been discovered is crucial to perform the most suitable follow-up observation, and by extension obtain the best constraints on its physics.
In this work we focus on the early detection of supernovae (SNe) by quickly discerning between SNe and various other confusing classes of astronomical objects. Photometric and spectroscopic observations carried out soon after the explosion are fundamental to put constraints on the progenitor systems and explosion physics.

In the case of thermonuclear explosions (Type Ia SNe), early observations probe the outermost part of the ejecta, where it is possible to detect the material present at the surface of the progenitor (e.g., \citealt{nugent_supernova_2011}), evaluate the degree of mixing induced by different explosion models (e.g., \citealt{piro_exploring_2016,jiang_hybrid_2017,noebauer_early_2017}), and estimate the size of the companion star (e.g., \citealt{kasen_seeing_2010}).

For core collapse (CC) SNe, observations carried out soon after the explosion can constrain the radius of the progenitor star, its outer structure and the degree of $^{56}$Ni mixing (e.g., \citealt{tominaga_shock_2011,piro_what_2013}), but also the immediate SN environment, providing a critical diagnostic for the elusive final evolutionary history of the progenitor and/or the progenitor system configuration (e.g., \citealt{moriya_supernovae_2011,gal-yam_wolf-rayet-like_2014,groh_early-time_2014,khazov_flash_2016, tanaka_rapidly_2016, yaron_confined_2017,morozova_unifying_2017,forster_delay_2018}).

We propose a method to quickly classify alerts among five different classes, four of which are astrophysical, and then use the predictions to find and report SNe. This work has been developed in the framework of ALeRCE\footnote{\url{https://alerce.online/}} (Automatic Learning for the Rapid Classification of Events; \citealt{forster_automatic_2020}). The ALeRCE system is able to read, annotate, classify and redistribute the data from large survey telescopes. Such efforts are commonly called \textit{Astronomical Broker Systems} (other examples include, e.g., ANTARES,  \citealt{narayan_machine-learning-based_2018}; Lasair, \citealt{smith_lasair_2019}).
Currently, ALeRCE is processing the alert stream generated by the Zwicky Transient Facility (ZTF; \citealt{bellm_zwicky_2018}) and its main goal is to reliably classify data of non-moving objects, and make these classifications available to the scientific community.

For the purpose of classifying astronomical objects or transients, one way to discriminate among them is by computing features from the light curve of each object (e.g., \citealt{richards_machine-learned_2011, pichara_meta-classification_2016, martinez-palomera_high_2018, boone_avocado_2019, sanchezsaez_alert_2020}), or using the light curve directly as input to a classifier (e.g., \citealt{mahabal_deep-learnt_2017, naul_recurrent_2018, muthukrishna_dash_2019, becker_scalable_2020}). In the case of an alert stream scenario such as for ZTF (whereby no forced photometry of past images is provided as of February 2021), the light curve is built by estimating the flux from the difference image for all alerts triggered at the same coordinates. 

Our model is dubbed the \emph{``stamp classifier''}, since it only uses the first alert of an astronomical object. ALeRCE also developed a \textit{light curve classifier} (\citealt{sanchezsaez_alert_2020}) based on light curves with $\geq6$ detections in $g$ or $\geq6$ detections in $r$ ZTF bands. The \textit{light curve classifier} is able to discriminate among a richer taxonomy of astronomical objects. Both the stamp and light curve classifiers are currently running through the ALeRCE frontend (\citealt{forster_automatic_2020}).

Our proposed \emph{stamp classifier} is based on a convolutional neural network (CNN) architecture that uses only the information available in the first alert of an astronomical object, which includes the images of the objects plus metadata regarding some of the object properties, observing conditions and information from other catalogs. The images included in the alert correspond to the \textit{science}, \textit{reference} and \textit{difference} images, which are shown in Figure~\ref{fig:alert_example} and described in Section \ref{sec:data_description}. The stamp classifier uses the first alert to discriminate between active galactic nuclei (AGN), supernovae (SNe), variable stars (VS), asteroids and bogus alerts. The architecture was designed to exploit the rotational invariance of astronomical images. The classifier was trained using an entropy regularizer that avoids the assignment of high probability to a single class, yielding softer output probabilities that give extra information to experts, useful for further analysis of candidates. To the best of our knowledge, this is the first classifier that discriminates among five classes using a single alert, allowing a rapid, reliable characterization of the data stream to trigger immediate follow-up. Previous work on stamp classification has focused instead on the classification of real objects vs. bogus detections (e.g., \citealt{goldstein_automated_2015, cabrera-vives_deep-hits_2017, reyes_enhanced_2018, duev_real-bogus_2019, turpin_vetting_2020}), galaxy morphologies (e.g., \citealt{dieleman_rotation-invariant_2015, perez-carrasco_multiband_2019, barchi_machine_2020}), or time domain classification (\citealt{carrasco-davis_deep_2019, gomez_classifying_2020}).

An associated contribution to the stamp classifier is the implementation of a visualization tool called \textit{SN Hunter}\footnote{\url{https://snhunter.alerce.online/}}, which allows experts to explore SN candidates to further filter alerts, and choose objects to request follow-up. This visualization tool is deployed online and provides a snapshot of the current ZTF data stream within minutes of receiving new alerts.

This work is structured as follows: In Section~\ref{sec:data_description} we describe the data used to train the proposed neural network model, as well as a brief description of each class and how we gathered labeled data. In Section~\ref{sec:methodology} we describe the data pre-processing, the neural network architecture, the entropy regularizer added to the optimization function and the experiments run to find the best architecture for the problem at hand. In Section~\ref{sec:results} we report and discuss our results in terms of the classification task, we also analyze the contribution to the classification performance of each one of the three images in the alert, as well as the metadata. In Section~\ref{sec:sn_hunter}, we describe the SN Hunter visualization tool and the visual criteria used by human experts to choose good candidates to report to the Transient Name Server (TNS)\footnote{\url{https://wis-tns.weizmann.ac.il/}}, along with an analysis of reported and confirmed SNe by ALeRCE using the proposed methodology since June 2019. We finally draw our conclusions and describe future work in Section~\ref{sec:conclusions}.

\section{Data} \label{sec:data_description}

\begin{figure*}[htbp]
\gridline{\fig{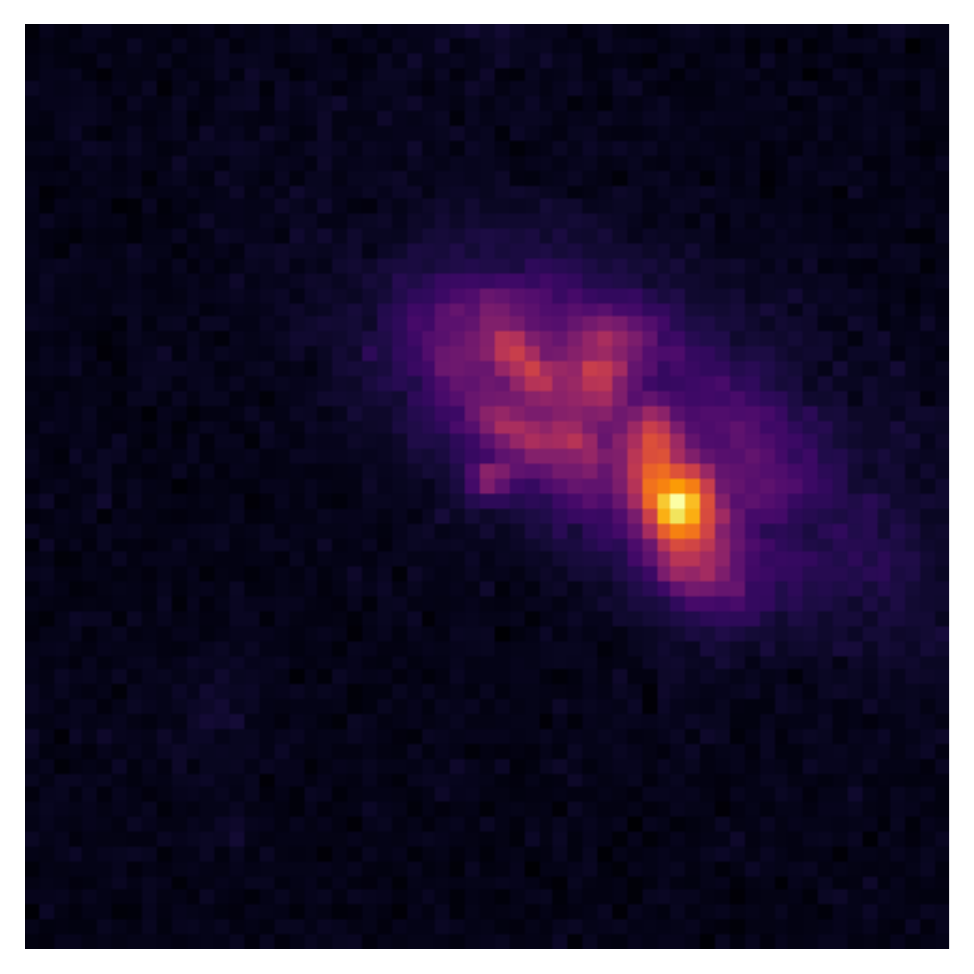}{0.24\textwidth}{(a) Science}
          \fig{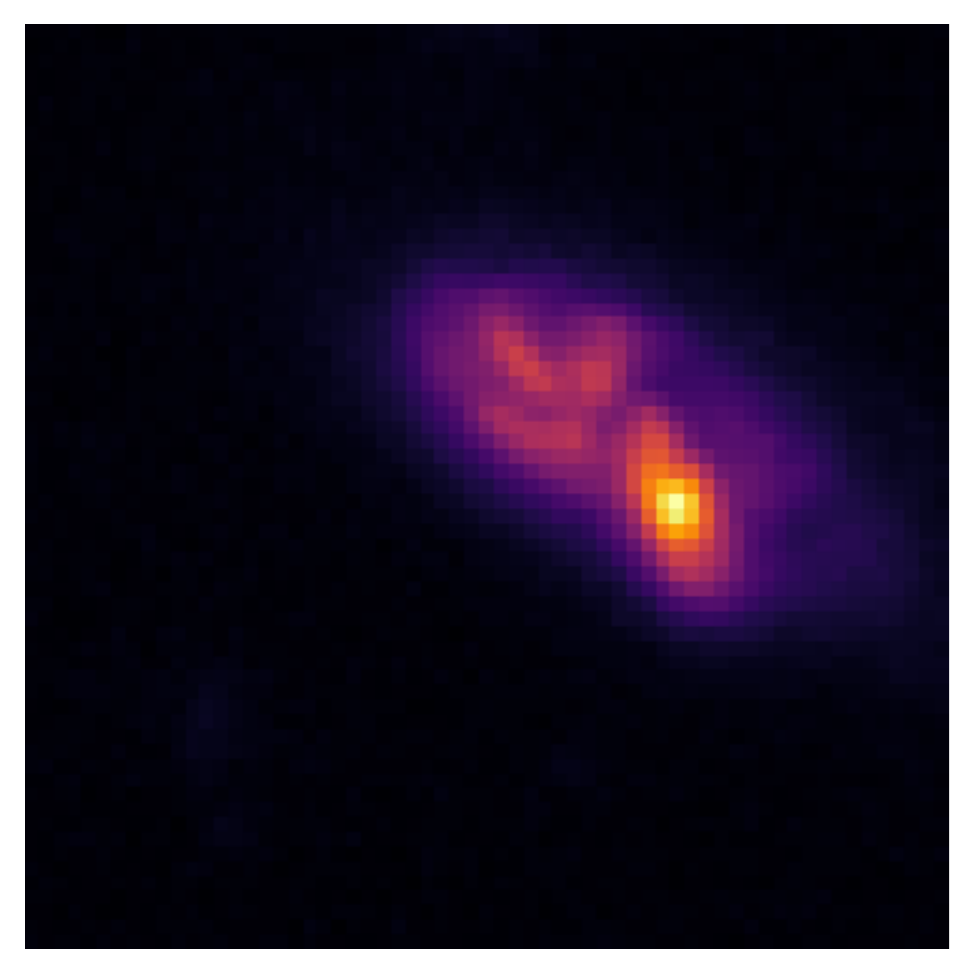}{0.24\textwidth}{(b) Reference}
          \fig{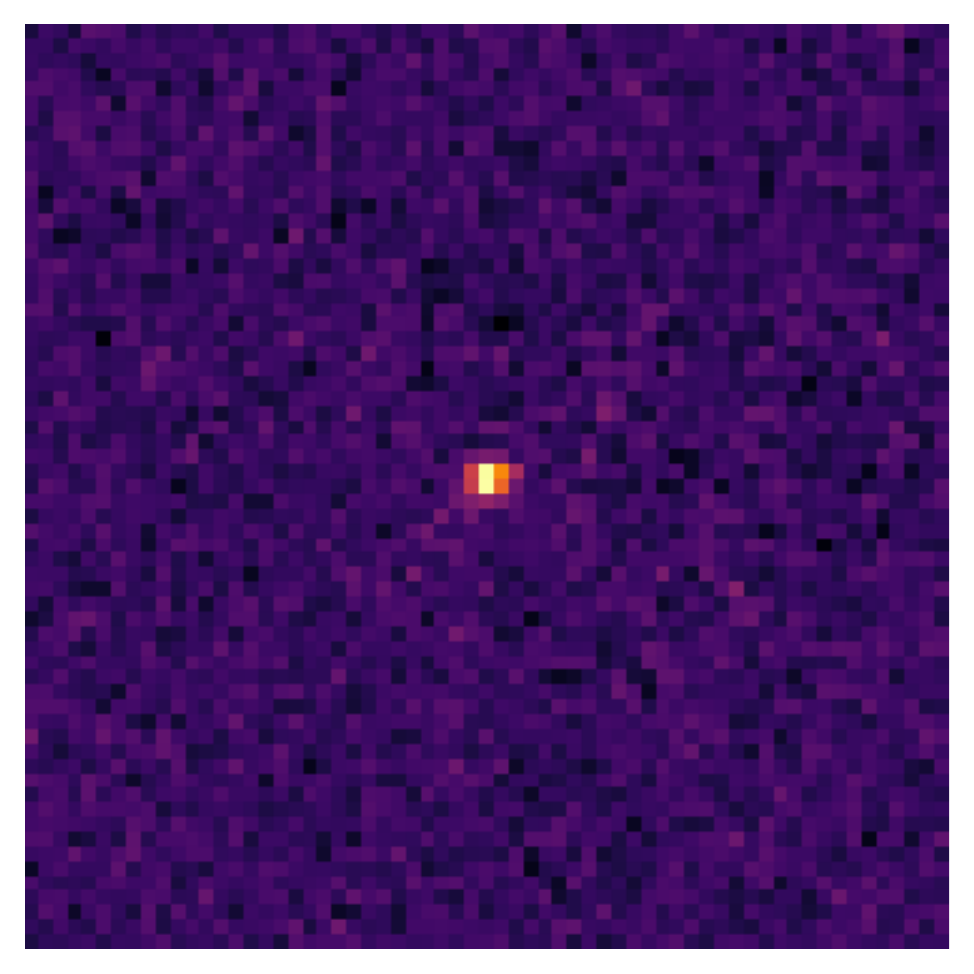}{0.24\textwidth}{(c) Difference}
          \fig{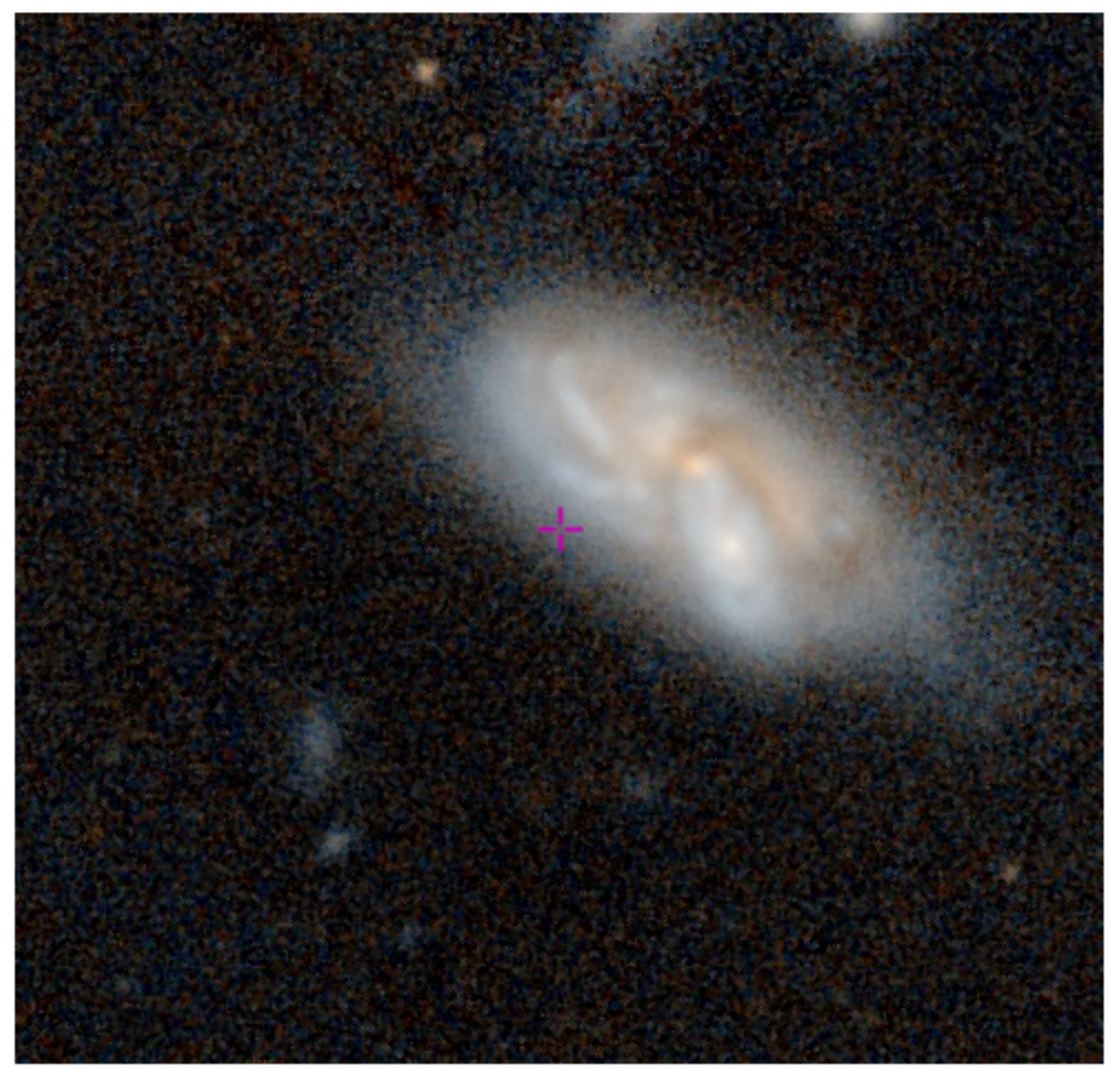}{0.24\textwidth}{(d) Colored Image}
          }
\caption{Example $g$-band images from a ZTF alert packet, in this case from a type Ia SN (ZTF19abmolyr) classified by our method. (a) The science image is the latest measurement of a source. (b) The reference image is usually a higher signal-to-noise measurement taken from an earlier epoch. (c) The third stamp is the \textit{difference} between the reference and science images. (d) For context, we also show the $gri$ color image from PanSTARRS, which is not part of the alert packet nor used in the current stamp classifier. Each image stamp is 63$\times$63 pixels, where 1 pixel = 1 arcsec. \label{fig:alert_example}}
\end{figure*}

An alert within the ZTF stream is defined as a source in the sky that produces a signal five standard deviations higher than the background noise (a five-$\sigma$ magnitude limit; \citealt{masci_zwicky_2018}), and which passes a real bogus filter designed by the ZTF collaboration (\citealt{mahabal_machine_2019}). When an alert is triggered, an \textit{alert packet} is generated with all the relevant information about the source that triggered the alert (\citealt{bellm_zwicky_2018}). The alert packet contains three images called stamps, which are cropped at 63 pixels on a side (1 pixel = 1 arcsec) from the original image and centered on the position of the source. In addition, the alert packet contains metadata related to the source, the observation conditions of the exposure and other useful information (\citealt{masci_zwicky_2018}). An example of the three stamps within an alert packet is shown in Figure~\ref{fig:alert_example}. The stamp in Figure~\ref{fig:alert_example}a is called the \textit{science image} and corresponds to the most recent measurement of the source. The stamp depicted in Figure~\ref{fig:alert_example}b is the \textit{reference image}, which is fixed for a given region and bandpass. It is usually based on images taken at the beginning of the survey and it is built by averaging multiple images to improve its signal-to-noise ratio. The stamp shown in Figure~\ref{fig:alert_example}c is the \textit{difference image}, between the science and reference images (\citealt{masci_zwicky_2018}), which shows the change in flux between those frames, removing other sources with constant brightness.

Each alert packet represents only 2 samples in time, the reference and science image exposures, and often is insufficient to correctly classify objects over the full taxonomy of different variable stars, transients or stochastic sources as in \cite{sanchezsaez_alert_2020}.

\begin{figure*}[htbp]
    \centering
    \includegraphics[width=1\textwidth]{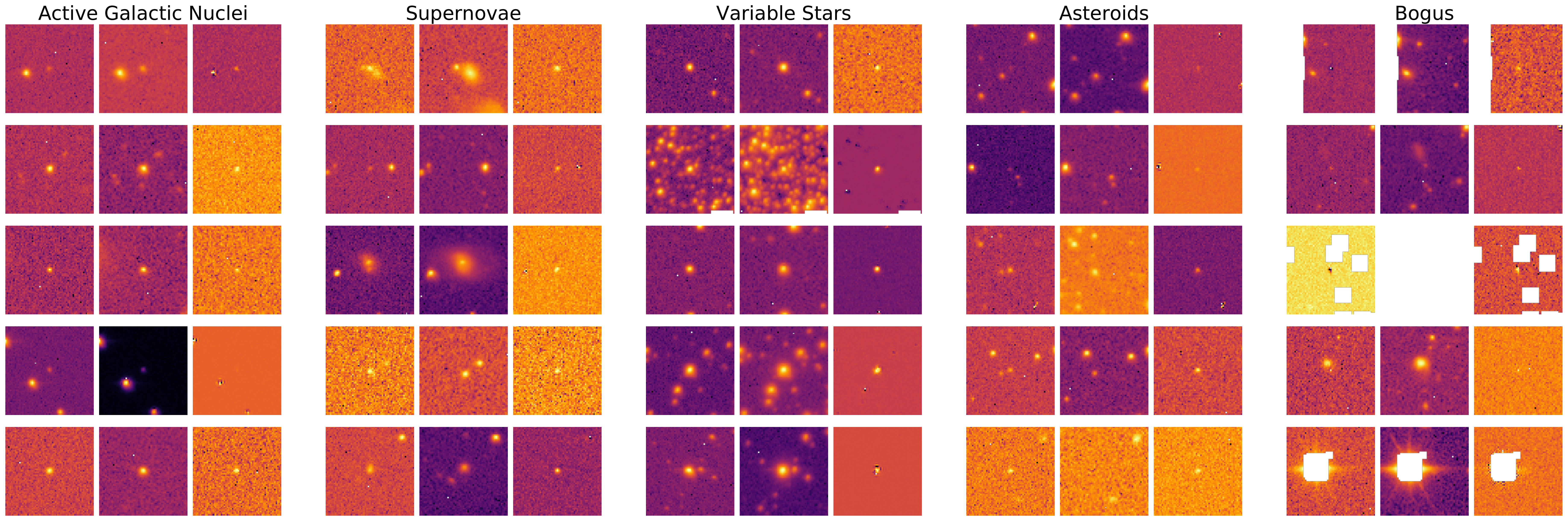}
    \caption{Examples of the five classes to be discriminated by using only the first detection. For each class, the triplet of images in each row are \textit{science, reference} and \textit{difference} images from left to right. Each row corresponds to a different candidate.}
    \label{fig:samples_per_class}
\end{figure*}

However, our hypothesis is that it is feasible to use the information included in a single alert packet to separate objects into several broad classes, namely AGN, SNe, VS, asteroids and bogus alerts. Each class presents distinctive characteristics within the image triplet of the first detection alert (see Figure~\ref{fig:samples_per_class}), which could be automatically learned by a CNN. In addition to the images, information in the metadata in the alert packet along with some derived features from the metadata are important to discriminate among the mentioned classes. The metadata used for the classification task is listed in Table~\ref{table:features}, and the distribution of values for each feature per class is shown in Figure~\ref{fig:features_grid} in Appendix \ref{sec:features_appendix}. Some of the distinctive characteristics and metadata features that help to distinguish between objects of each class, are the following:

\begin{itemize}[leftmargin=*]
    \item \textbf{AGN}: Being stochastically variable objects, an alert generated by an AGN should have flux from the source in both the \textit{reference} and \textit{science} stamps. Considering this feature alone, it is difficult to discriminate AGNs from other variable sources. Nevertheless, AGNs should lie at the centers of their host galaxies (based on dynamical friction arguments), or appear as (quasi-)stellar objects, in relatively lower stellar density fields. Thus, a change in flux will appear as a variable source, which may lie at the center of a galaxy, or even when the galaxy is not visible they tend to be in lower stellar density fields. In these cases, the alert is likely to be triggered by an AGN. In addition, AGNs are commonly found outside the Galactic plane, as shown in Appendix~\ref{sec:features_appendix}. The important metadata features that characterize AGNs are the Star/Galaxy score, or \texttt{sgscores} of the first, second and third closest source from PanSTARRS1 catalog which tend to have values closer to 0 (i.e., extended), since AGNs occur in the center of extended galaxies, and the distance of the first, second and third closest sources in PanSTARRS1 catalog, which should have \texttt{distpsnr1} values consistent with zero since the nearest source should be the AGN itself, combined with large \texttt{distpsnr2} and \texttt{distpsnr3} values due to the lower source density outside of the Galactic plane. The \texttt{classtar} is also useful as more weakly accreting AGN candidates tend to be classified as galaxy-shaped sources by the SExtractor classifier (\citealt{bertin_sextractor_1996}).
    
    \item \textbf{Supernovae (SNe)}: An alert generated by a SN should appear as a change in flux where no unresolved sources were present. These transients tend to appear near their host galaxies, and their location should be consistent with the underlying host stellar population distribution (e.g., a SN will have a higher probability of arising from a location aligned with the disk than perpendicular to it). As such, most SN detections exhibit a visible host galaxy in both the \textit{science} and \textit{reference} stamps, with the flux from the SN arising only in the \textit{science} and \textit{difference} images. SN candidates tend to appear outside the Galactic plane, and so the \texttt{sgscores}, \texttt{distpsnr}, and Galactic latitude features have similar distributions to AGN candidates. However, there are other features that might help to classify SN candidates correctly. For instance, the \texttt{chi} and \texttt{sharp}parameters from DAOPhot (\citealt{stetson_daophot_1987}), or chinr and sharpnr, of the nearest source in reference image PSF-catalog within 30 arcsec, have different distributions for the SN class, compared to the other classes (see Appendix~\ref{sec:features_appendix}). Furthermore,the \texttt{isdiffpos} value, which measures whether the candidate is positive or negative in the science minus reference subtraction,  should always be 1 for new SN candidates.
    
    \item \textbf{Variable Stars (VS)}: The flux coming from variable stars usually appears in both the \textit{reference} and \textit{science} stamps. With ZTF's sensitivity, variable stars can be detected within the Milky Way or the Local Group, and thus the alert will typically not be associated with a visible host galaxy in the stamp, but rather with other point-like sources. In addition, such alerts will have a higher probability of residing at lower Galactic latitudes and in crowded fields with multiple point sources within the stamps, given the high concentration of stars in the disk and bulge of our Galaxy. Therefore, VS candidates present a distribution of higher \texttt{sgscores}, lower \texttt{distpsnr} and Galactic latitude closer to 0 compared to AGN and SN candidates (see Figure~\ref{fig:features_grid}).

    \item \textbf{Asteroids}: Alerts from moving Solar-system objects will appear only one time at a given position, and thus will show flux only in the \textit{science} and \textit{difference} images. Depending on their distance and apparent speed, they may appear elongated in the direction of motion. In addition, such alerts should have a higher probability of residing at lower ecliptic latitudes as shown in Figure~\ref{fig:features_grid}. Also, new asteroid candidates should always have an \texttt{isdiffpos} feature equal to 1.
    
    \item \textbf{Bogus alerts}: Camera and telescope optics effects, such as saturated pixels at the centers of bright sources, bad columns, hot pixels, astrometric misalignment in the subtraction to compute the \textit{difference} image, unbaffled internal reflections, etc., can produce bogus alerts with no interesting real source. Bogus alerts are characterized by the presence of NaN pixels due to saturation, single or multiple bright pixels with little or no spatial extension (i.e., smaller than the telescope point spread function PSF and nightly seeing), or lines with high or low pixel values that extend over a large portion of the stamp (hot or cold columns/rows). We are currently working to include satellites in this class. However, they may share some image traits with asteroids, but are not confined to the ecliptic. According to the estimates shown in section \ref{sec:bogus_analysis} of the Appendix, bogus alerts comprise a big portion of the total of alerts generated each night by ZTF, with $\sim$20\% of all alerts being bogus, and a $\sim$60\% of them having a single detection. These estimates were carried out by applying the stamp classifier over 176,376 alerts generated by ZTF's stream. In the same section, we present a more thorough characterization and definition of the nine types of bogus we have found in ZTF's alert stream.
\end{itemize}

\begin{table}[h!] 
\begin{center}
\caption{List of metadata of the alert used as features by the classifier. The definitions are from the ZTF avro schemas$^{\text{a}}$.}
\label{table:features}
\footnotesize{
\begin{tabular}{>{\centering\arraybackslash}p{0.08\textwidth}|p{0.36\textwidth}}

\textbf{Feature}   & \textbf{Description} [units] \\ \hline \hline

\footnotesize{\texttt{sgscore\{1, 2, 3\}}} & \footnotesize{Star/Galaxy score of the \{first, second, third\} closest source from PanSTARRS1 catalog 0 $\leq$ \texttt{sgscore} $\leq$ 1 where a value closer to 1 implies higher likelihood of being a star, -999 when there is no source}. \\ \hline

\footnotesize{\texttt{distpsnr \{1, 2, 3\}}} & Distance of the \{first, second, third\} closest source from PanSTARRS1 catalog, if one exists within 30 arcsec, -999 if there is no source [arcsec]. \\ \hline

\texttt{isdiffpos} &  t (converted to 1) if the candidate is from positive (science minus reference) subtraction; f (converted to 0) if the candidate is from negative (reference minus science) subtraction. \\ \hline

\texttt{fwhm}  & Full Width Half Max assuming a Gaussian core of the alert candidate in the science image from SExtractor (\citealt{bertin_sextractor_1996}) [pixels]. \\ \hline

\texttt{magpsf}  & magnitude from PSF-fit photometry of the alert candidate in the difference image. [mag]. \\ \hline

\texttt{sigmapsf}  & 1-sigma uncertainty in \texttt{magpsf} [mag].
 \\ \hline

\texttt{ra}, \texttt{dec}  & Right ascension and declination of candidate; J2000 [deg]. \\ \hline

\texttt{diffmaglim}  & 5-sigma mag limit in difference image based on PSF-fit photometry [mag]. \\ \hline

\texttt{classtar}  & Star/Galaxy classification score of the alert candidate in the science image, from SExtractor. \\ \hline

\texttt{ndethist}  & Number of spatially-coincident detections falling within 1.5 arcsec going back to beginning of survey; only detections that fell on the same field and readout-channel ID where the input candidate was observed are counted. All raw detections down to a photometric Signal/Noise$\approx$3 are included. \\ \hline

\texttt{ncovhist} & Number of times input candidate position fell on any field and readout-channel going back to beginning of survey. \\ \hline

\texttt{chinr}, \texttt{sharpnr} & DAOPhot (\citealt{stetson_daophot_1987}) chi, sharp parameters of nearest source in reference image PSF-catalog within 30 arcsec. \\ \hline

Ecliptic coordinates & ecliptic latitude and longitude computed from the \texttt{ra}, \texttt{dec} coordinates of the candidate [deg]. \\ \hline

Galactic coordinates & Galactic latitude and longitude computed from the \texttt{ra}, \texttt{dec} coordinates of the candidate [deg]. \\ \hline

approx non-detections & \texttt{ncovhist} minus \texttt{ndethist}. Approximate number of observation in the position of the candidate, with a signal lower than Signal/Noise$\approx$3. \\ \hline

\end{tabular}}
\end{center}
\footnotesize{
$^{\text{a}}$ \url{https://zwickytransientfacility.github.io/ztf-avro-alert/}}

\end{table}

We built a training set of ZTF alerts using the labeled set from \cite{sanchezsaez_alert_2020}, which is a result of cross-matching with other catalogs, such as the ASAS-SN catalogue of variable stars (\citealt{jayasinghe_asas-sn_2018, jayasinghe_asas-sn_2019, jayasinghe_asas-sn_2019-1, jayasinghe_asas-sn_2020}),  the Roma-BZCAT Multi-Frequency Catalog of Blazars (\citealt{massaro_5th_2015}), the Million Quasars Catalog (version June 2019, \citealt{flesch_half_2015, flesch_million_2019}), the New Catalog of Type 1 AGNs (Oh2015; \citealt{oh_new_2015}), the Catalina Surveys Variable Star Catalogs (\citealt{drake_catalina_2014, drake_catalina_2017}), the LINEAR catalog of periodic light curves (\citealt{palaversa_exploring_2013}), {\em Gaia} Data Release 2 (\citealt{mowlavi_gaia_2018, rimoldini_gaia_2019}), the SIMBAD database (\citealt{wenger_simbad_2000}), and spectroscopically classified SNe from the TNS database. The asteroid subset was built by selecting the alerts that were near a Solar-system object, requiring that the \texttt{ssdistnr} field in the alert metadata exists.
Each sample corresponds to the triplet of \textit{science, reference}, and \textit{difference} images of the first detection. The number of samples of AGN, SN, VS, asteroid, and bogus are 14,966 (29\%), 1620 (3\%), 14,996 (29\%), 9899 (19\%), and 10,763 (20\%), respectively, with a total of 52,244 examples (undersampling the labeled set from \cite{sanchezsaez_alert_2020} for better balance between classes since 3\% SNe would not exactly be considered balanced compared to the rest). The bogus class was built in two steps: We first used \emph{step 1 bogus}, composed by 1980 bogus examples reported by ZTF (based on human inspection) and ran an initial iteration of the proposed classifier detailed in Section~\ref{sec:classifier}. Then, we added \emph{step 2 bogus}, where another 8783 bogus samples were labeled by our team of experts using the SN Hunter and added to the training set by manually inspecting the samples predicted by an early version of the model as SNe. 

Appendix \ref{sec:bogus_analysis} contains an analysis of bogus alerts present in the training set. Briefly, Figure~\ref{fig:bogus_type_distribution} shows the distribution of different types of bogus alerts in our labeled set, whereby an expert manually assigned type labels to a representative subset of 1,000 bogus samples. We stress here that bogus class generation is an ongoing process with different stages that involves labeling by hand. To highlight the current state, we made a 2D U-MAP projection of the bogus samples alongside SNe, differentiating both stages of a 2-step bogus labelling system. This projection, shown in Figure~\ref{fig:umap_projection_bogus}, groups alerts with similar triplet images as neighboring or adjacent points. Bogus alerts categorized by step 2 bogus are within a big cluster that mainly overlaps with SNe, which reflects the bias on how step 2 bogus alerts are selected samples that were confused with SNe by early versions of the stamp classifier. We continue adding new bogus alerts in this way. In Appendix \ref{sec:bogus_analysis}, we analyze in greater detail which type of bogus alerts are the most representative of each of the 3 clusters present in the U-MAP of Figure~\ref{fig:umap_projection_bogus}. 

One final point to stress is that a key aim of the stamp classifier is the fast detection of SNe, and therefore the training set consists only of the initial alert from each object, which allows us to estimate probabilities of objects as soon as we receive the alert.

\section{Methodology} \label{sec:methodology}
\subsection{Data Pre-Processing} \label{sec:pre-processing}

The standard shape for each stamp within an alert is 63$\times$63 pixels; 650 non-square shaped stamps were removed from the dataset. After removing misshaped stamps, we obtained 14,742 (29\%) AGN, 1596 (3\%) SN, 14,723 (29\%) VS, 9799 (19\%) asteroids and 10,734 (20\%) bogus alerts, with a total of 51,594 examples. Some pixels have NaN values due to pixel saturation, bad columns or stamps from the edges of the camera; all NaN pixels were replaced by a value of 0, giving information about NaNs content within the stamp to the classifier. Preliminary tests showed that smaller images for training led to better results, therefore we cropped all the stamps at the center getting 21$\times$21 pixels images; this size was selected by the hyperparameter random search discussed in Section~\ref{sec:expertiments}. Better results with a small stamp size may be explained by the fact that smaller stamps means a dimensionality reduction with respect to the original image size in the input of the CNN, and this may be easier to handle by the model. Further analysis of the optimal stamp size for the classification task at hand must be carried out since it might be important for the design of future alert stream based surveys. Each stamp was normalized independently between 0 and 1 by subtracting the minimum pixel value in the image, then dividing by the maximum pixel value. Finally, a 3-channel cube is assembled as input to the classifier, built by stacking the resulting \textit{science, reference} and \textit{difference} images as separate channels, resulting in a $21\times21\times3$ image. The metadata are clipped differently for each feature following the values in Table \ref{table:clipping_values}, then each feature is normalized by subtracting the mean value of the training set and dividing by the standard deviation.

\subsection{Classifier Architecture}\label{sec:classifier}

The classification model is a CNN based on the real-bogus classifier proposed by \cite{reyes_enhanced_2018}, which is an improvement over Deep-HiTS (\citealt{cabrera-vives_deep-hits_2017}) by adding rotational invariance to the CNN and analyzing the predictions of the model using Layer-wise Relevance Propagation (LRP;  \citealt{bach_pixel-wise_2015}). The specific CNN architecture used in this work is shown in Table~\ref{table:architecture}. In these previous works, metadata were not included for classification.

The input of the neural network has a shape of $21\times21\times3$ as explained in Section~\ref{sec:pre-processing}. Following the architecture of \cite{reyes_enhanced_2018}, a zero padding is applied to the input, to then augment the batch with rotated versions of itself as described in Section~\ref{sec:rotational_invariance}. For the convolutional layers, the parameters shown in Table~\ref{table:architecture} are the filter dimensions and number of output channels. All convolutional layers, except for the first one, have zero padding (filling the edges of the images with zeros) that preserves the input shape after the convolution. Moroever, all the convolutional layers and fully connected layers have a Rectified Linear Unit (ReLU; \citealt{nair_rectified_2010}) activation function (except for the last fully connected one that has a softmax output). The output of the last max-pooling layer, which reduces the dimensionality of the image by selecting the largest values of non-overlapping windows of 2$\times$2 pixels, is re-arranged (flattened) to a single dimension array for each sample in the batch, to feed the fully connected layers. Then, the rotation concatenation step takes place, stacking the fully connected output representation of the rotated versions of each sample, and passing them through the cyclic pooling layer, where an average is applied in the stacked dimension. The metadata features are first processed by a batch normalization layer that learns an optimal bias and scale to normalize the data. The normalized features are concatenated to the output of the cyclic pooling. The concatenated representation passes through two fully connected layers. Finally, a softmax function is applied to the output of the last fully connected layer to obtain the estimated probabilities for each of the five classes.
A glossary about CNNs and its training is presented in Appendix~\ref{sec:machine_learning_appendix}.

\begin{table}[h!]
\begin{center}
\caption{Convolutional neural network architecture.}
\label{table:architecture}
\begin{tabular}{>{\centering\arraybackslash}p{0.13\textwidth}|>{\centering\arraybackslash}p{0.14\textwidth}|>{\centering\arraybackslash}p{0.13\textwidth}}
Layer                 & Layer Parameters & Output Size          \\ \hline \hline
Input                 & -                & $21\times21\times3$  \\ \hline
Zero padding          & -                & $27\times27\times3$  \\ \hline
Rotation augmentation & -                & $27\times27\times3$  \\ \hline
Convolution           & $4\times4, 32$   & $24\times24\times32$ \\ \hline
Convolution           & $3\times3, 32$ & $24\times24\times32$\\ \hline
Max-pooling           & $2\times2$, stride 2 & $12\times12\times32$ \\ \hline
Convolution &$3\times3, 64$ &$12\times12\times64$ \\ \hline
Convolution &$3\times3, 64$&$12\times12\times64$ \\ \hline
Convolution &$3\times3, 64$&$12\times12\times64$ \\ \hline
Max-pooling &$2\times2$, stride 2, & $6\times6\times64$ \\  \hline
Flatten & - & 2304 \\ \hline
Fully connected & $2304\times64$ & $64$\\ \hline

Rotation concatenation & - & $4\times64$\\ \hline
Cyclic pooling & - & 64 \\ \hline
Concat with BN$^{\text{a}}$ features & - & 64 + 23 \\ \hline
Fully connected with dropout & $90\times64$ & $64$\\ \hline
Fully connected & $64\times64$ & $64$\\ \hline
Output softmax & $64\times5$ & 5 (n$^{\circ}$ classes) \\ \hline
\end{tabular}
\end{center}
\footnotesize{$^{\text{a}}$ BN stands for batch normalization}
\end{table}

\subsection{Rotational Invariance}\label{sec:rotational_invariance}

\begin{figure*}[htbp]
    \centering
    \includegraphics[width=0.95\textwidth]{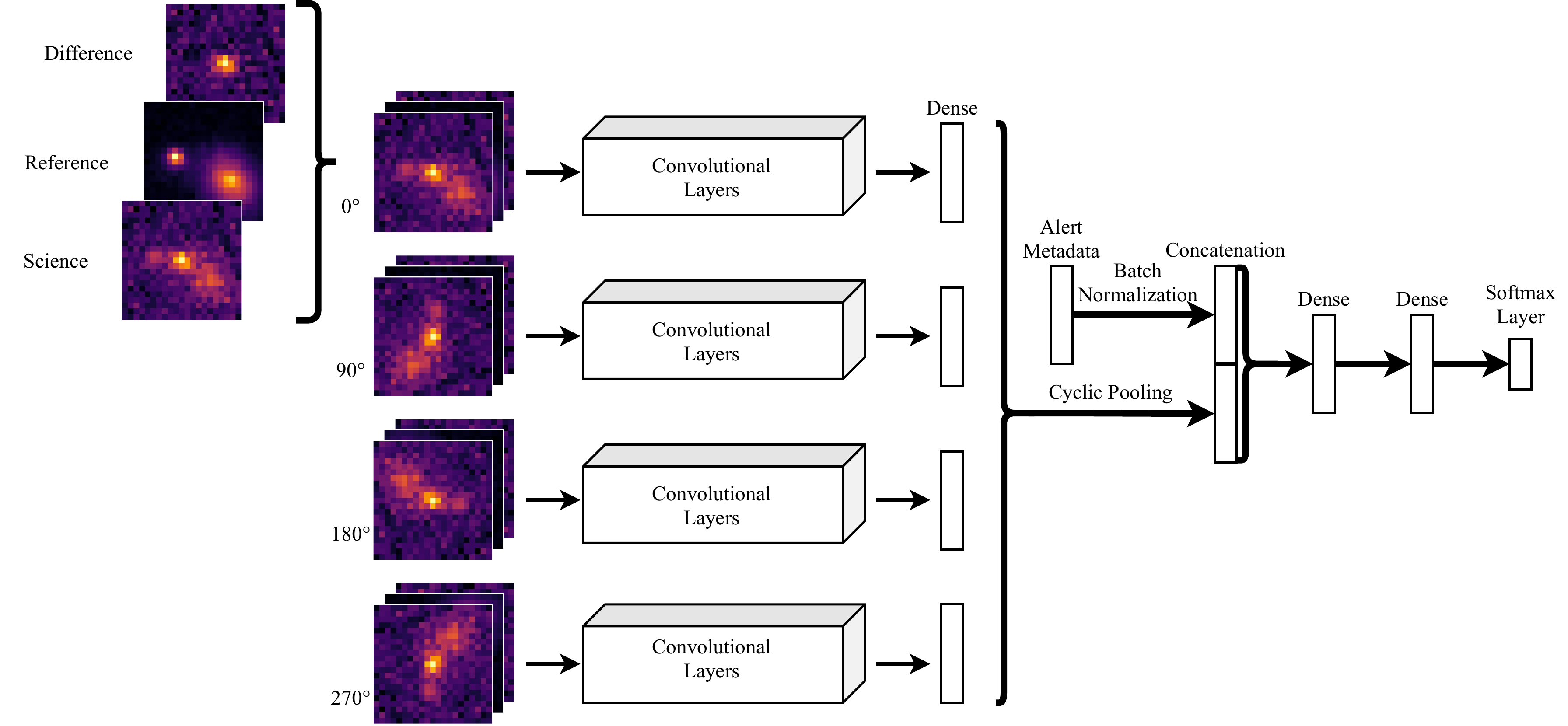}
    \caption{CNN enhanced with rotational invariance. The box \textit{Convolutional Layers} refers to those described in Table~\ref{table:architecture}, from the first convolutional layer to the last pooling layer. For each sample, the \textit{science, reference} and \textit{difference} images are concatenated in the channel dimension, obtaining an image input of dimension $21 \times 21 \times 3$. For each sample within the sampled batch, rotated versions are generated as described in Section~\ref{sec:classifier} and fed to the CNN. After the first dense layer, the Cyclic pooling is performed. The metadata features are passed through a batch normalization layer, and its output is concatenated with the cyclic pooling output. Then, the concatenation goes through 2 fully connected layers, and finally a softmax function is applied to estimate the output probabilities.}
    \label{fig:early_classifier}
\end{figure*}

Astronomical objects present within a stamp usually have a random orientation. It has been shown that imposing rotational invariance to a classifier improves its cumulative accuracy for some classification problems (e.g., \citealt{dieleman_rotation-invariant_2015, dieleman_exploiting_2016, cabrera-vives_deep-hits_2017, reyes_enhanced_2018}). In this work, rotational invariance is achieved by feeding the neural network with $90^{\circ}$, $180^{\circ}$ and $270^{\circ}$ rotated versions of the original input batch $x$. Defining $r$ as a $90^{\circ}$ rotation operation, then the samples within the extended batch will be $B(x) = [x, rx, r^{2}x, r^{3}x]$ after applying the rotations. At the last step of the architecture before the softmax output layer, a cyclic pooling operation is performed, which is basically an average pooling over the representation of the dense layer for each rotated example. 
A scheme of the procedure described in this section is shown in Figure~\ref{fig:early_classifier}.

\subsection{Entropy Regularization}\label{sec:entropy}

When the CNN model is trained using cross-entropy as the loss function to be minimized, the classification confidence of the model is very high, resulting in a distribution of output probabilities with saturated values of 0s and 1s without populating the values in between, even for wrong classifications. In this case there is no insight of certainty (relative probabilities between classes) of the prediction because most estimated probabilities for each class were either 0 or 1. In order to provide more granularity to the astronomers, who revise SN candidates based on the probability of the classification reported by the model to later request follow-up observing time, we added the entropy of the predicted probabilities of the models as a regularization term, to be maximized during training (\citealt{pereyra_regularizing_2017}). By maximizing the entropy of the output probabilities, we penalize predictions with high confidence, in order to get better insight in cases where the stamps seem equally likely to belong to more than one class. The loss function $\mathcal{L}$ per sample is as follows:
\begin{equation}\label{eq:loss_function}
    \mathcal{L} = \underbrace{-\sum_{c=1}^{N}y_{c}\log{(\hat{y}_{c})}}_{\text{cross-entropy}} + \underbrace{\beta \sum_{c=1}^{N}\hat{y}_{c}\log{(\hat{y}_{c})}}_{\text{entropy regularization}},
\end{equation}
where $N$ is the number of classes, $y_{c}$ is the one-hot encoding label (a value of 1 in the corresponding index of class, and 0 for the rest)  indexed by $c$, $\hat{y}_{c}$ is the model prediction for class $c$, and $\beta$ controls the regularization term in the loss function. Further explanation on the role of the loss function in the training process of a neural network is given in Appendix \ref{sec:machine_learning_appendix}.

\subsection{Experiments}\label{sec:expertiments}

A hyperparameter search was done by randomly sampling 133 combinations of the parameters shown in Table~\ref{table:hyperparam_search}. For each combination of hyperparameters, we trained 5 networks with different initial random weights. The initial maximum number of iterations (presenting a single batch per iteration) was 30,000, evaluating the loss in the validation set every 10 iterations to save the best model thus far. After the first 20,000 iterations, if a lower loss is found on the validation set, 10,000 more iterations are performed. The validation and testing subsets were sampled randomly only once, taking 100 samples per class from the whole dataset, obtaining 500 samples for each of the mentioned subsets. The remaining samples were used in the training set. For each training iteration, the batch was built to contain roughly the same number of samples per class. We used Adam (\citealt{kingma_adam_2017}) as the updating rule for the network parameters during training, with $\beta_{1}=0.5$ and $\beta_{2}=0.9$. Further details on the updating rules of a neural network and the Adam optimizer are described in Appendix~\ref{sec:machine_learning_appendix}.

\begin{table}[h!]
\begin{center}
\caption{Hyperparameter random search values.}
\label{table:hyperparam_search}
\begin{tabular}{c|c}
Hyperparameter           & Random Search Values                  \\ \hline \hline
Learning rate            & $5e\text{-}3, 1e\text{-}3, 5e\text{-}4, 1e\text{-}4
, 5e\text{-}5$ 
\\ \hline
Regularization parameter ($\beta$) & $0, 0.3, 0.5, 0.8, 1.0$        \\ \hline
Batch size               & $16, 32, 64$                   \\ \hline
Image size               & $21, 41, 63$                   \\ \hline
Dropout rate             & $0.2, 0.5, 0.8$                \\ \hline
CNN kernel size          & $3, 5, 7$                      \\ \hline
\end{tabular}
\end{center}
\end{table}

To account for the relevance of each part of the input, which is comprised of the three images and metadata features, we trained several versions of the stamp classifier, each of them with a different combination of images and metadata features. First, we trained the stamp classifier using combinations of the three images (without features). Second, we trained a random forest (\citealt{breiman_random_2001}) to classify our training set but using the features only, in order to obtain a feature importance ranking. Once we got the feature importance, we trained different stamp classifier models (with the three images) by adding one feature at a time, from the most important to the least important according to the ranking, and measured the accuracy for the corresponding model with the aggregated feature. For each of these models, we trained the model 5 times to account for variance due to random initialization parameters.

\section{Results} \label{sec:results}

In this section, we first describe our results in terms of the classification task for the five classes. Then, we change our focus to the detection of SN candidates, since our main interest in this work is to discover extremely young transient candidates to be observed with follow-up resources. Further applications of this early classification system might include rapid detection of extreme variability in AGN or tracking solar system objects.

The following results correspond to the best model (including metadata) in the search for hyperparameters, which adopts a batch size of 64 samples, learning rate of $1e\text{-}3$, dropout rate of 0.5, CNN kernel size of 5, image size of $21\times21$ pixels, and regularization parameter of $\beta=0.5$. Appendix~\ref{sec:random_search_appendix} contains the details on how this model was selected.
    We use accuracy\footnote{$\text{accuracy} = \frac{\text{N}^{\circ} \text{ correct classifications}}{\text{Total N}^{\circ} \text{ of samples}}$} to compare models since the validation and test sets are balanced; achieving $0.95 \pm 0.005$ in the validation set and $0.941 \pm 0.004$ in the test set. 

\begin{figure}[htbp]
    \centering
    \includegraphics[width=0.47\textwidth]{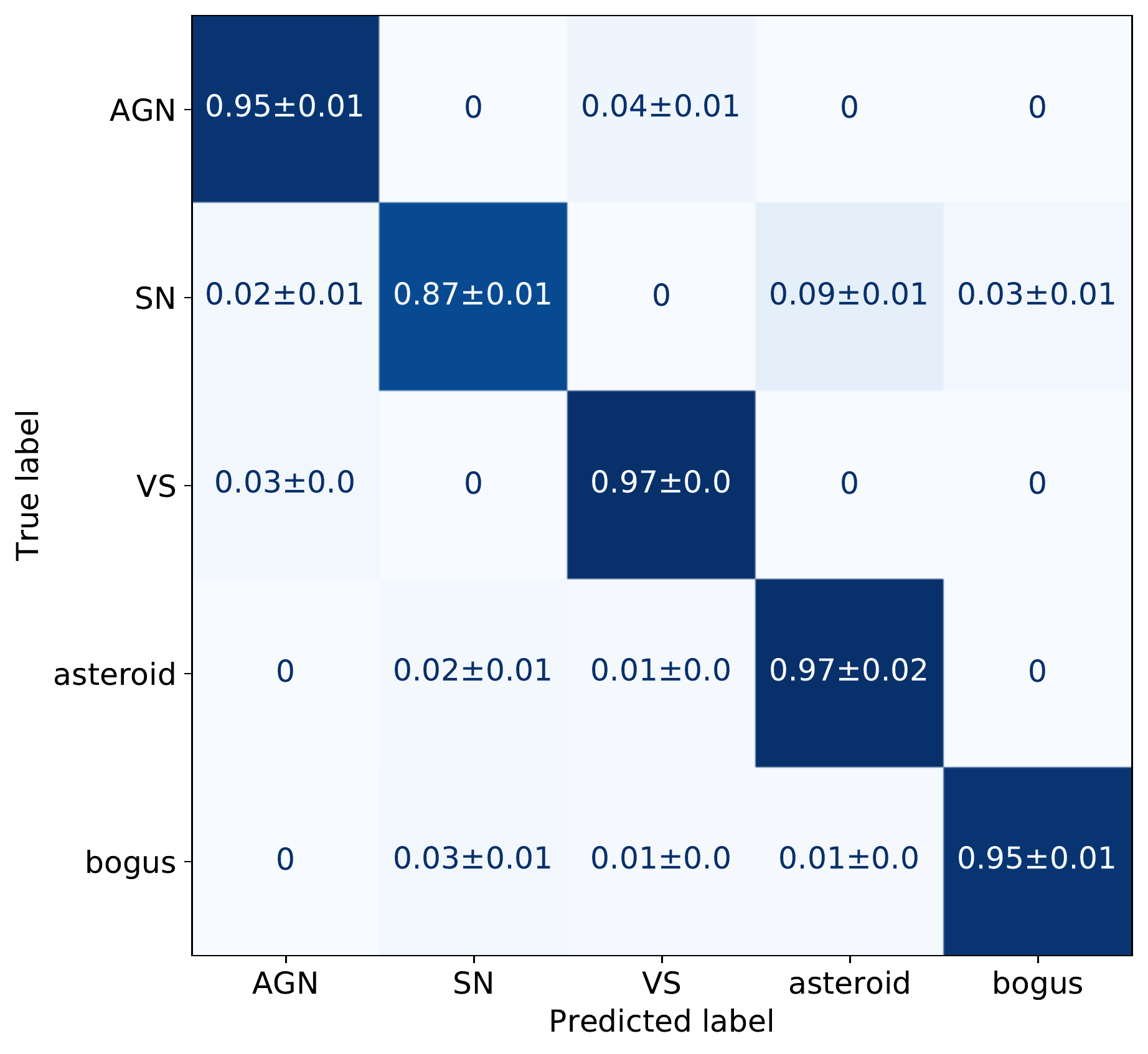}
    \caption{Average confusion matrix for the test set using 5 different realizations of the stamp classifier with metadata.}
    \label{fig:cm_test}
\end{figure}

Figure~\ref{fig:cm_test} shows the confusion matrix for the test set consisting of using five realizations of the proposed model. With our five class model, we recover $87 \pm 1\%$ of the SNe, with only $5 \pm 2\%$ of false positives. For completeness, we also report the confusion matrix of the stamp classifier when no metadata features are included in the fully connected layers (see Figure~\ref{fig:cm_test_no_feat}), which has a test-set accuracy of $0.883 \pm 0.006$, recovering $80 \pm 2\%$ of the SNe in the test set, with $10 \pm 4\%$ of false positives.

\begin{figure*}[htbp]
\gridline{\fig{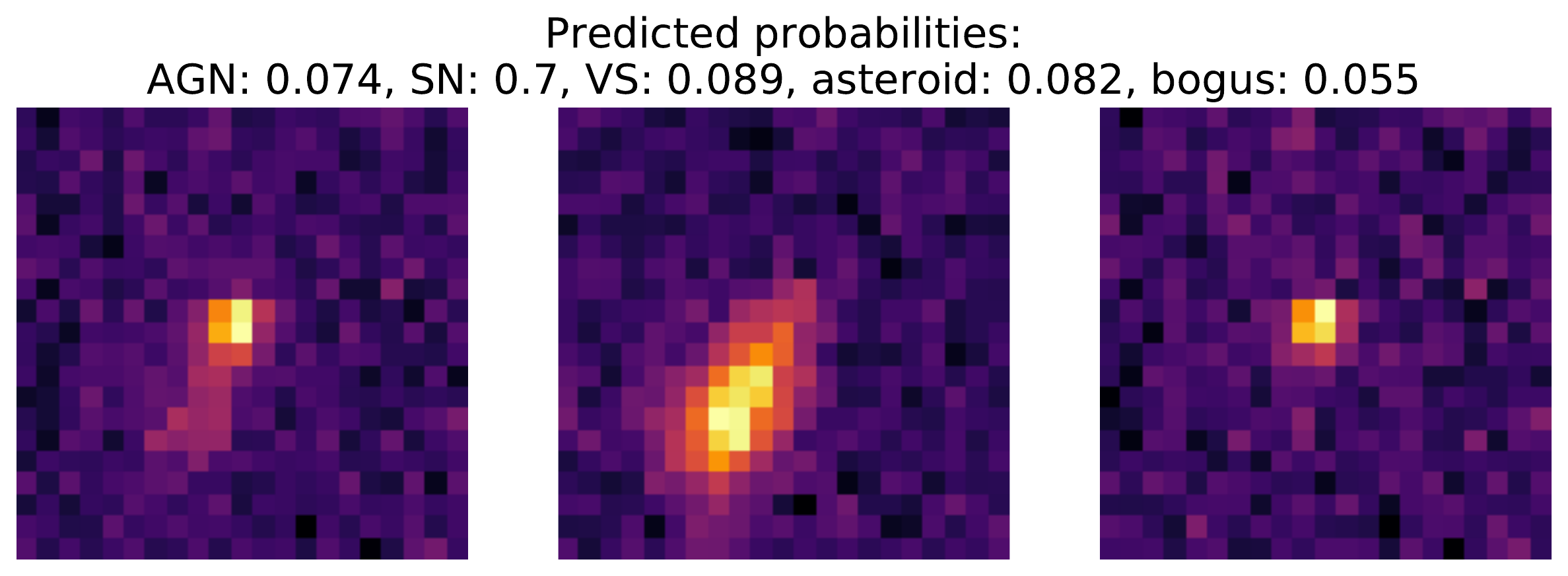}{0.45\textwidth}{(a) Object ID: ZTF19abfdsbu}
          \fig{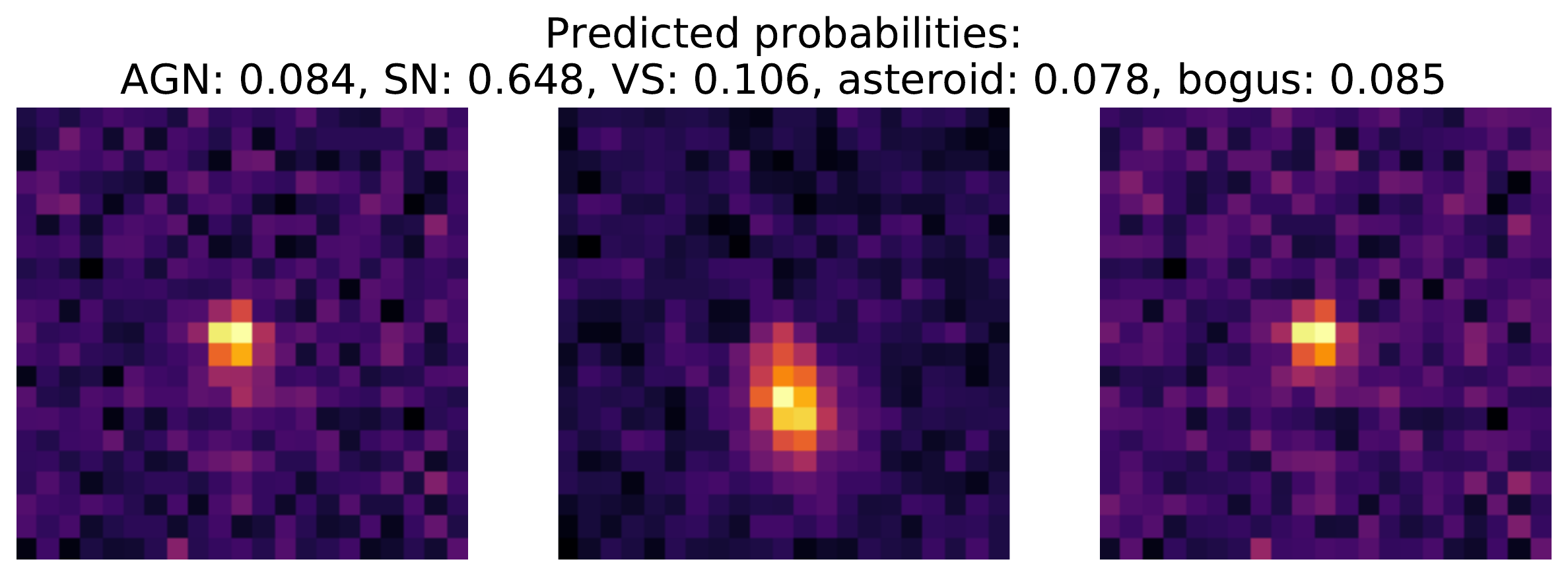}{0.45\textwidth}{(b) Object ID: ZTF18acrdwcf}}
\gridline{\fig{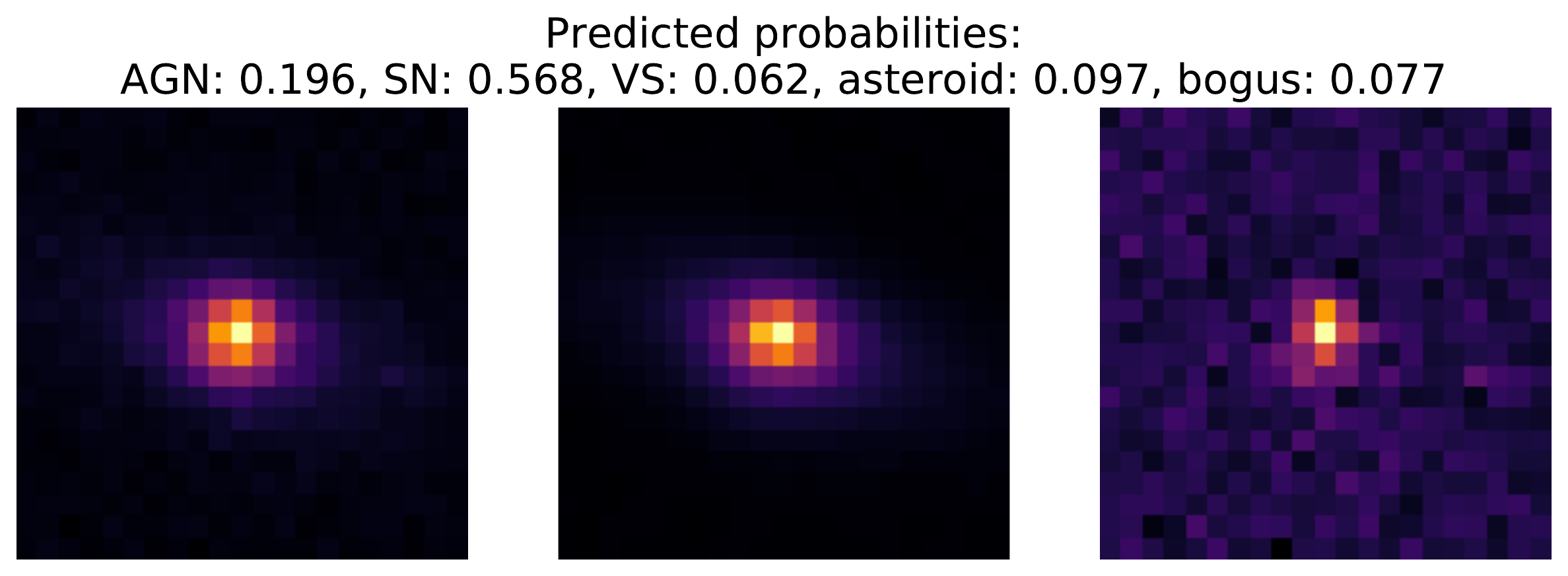}{0.45\textwidth}{(c) Object ID: ZTF18abuhzfc}
          \fig{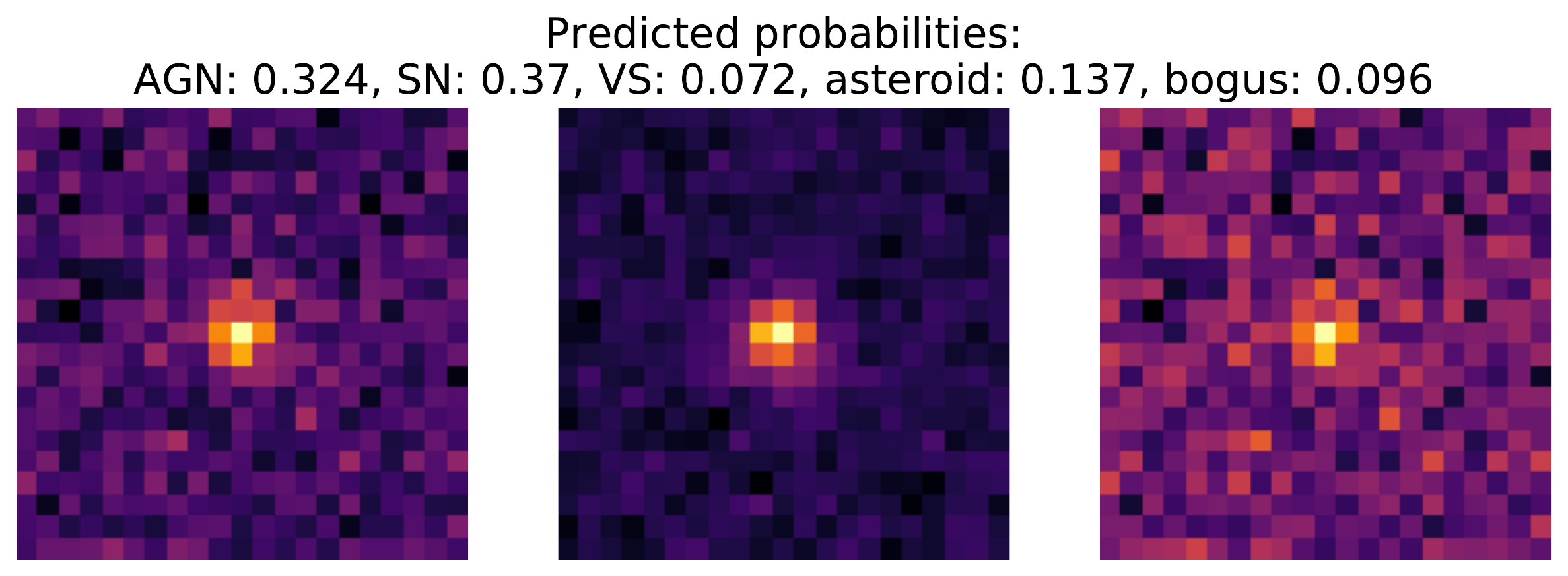}{0.45\textwidth}{(d) Object ID: ZTF19abrirdm}}
\caption{Correctly classified SN examples, with their respective predicted probabilities according to the proposed model. Panels (a) and (b) show typical examples of well-classified SNe, where the presence of a host galaxy within the stamps increases the chances of a SN alert being triggered. Panels (c) and (d) show small confusions between SN and AGN, due to the spatial coincidence of the transient with the center of the host galaxy. \label{fig:correct_classifications}}
\end{figure*}

By inspecting the predictions made by our model for each SN sample in the test set, we found that the results are in agreement with our initial expectations regarding the class discrimination described in Section~\ref{sec:data_description}, and the characteristics presented within the three stamps for each sample. Figure~\ref{fig:correct_classifications} shows SNe examples from TNS that have been correctly classified by our model, where in most cases a host galaxy is present, which is a good indicator of an alert triggered by a SN. In the examples shown in Figures~\ref{fig:correct_classifications}c and \ref{fig:correct_classifications}d, the second most likely class is AGN, due to the spatial coincidence of the transient with the center of the host galaxy.

\begin{figure*}[htbp]
\gridline{\fig{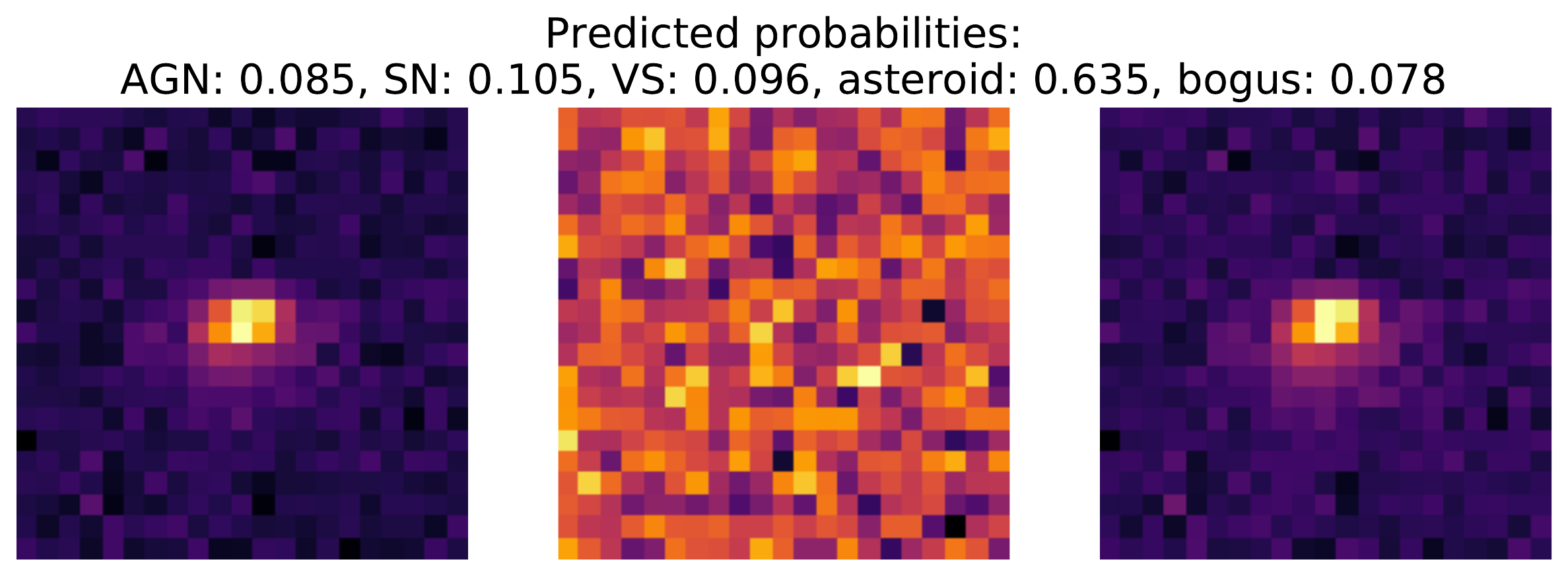}{0.45\textwidth}{(a) Object ID: ZTF18absoomk}
          \fig{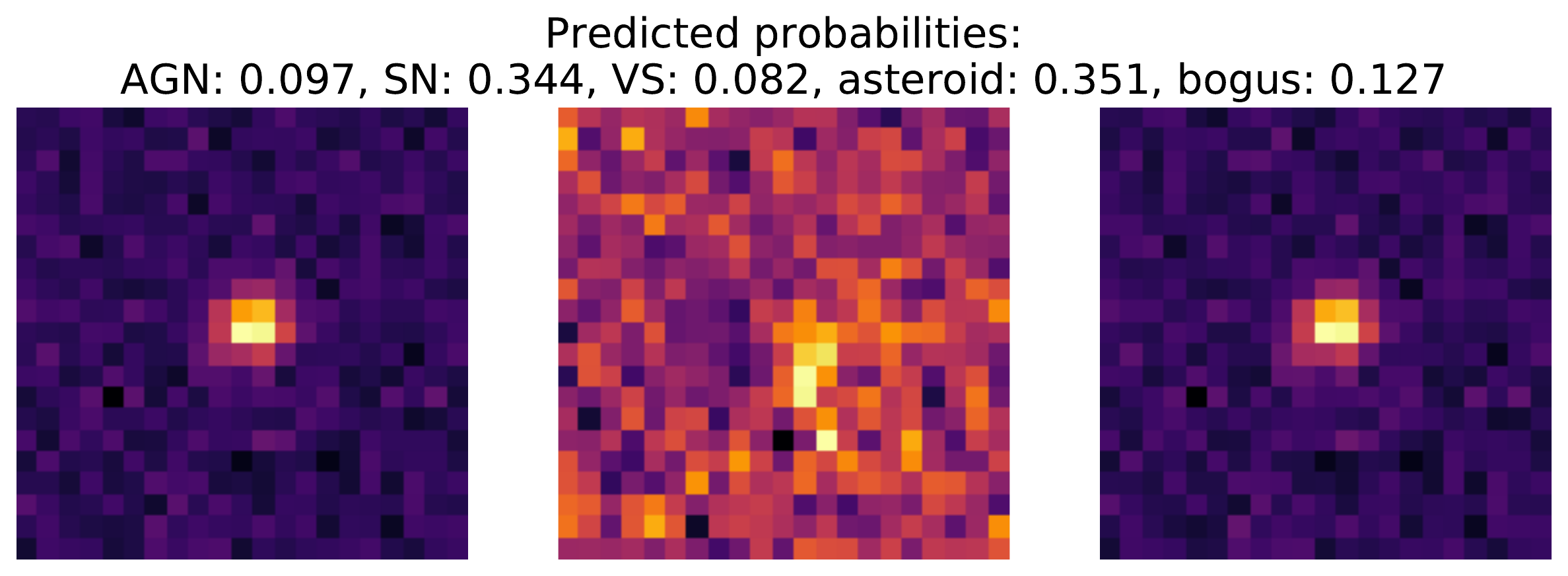}{0.45\textwidth}{(b) Object ID: ZTF19abmqasg}}
\gridline{\fig{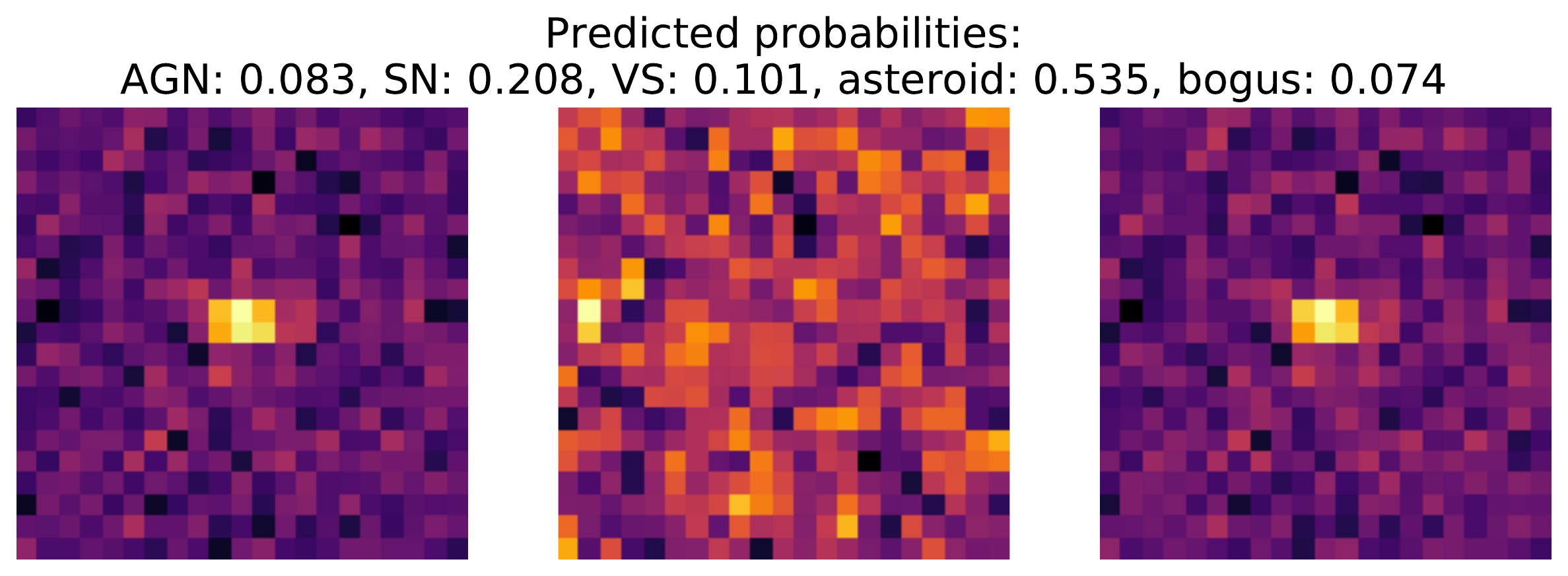}{0.45\textwidth}{(c) Object ID: ZTF19aazlsfj}
          \fig{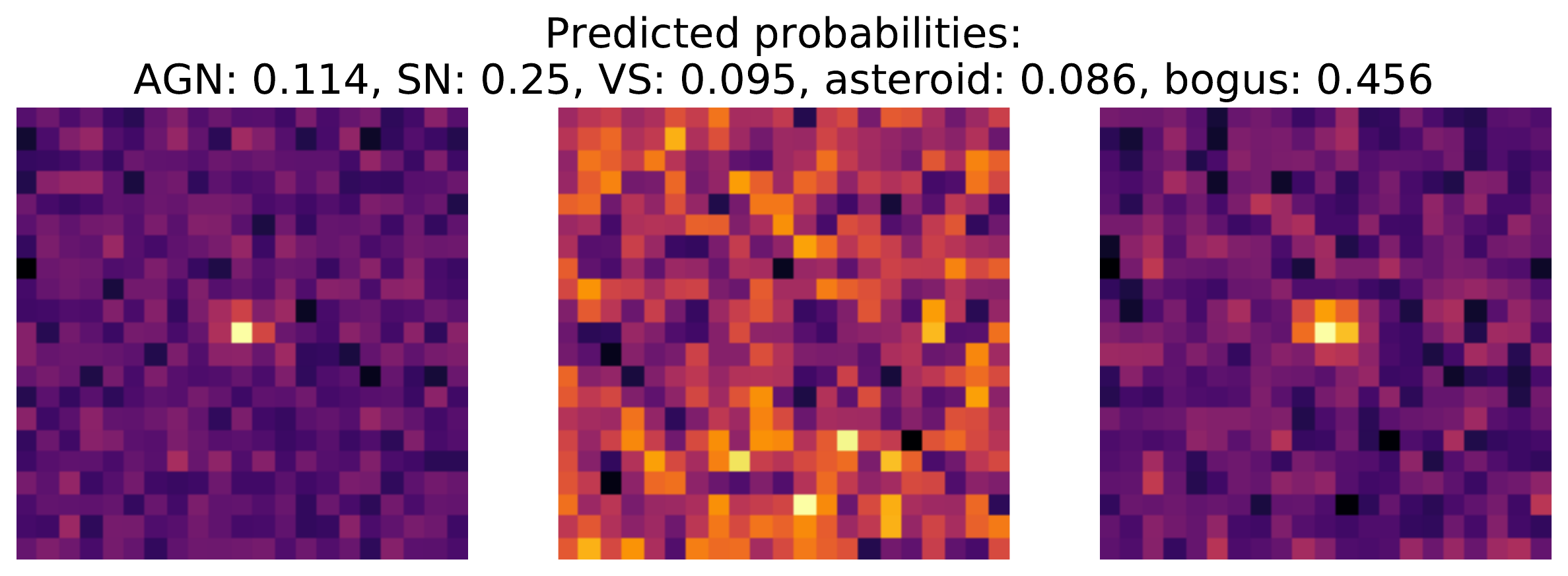}{0.45\textwidth}{(d) Object ID: ZTF19abpbvsk}}
\caption{Incorrectly classified SN examples, with their respective predicted probabilities by the proposed model. In Panels (a), (b) and (c), the SNe are classified as asteroids. The SN in panel (d) is classified as a bogus alert, which might be caused by the small size of the PSF, confusing the classifier with a hot pixel or a cosmic ray, which usually occupies a very narrow portion of the stamp at the center. In all cases, the absence of a clear host-galaxy within the stamps reduces the probability of a SN alert being triggered. \label{fig:incorrect_classifications}}
\end{figure*}

In Figure~\ref{fig:incorrect_classifications}, incorrectly classified examples are shown. The examples in Figures~\ref{fig:incorrect_classifications}a, \ref{fig:incorrect_classifications}b and \ref{fig:incorrect_classifications}c are SNe from TNS classified as Asteroids by our model. The absence of a host galaxy in these cases reduces the probability of an alert to be triggered by a SN. In the samples shown in Figures~\ref{fig:incorrect_classifications}d, the confusion occurs between SN and bogus alerts. In this case, the confusion is likely due to the small size of the point spread function (PSF) for some observations, which leads to confusion between true variables and hot pixels or cosmic rays (these often appear as single or a few adjacent bright pixels on the image).

It is worth highlighting again that the results of our model are achieved using the first alert only. According to the confusion matrix, the most probable missclassification for SN candidates are asteroid and bogus classes. This confusion between SN, asteroids, and bogus could be fixed by looking at the second alert of the same object. If the second alert exists, it is safe to discard the bogus and asteroid classes, since it is extremely unlikely that the same bogus error or a moving object will appear in the exact same location in consecutive images, unless the alert is near a bright star that produces pixel errors due to saturation.

\begin{figure*}[htbp]
    \centering
    \includegraphics[width=1\textwidth]{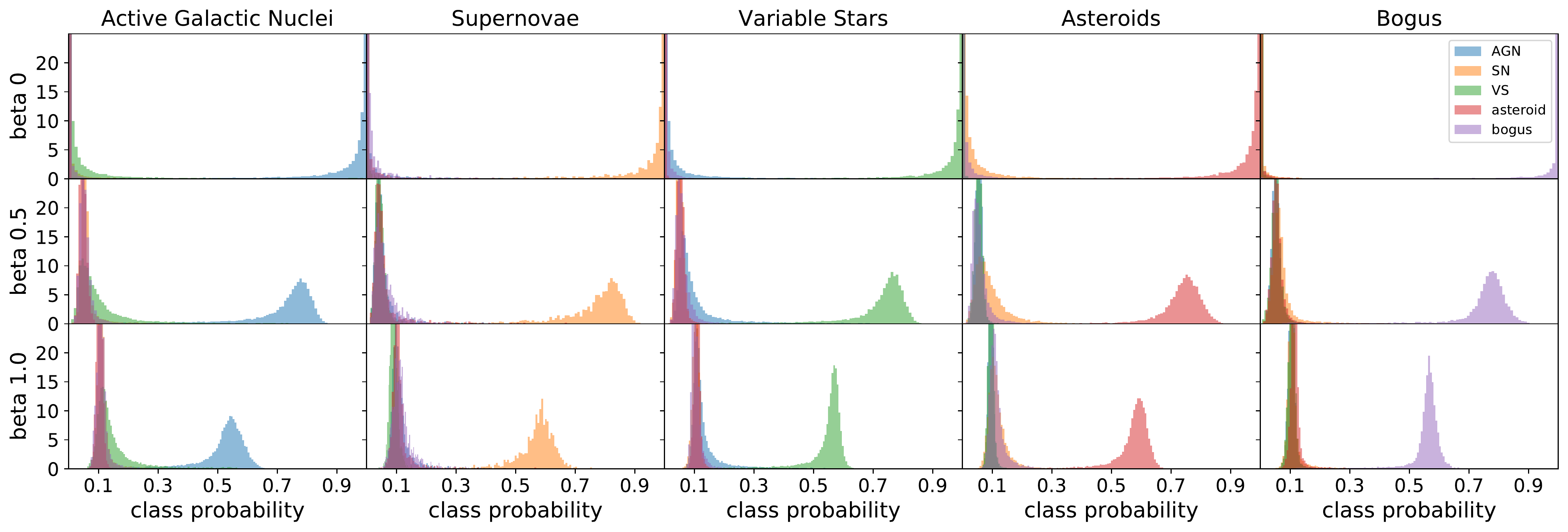}
    \caption{Probability distribution for each of the classes in the training set, for different values of the regularization constant $\beta=\{ 0, 0.5, 1.0 \}$. For the model without regularization ($\beta = 0$ shown on the top plot), the probability distribution saturates to 1 or 0. Increasing $\beta$ to 0.5 or 1.0 decreases the saturation and spreads the distribution of predictions made by the model (mid and bottom plots).}
    \label{fig:regularization_effect}
\end{figure*}

An example of the effect of the regularization term discussed in Section \ref{sec:entropy} is depicted in Figure~\ref{fig:regularization_effect}. Considerable differences in the distribution of the predicted probability for each class can be observed by varying $\beta$ between 0 and 1, since both terms in eq.~\ref{eq:loss_function} are expected values of log probabilities. In the case of $\beta=0$, the predictions are mostly saturated around 0 or 1 for the SN, VS, Asteroids and Bogus alert classes, creating difficulties to identify stamps that seem equally likely to belong to more than one class, because every sample is mapped to similar levels of high certainty. As the value of $\beta$ increases, the saturation of predicted values decreases, spreading the predicted probability distributions and emphasizing the different levels of certainty between predictions of different samples. The order of predicted probabilities for each sample does not change significantly by varying $\beta$, achieving 99\% of accuracy in the test set by checking whether the correct label lies in the highest two predicted probabilities for different $\beta$. The use of regularization to find noticeable differences in the predicted probabilities could be helpful to an expert for evaluating the output of the classifier, gaining better insight into how reliable the classifications are.

\begin{figure*}[htbp]
    \centering
    \includegraphics[width=1\textwidth]{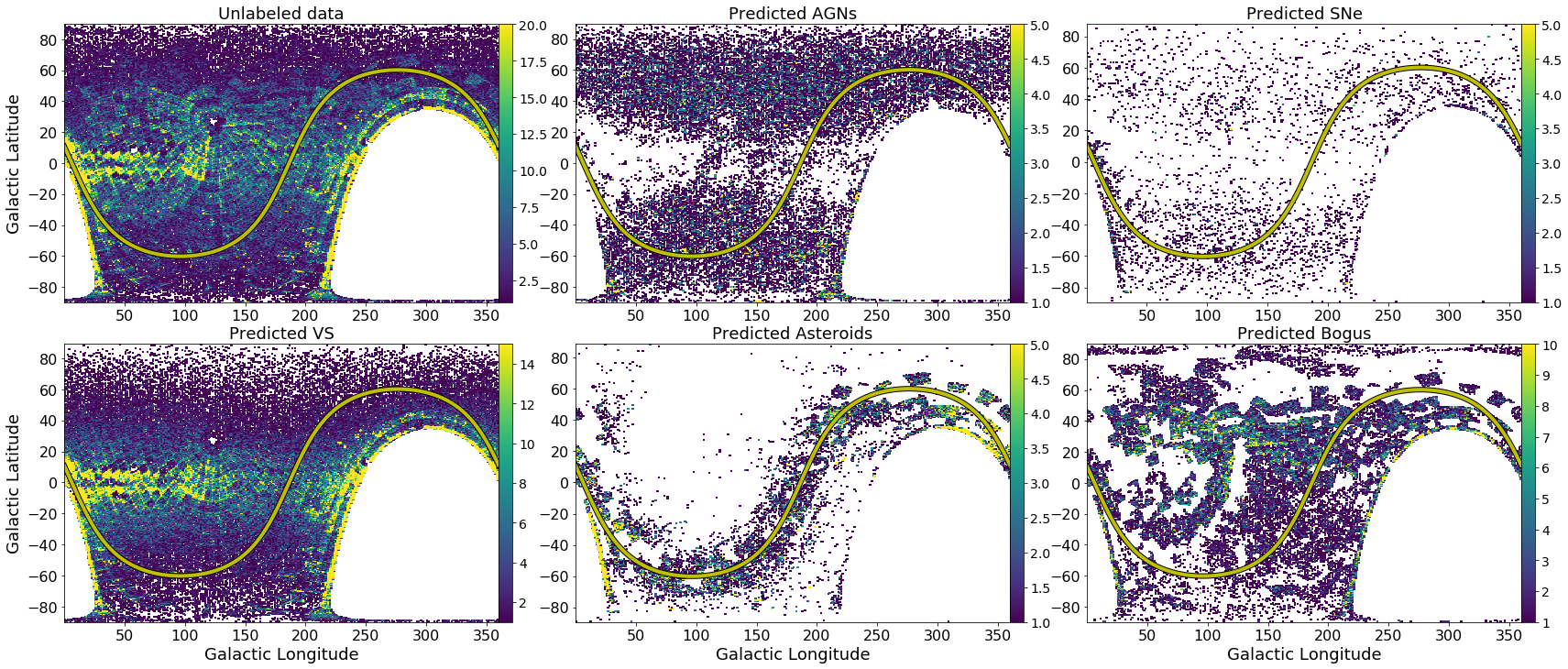}
    \caption{Spatial distribution for the unlabeled data, and distribution of predictions per class. The colorbar indicates the density of points. The ecliptic is shown with a yellow line with black edges. The distributions are shown as a 2d histogram of density of alerts. Extragalactic sources (SNe and AGNs) are found outside the Galactic plane. On the contrary, VS are concentrated in the Galactic plane. Asteroids are near the ecliptic.}
    \label{fig:space_distribution}
\end{figure*}

As a consistency check, we predicted the classes of unlabeled candidates using the stamp classifier, in order to compare their spatial distribution to the expected spatial locations for each class as mentioned in Section~\ref{sec:data_description}. To gather the unlabeled candidates, we queried objects using the ALeRCE API\footnote{\url{https://alerceapi.readthedocs.io/en/latest/}} by selecting 390,498 first alerts of different objects, chosen to be uniformly distributed over the full sky coverage of ZTF, where 325,582 of the alerts come from objects with $>1$ alert (SNe, AGN, and VS) and 64,916 come from objects with only 1 alert to have a better representation of asteroids and bogus candidates. Figure~\ref{fig:space_distribution} shows the spatial distribution of the predictions made by our model over the unlabeled data. As expected,due to extinction of extragalactic sources (SNe and AGNs) in the Galactic plane, the spatial distribution for these sources has a lower density of predicted candidates at low Galactic latitudes. % FROM BOTTOM TO HERE FRANZ COMMENTS WERE REVIEWED, EXCEPT FOR ADDING TEXT TO LONELY FIGURES IN APPENDICES: additional results - EXPLORING THE RELATIONSHIP BETWEEN FEATURES AND CLASSES 
On the contrary, the spatial distribution of VS candidates is more concentrated toward the Galactic plane. In the case of asteroids, these are found near the ecliptic. It is also possible to see a slight trend of predicted SNe near the ecliptic due to the confusion with asteroids class when there is no apparent host galaxy in the stamp, as shown in Figure~\ref{fig:incorrect_classifications}. In addition, we show in Figure~\ref{fig:space_distr_no_feat} in Appendix \ref{sec:additional_results}, the distribution for the same unlabeled data classified by the CNN without including features. It is noticeable the presence of predicted extragalactic objects (SNe and AGNs) within the Galactic plane, and a higher density of predicted asteroids far from the ecliptic. Even though the images alone have important information to classify the five classes, the metadata features are essential to improve the accuracy of the classifier in the labeled dataset as shown in the Appendix~\ref{sec:additional_results}, in addition to obtain the expected spatial distribution for each class.

We further extended our analysis by comparing the predictions of our model in the unlabeled dataset, with the ones made by the feature based light-curve classifier from \cite{sanchezsaez_alert_2020}, which is able to classify a finer taxonomy of objects, but it requires at least 6 detections in one of the two bands. For more details, see \cite{forster_automatic_2020}, Section 3.6. Here we summarize the important results. The stamp classifier predictions strongly agree with the ones of the light curve classifier. The stamp classifier finds 78\% of the SN classified by the light curve classifier, 85\% for AGNs and 96\% for VS. The main confusions in the SN class (false positives) are 9\% of AGNs, 6\% of VS, 4\% of asteroids and 3\% of bogus. The false positives of AGNs are 4\% of SN and 1\% of VS, and the confusion of VS is only 3\% with AGNs. Classifications of objects in the unlabeled set comparing the light curve classifier and stamp classifier are shown in Table \ref{table:lc_comparisonn}. To further account for the performance of the stamp classifier in the asteroids and bogus classes, we randomly selected 20000 asteroids and 20000 bogus objects predicted by the stamp classifier. The proportion of objects with a single detection as of February 2021 is 98\% and 96\%, for asteroids and bogus respectively for which we expect a single detection (except for a small proportion of bogus).

To understand how the stamp classifier performs in different conditions, we computed the probability assigned to alerts of each class in the training set as a function of the specific value of each of the features given to the classifier; these probabilities are shown in Figures~\ref{fig:model_probabilities_part1} and \ref{fig:model_probabilities_part2} in Appendix \ref{sec:additional_results}. Here we can inspect for which feature values does the probability assigned to the correct class decreases, showing the classifier performance in different regimes. In what follows, we remark some examples that support our hypotheses about class separability according to features mentioned in Section~\ref{sec:data_description}. For example, variable stars with low \texttt{sgscore1} are less likely to be classified correctly, as well as variable stars with higher \texttt{distpsnr} (Figures \ref{fig:model_probabilities_part1}b, \ref{fig:model_probabilities_part1}d and \ref{fig:model_probabilities_part1}f), since we anticipate that variable stars are in higher stellar density regions (i.e., in the plane or bulge of the Milky Way). In the case of photometry measurements, variable stars have a lower probability of being correctly classified at higher magnitudes, while AGN have a higher probability of getting correctly classified. The inverse is true for \texttt{sigmapsf} (Figure~\ref{fig:model_probabilities_part1}i and \ref{fig:model_probabilities_part1}j respectively), which is probably due to the higher error in the estimated magnitude owing to the host galaxy in the case of AGNs, which is not present in variable stars. We can see that the assigned probability to asteroids decays far from the ecliptic in Figure~\ref{fig:model_probabilities_part2}e in line with their known distribution, while the probability assigned to AGN is lower near the Galactic plane in Figure~\ref{fig:model_probabilities_part2}g due to dust extinction. In Figure~\ref{fig:model_probabilities_part2}l the probability assigned to each class is shown for different values of SNR. Higher values of SNR means less probability of being an AGN, since these objects are distant and are usually found within a host galaxy, the SNR for the AGN source is smaller. Note that the SNR is not added as a feature to the classifier, but it is encoded in the photometry features \texttt{magpsf} and \texttt{sigmapsf} from which we computed the SNR.

\begin{figure}[h]
    \centering
    \includegraphics[width=0.49\textwidth]{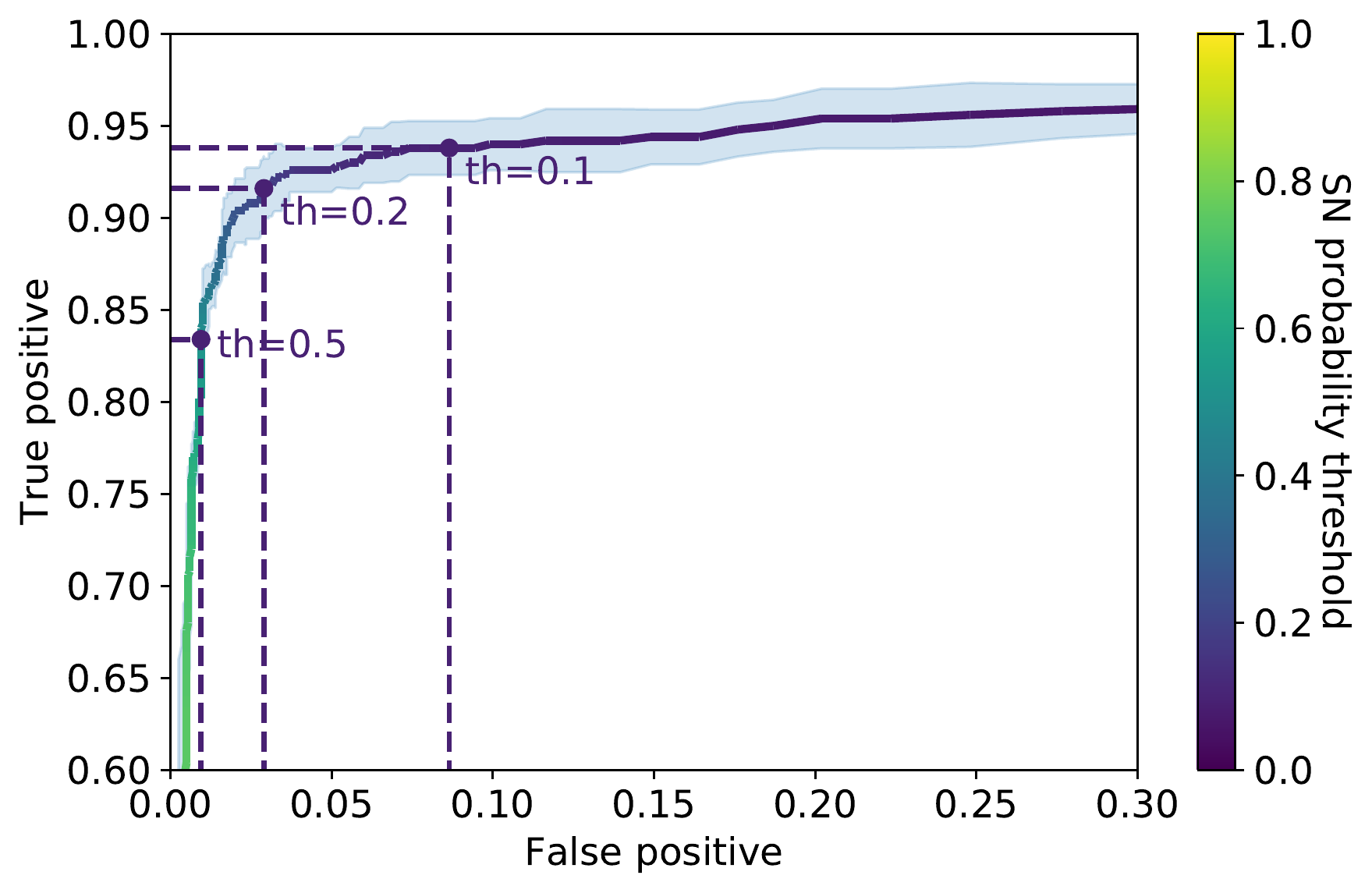}
    \caption{ROC curve with SN detection threshold. The colorbar shows the threshold that a sample's predicted SN probability must surpass in order to be assigned as a member of the SN class. For a SN probability threshold of 0.1, 0.2 and 0.5, the false positive ratio is 0.87, 0.03 and 0.01, respectively, while the true positive ratio is 0.94, 0.92 and 0.83, respectively.}
    \label{fig:roc_curve_th}
\end{figure}

We use the stamp classifier on a daily basis to filter suitable SN candidates to report for follow-up. The filtered candidates are inspected by experts\footnote{For a full list of reporters, please check \href{https://www.wis-tns.org/search?&discovered_period_value=1&discovered_period_units=months&unclassified_at=0&classified_sne=0&include_frb=0&name=&name_like=0&isTNS_AT=all&public=all&ra=&decl=&radius=&coords_unit=arcsec&reporting_groupid\%5B\%5D=74&groupid\%5B\%5D=null&classifier_groupid\%5B\%5D=null&objtype\%5B\%5D=null&at_type\%5B\%5D=null&date_start\%5Bdate\%5D=&date_end\%5Bdate\%5D=&discovery_mag_min=&discovery_mag_max=&internal_name=&discoverer=&classifier=&spectra_count=&redshift_min=&redshift_max=&hostname=&ext_catid=&ra_range_min=&ra_range_max=&decl_range_min=&decl_range_max=&discovery_instrument\%5B\%5D=null&classification_instrument\%5B\%5D=null&associated_groups\%5B\%5D=null&official_discovery=0&official_classification=0&at_rep_remarks=&class_rep_remarks=&frb_repeat=all&frb_repeater_of_objid=&frb_measured_redshift=0&frb_dm_range_min=&frb_dm_range_max=&frb_rm_range_min=&frb_rm_range_max=&frb_snr_range_min=&frb_snr_range_max=&frb_flux_range_min=&frb_flux_range_max=&num_page=50&display\%5Bredshift\%5D=1&display\%5Bhostname\%5D=1&display\%5Bhost_redshift\%5D=1&display\%5Bsource_group_name\%5D=1&display\%5Bclassifying_source_group_name\%5D=1&display\%5Bdiscovering_instrument_name\%5D=0&display\%5Bclassifing_instrument_name\%5D=0&display\%5Bprograms_name\%5D=0&display\%5Binternal_name\%5D=1&display\%5BisTNS_AT\%5D=0&display\%5Bpublic\%5D=1&display\%5Bend_pop_period\%5D=0&display\%5Bspectra_count\%5D=1&display\%5Bdiscoverymag\%5D=1&display\%5Bdiscmagfilter\%5D=1&display\%5Bdiscoverydate\%5D=1&display\%5Bdiscoverer\%5D=1&display\%5Bremarks\%5D=0&display\%5Bsources\%5D=0&display\%5Bbibcode\%5D=0&display\%5Bext_catalogs\%5D=0}{the list of reported objects by ALeRCE in TNS}, such as \url{https://www.wis-tns.org/object/2021mfa}.}  %(F. F\"orster, F. E. Bauer, G. Pignata, J. Silva-Farf\'an, E. Camacho-I\~{n}iguez, L. Galbany, \textbf{A. Mouras, A. Maurao, A. Muñoz-Arancibia}) 
to choose the most reliable candidates to report among the ones indicated by the classifier. Therefore, it is important to control the false positive ratio and the amount of classified SNe events. To understand this trade-off, we computed the Receiver Operating Characteristic (ROC) curves depicted in Figure~\ref{fig:roc_curve_th}. To build the ROC curve, we converted the classification problem into a detection problem by making a binary classification between SN vs. the rest of the classes (AGN, VS, asteroids and bogus alerts). Using the predicted probabilities in the test set of each alert being a SN, we varied the threshold value (minimum probability) necessary to assign the SN class to an alert and change the operation regime of the model. By choosing a high SN probability threshold, the false positive ratio can be reduced in order to decrease the number of false candidates in the list for inspection by experts, while keeping a high true positive ratio. For instance, for a SN probability threshold of 0.1, 0.2 and 0.5, the false positive ratio is 0.87, 0.03 and 0.01 respectively, while the true positive ratio is 0.94, 0.92 and 0.83, respectively. Our model is suitable to be used to process large volumes of alerts, when limited resources for manual inspection and confirmation by means of follow-up observations are available.

\subsection{Images and Metadata Feature relevance}

\begin{figure*}[htbp]
    \centering
    \includegraphics[width=0.6\textwidth]{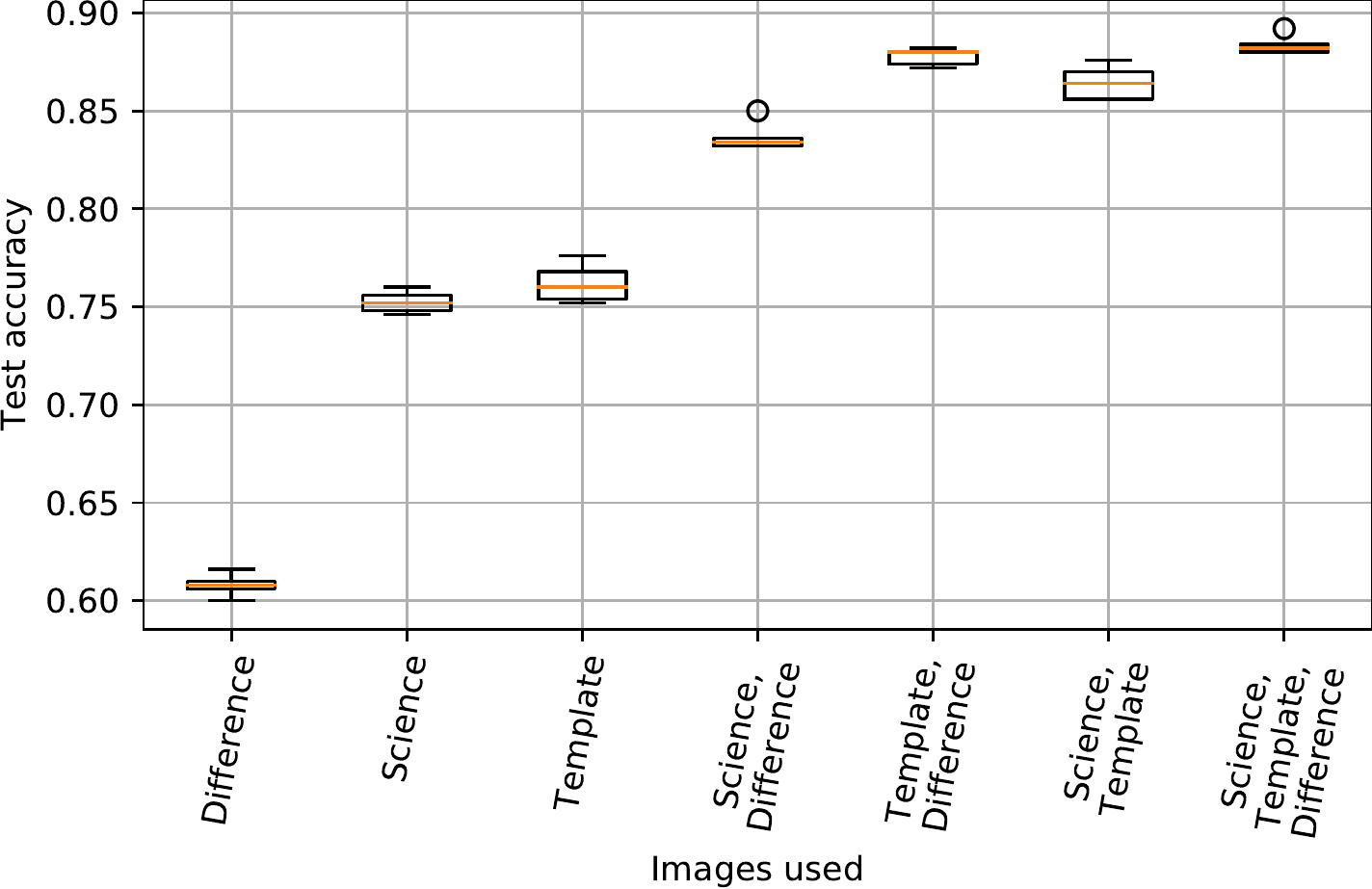}
    \caption{Accuracy of stamp classifier model when varying the images available at the input, isolated points are outliers. As we can observe, the most important image is the template, which gives information about nearby objects and context to the classifier, as well as the science image, which is slightly less informative according to the test accuracy, probably due to the higher noise compared to the template. Using the three images at the same time is better, but surprisingly not much better compared with using the science plus template, or difference plus template. This might be important to consider for classification purposes when designing alert based surveys such as LSST.}
    \label{fig:image_importance}
\end{figure*}

In this section, we explain the results of the experiments designed to account for the relevance of each image in the input (science, difference, and template) and the importance of each feature in the classification. Figure~\ref{fig:image_importance} shows the results of training the stamp classifier with a different combination of the input images. Here we highlight the main conclusions from these experiments. When using a single image as input, the image that gives the most accuracy is the template image. We hypothesize that the template image gives information about the context of the alert (e.g., stellar density, host galaxy, bright star, no counterpart), which is valuable information for the specific classes we are trying to correctly classify, as explained in Section \ref{sec:data_description}. Notably, the template image has a better signal to noise ratio than the science image (which also contains contextual information), explaining the difference in the accuracy of the stamp classifier model when trained on each of these. Surprisingly, using two images (science and template, or difference and template) as inputs, considerably improves the classification accuracy, almost reaching the accuracy of using the three images at the same time. This result could be considered when designing alert based surveys where bandwidth or storage of the alert streaming is an important restriction, but still the best accuracy is achieved when using the three images combined to train the model.

\begin{figure*}[htbp]
\gridline{\fig{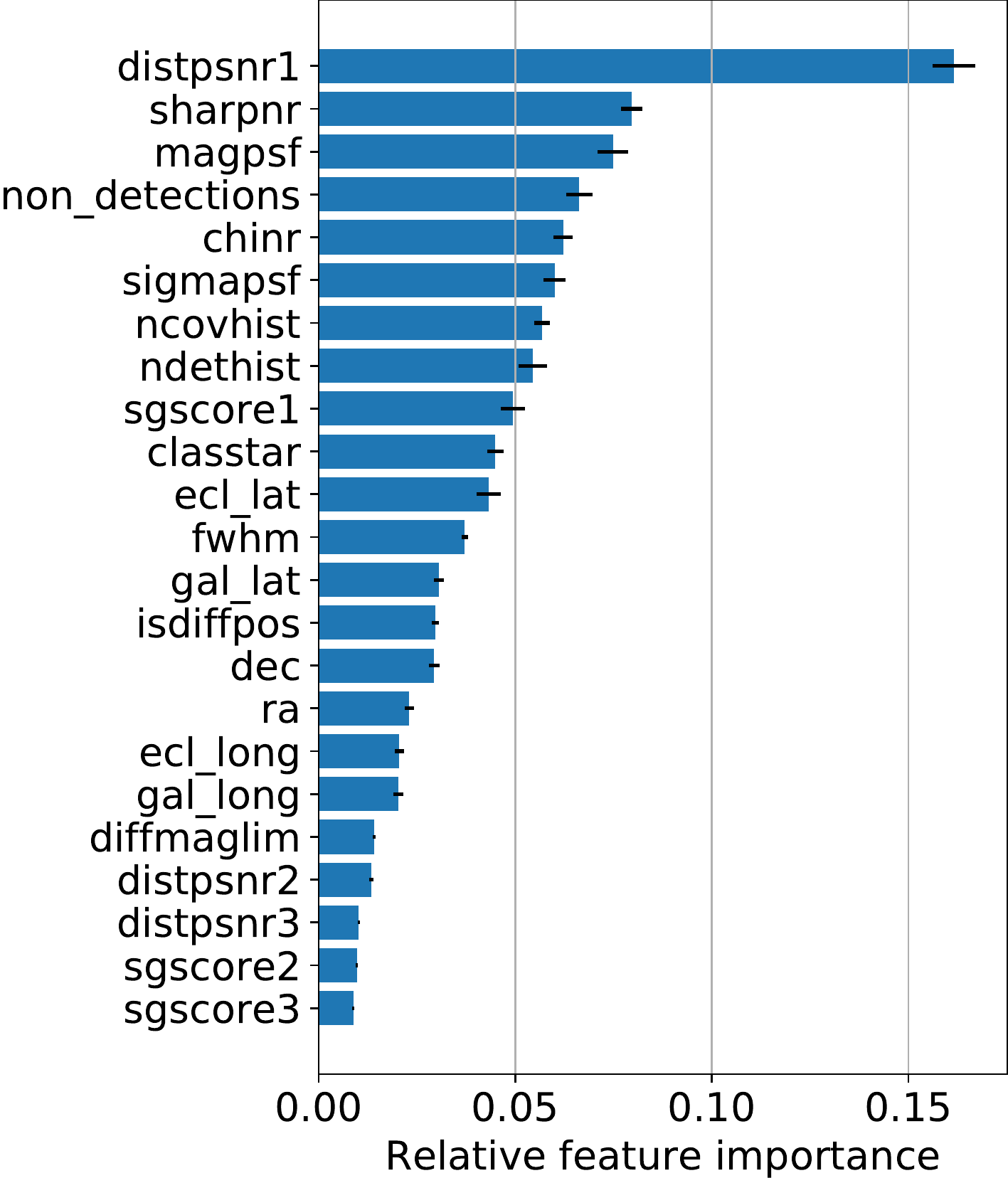}{0.26\textheight}{(a) Average feature importance according to 10 random forest classifiers trained to predict the label only using the metadata features.}
          \fig{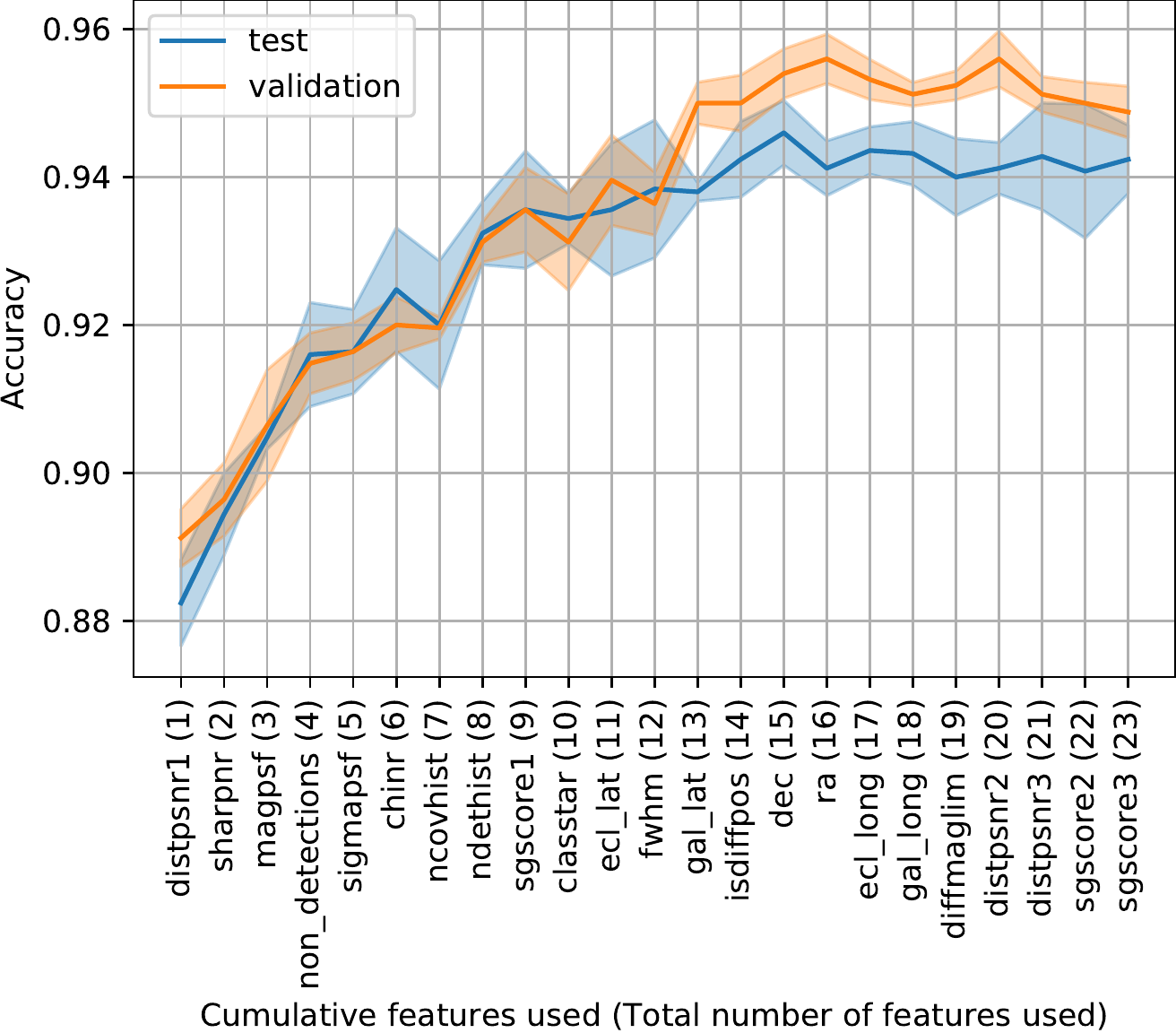}{0.45\textwidth}{(b) Accuracy of stamp classifier models when cumulative adding one feature at a time according to the feature ranking in (a).}}
\caption{Feature relevance analysis. The feature ranking in (a) was used to built stamp classifier models by aggregating one feature at a time and evaluate its accuracy (b). We can see, for instance, that the galactic and ecliptic \textbf{latitudes} are more important than longitude, since the former indicates the distance to the Galactic plane and ecliptic, which is useful for classification. Also, we see that \texttt{dispsnr1} is the most important feature as mentioned in Section \ref{sec:data_description}, and also \texttt{sharpnr} which helps to separate the SN class from the rest. This feature analysis is useful when using classification models that do not provide the relevance of each input dimension explicitly.}
\label{fig:ac_vs_features}
\end{figure*}

For analyzing the relevance of each metadata feature for classification of the alerts, we trained a random forest to classify the alerts only using the features, as mentioned in Section \ref{sec:expertiments} to obtain a feature importance ranking. The feature importance ranking is shown in Figure~\ref{fig:ac_vs_features}a, where the highest score feature is \texttt{distpsnr1}, which indicates the distance to the first closest source from PanSTARRS1 catalog, giving a measure of the density of objects near the alert and providing important context information. Another relevant feature is \texttt{sharpnr}, which is useful to discriminate the SN class among the others. Notice the relevance of the Galactic and ecliptic latitudes. The former provides context for stars (which have a higher probability of lying at low Galactic latitude) and extragalactic sources (which have a lower source density at low Galactic latitudes due to extinction), while the latter provides context for asteroids (which have a higher probability of lying at low ecliptic latitude). To assess the impact of each feature in the stamp classifier accuracy, we trained different models by adding one feature at a time in the order given by the feature ranking (from more important to less important) as detailed in Section~\ref{sec:expertiments}. The change in accuracy for each combination of accumulated features is shown in Figure~\ref{fig:ac_vs_features}b. The model gets higher accuracies for both, validation and test sets, by adding more features up to the galactic latitude feature, where the accuracy in the validation set goes further up compared to the test set. We interpret this as a sort of overfitting to the validation set, which in this particular case is not harmful to the performance of the model because the accuracy on the test set still increases and converges to a value when including additional features, without any statistically significant drop. We argue that while the accuracy on the test set does not drop, adding a new feature might be valuable extra information to the classifier, but further checking is needed in a larger set. For instance, using the predictions by the light curve classifier by \citealt{sanchezsaez_alert_2020} in the cases of SN, AGN and VS classes, or number of detections in the asteroid and bogus classes, as mentioned in Section~\ref{sec:results}.

\section{Model deployment and SN Hunter}\label{sec:sn_hunter}

\begin{figure*}[htbp]
    \centering
    \includegraphics[width=1\textwidth]{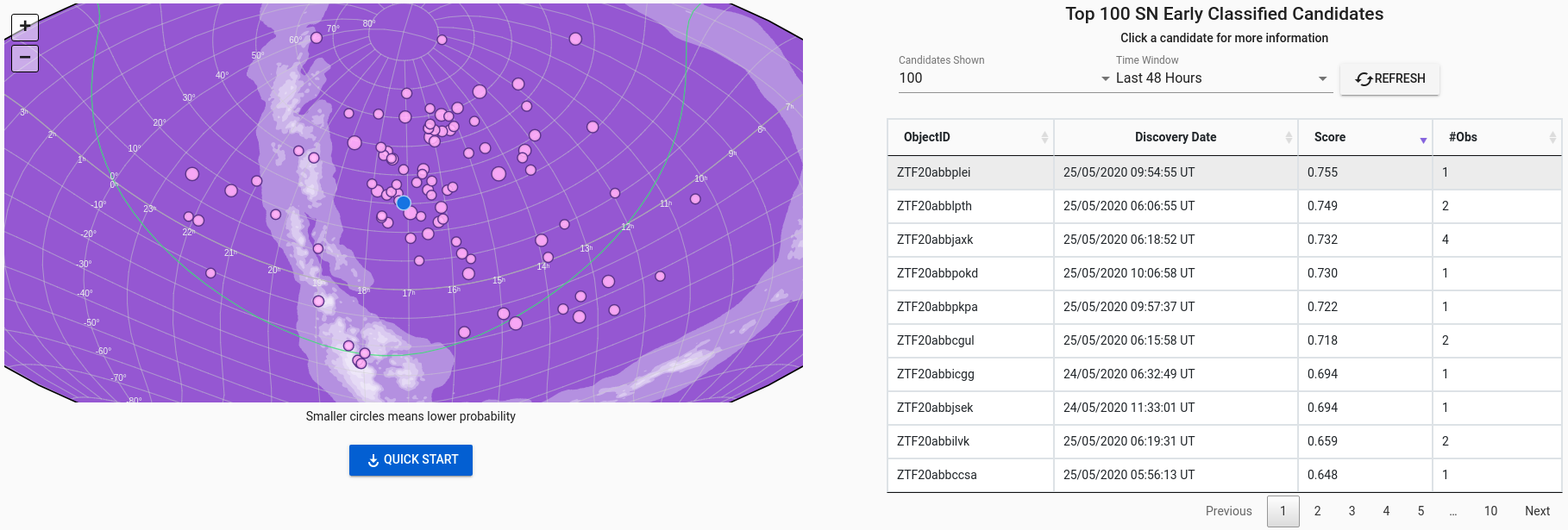}
    \caption{SN Hunter, a tool for the visualization of SNe candidates. On the left side, the location of each candidate in sky coordinates with respect to the Galactic plane and the ecliptic are depicted. On the right side, a selection of the top candidates is listed, initially ordered by SN probability score from the stamp classifier. The list of candidates can be sorted by other parameters, and updated/refreshed to include newly ingested alerts.}
    \label{fig:snhunter1}
\end{figure*}

The SN Hunter (\url{https://snhunter.alerce.online}) is a visualization tool that allows the user to inspect SN candidates classified by the model in real time, in order to select good targets for follow-up observations. The interface of the SN Hunter is shown in Fig.~\ref{fig:snhunter1}. At the left of the interface, a celestial map shows the position of each candidate with a circle, where the size of the circle is proportional to the class probability assigned by our model, with the map centered on the right ascension (\texttt{ra}) and declination (\texttt{dec}) coordinates of the alert. The Milky Way plane is highlighted by the regions with lighter shades of purple. The green curve in the map represents the ecliptic, where SN candidate alerts are more likely to be triggered by asteroids instead of real SNe. The right side of the interface provides a table where the highest probability SN candidates are listed. The table shows the ZTF Object ID which uniquely identifies each astronomical alert, the discovery date specifying day, month, year and time where the first alert was triggered, the corresponding SN probability (score) from the stamp classifier, and the number of available alerts in the $r$ and $g$ bands (\#Obs) since the discovery date. The list can be sorted by object, discovery date, score or number of alerts. The total number of high probability candidates shown in the table and maximum age of the candidates can be modified by the user. By clicking on a given candidate row, a new visualization panel is deployed as shown in Figure~\ref{fig:snhunter2}, with detailed information for the selected candidate.

The visualization panel contains some metadata on the left side (see Figure~\ref{fig:snhunter2}), including the ZTF Object ID (which is a clickable link to open the objects as a new tab within the ALeRCE online main frontend, \url{https://alerce.online/}), \texttt{ra} and \texttt{dec} coordinates of the alert, the filter in which the first detection was made, the PSF magnitude and the observation date. Below comes the PanSTARRS cross-match information, containing the Object ID, distance to the first closest known object, and the \texttt{classtar} score of the first closest known object, where a score closer to 1 implies higher likelihood of it being a star. The buttons below this information, from left to right, correspond to queries with the ALeRCE frontend, the NASA Extragalactic Database (NED\footnote{The NASA/IPAC Extragalactic Database (NED) is operated by the Jet Propulsion Laboratory, California Institute of Technology, under contract with the National Aeronautics and Space Administration. \url{https://ned.ipac.caltech.edu/}}), TNS, and the Simbad Astronomical Database (\citealt{wenger_simbad_2000}) around the position of the candidate. Finally, the full metadata associated with the first alert of the SN candidate is linked below these buttons. The middle panel of Figure~\ref{fig:snhunter2} contains an interactive color image from PanSTARRS DR1 (\citealt{chambers_pan-starrs1_2019}), centered around the source using Aladin (\citealt{bonnarel_aladin_2000, boch_aladin_2014}); this image is also available in the main frontend of ALeRCE. The right panel of Figure~\ref{fig:snhunter2} provides the science, reference and difference stamps of the first detection. It is also possible to sign in with a user account and label candidates as either a possible SN or bogus by clicking the corresponding buttons below the image stamps. These can be used to build up larger training sets, as well as select candidates for the Target and Observation Managers (TOMs).

We implemented the CNN stamp classifier using TensorFlow 1.14 (\citealt{martin_abadi_tensorflow_2015}) and deployed it to classify the streaming alerts from ZTF's Kafka server\footnote{\url{https://kafka.apache.org/}}. The timespan between a ZTF exposure and its first arrival as an alert from the stream is 14.6 $\pm$ 4.5 minutes. Once the alert is received by ALeRCE, it takes a few seconds for the candidate to be listed in the Supernova Hunter tool for expert inspection. Further details about the complete processing pipeline are described in \cite{forster_automatic_2020}.

\subsection{Additional Visual Selection Criteria} \label{sec:visual_selection}

We note that the SN candidate sample presented in this and the following subsections resulted from an older version of the Stamp Classifier which relied only on the three images within the first alert and did not use features for SN classification. Moreover, some of the filtering steps we applied manually are no longer necessary now that features are included (we note these below). As shown in Appendix \ref{sec:additional_results}, even without the metadata features, the classifier provides reasonably high accuracy (only ~6\% worse than the model with features). Regardless of whether features are included or not, we found it critical to visually inspect the predicted SNe candidates in order to weed out misclassifications and submit more reliable candidates to TNS. 

\begin{figure*}[htbp]
    \centering
    \includegraphics[width=1\textwidth]{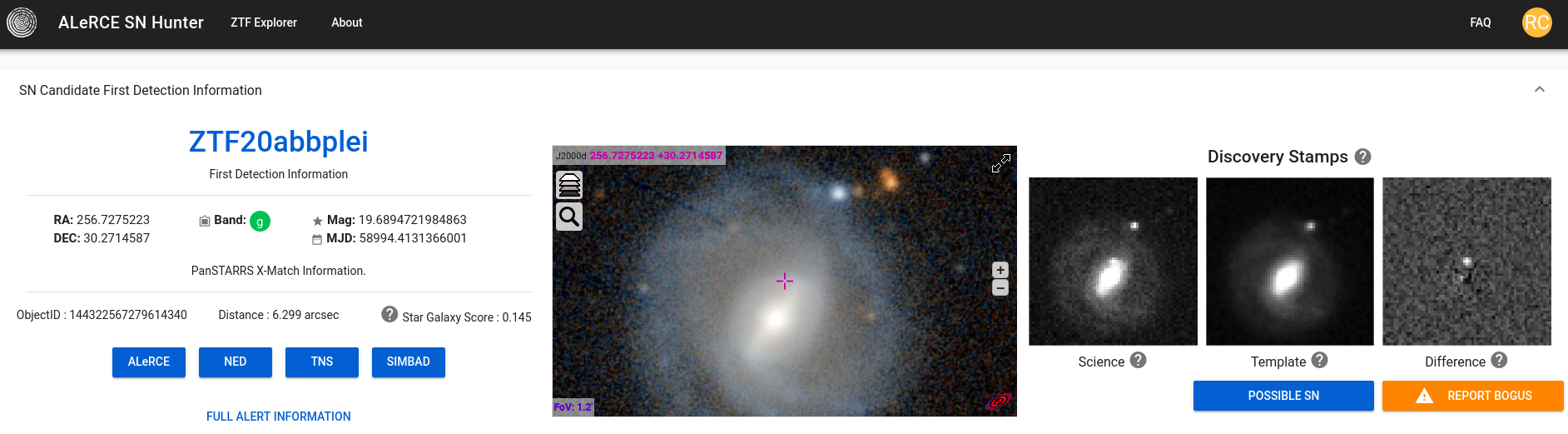}
    \caption{Candidate visualization in the Supernova Hunter tool. On the left side, the SN candidate ID is shown as a clickable link to the ALeRCE frontend, with relevant metadata such as \texttt{ra}, \texttt{dec}, magnitude, date, etc. At the bottom there are links to other sources of information, including ALeRCE, NED, TNS, and the Simbad Astronomical Database. In the middle of the figure there is a colored image from Aladin. On the right side, the stamps of the first detection are shown, along with buttons for reporting the candidate as eventual bogus or as a possible SN.}
    \label{fig:snhunter2}
\end{figure*}

There are some common characteristics among the higher confidence SN candidates. As mentioned in Section~\ref{sec:data_description}, most confirmed SNe are located on top or near an extended galaxy. If there is no galaxy within the stamps, then it is more likely that is a variable star or asteroid, when the candidate is located near the ecliptic or the Milky Way, respectively, or bogus. In some cases, it is difficult to tell if the nearest source to the alert in the science image is an extended galaxy or star; for these, a search of archival catalogs and/or an assessment of the spectral energy distribution can further aid classification. Therefore, the star galaxy score from PanSTARRS in Figure~\ref{fig:snhunter2} should be closer to 0, indicating that the extended source is more likely to be an extended galaxy. Real SN should have a positive flux in the difference image, so we removed candidates that have negative flux in the latter, by checking if the field \texttt{isdiffpos} value in the metadata is false, this is automatically done in the current pipeline. It is also important to check that the object is visible in the difference image. 

Another relevant feature is the shape of the candidate, which should be similar to other stars with fuzzy edges and generally symmetric in shape. If the shape of the candidate is sharp (pixelized) or very localized, it might be a cosmic ray or a defect of the CCD camera. Alternatively, if it is elongated, it could be an asteroid or a satellite (often seen as a streak or multiple small dashes due to rotational reflections). After doing all of these checks, if the candidate is not convincing enough, then it is helpful to look at the next detections when available and search for the characteristics mentioned, which can be done using the ALeRCE frontend by clicking the ALeRCE button in the SN Hunter tool and querying directly that specific candidate's data. The 100 highest probability SNe candidates each day are manually inspected by astronomers of the ALeRCE team, and all of them must be in agreement before a candidate is reported to TNS. As a qualitative analysis, we report that the confusion of the SN class will depend on the weather. In optimal weather conditions, the point spread function size could be a few pixels long and the classifier confuses SN with bogus samples of type cosmic rays. In regular nights, asteroids and satellites are a key source of contamination, and then a small fraction of image subtraction issues. Alerts triggered near known variable stars or asteroids and classified as SNe by the SN Hunter are removed from the list of 100 candidates to be visually inspected by astronomers.

\subsection{Reported and Confirmed Supernovae}\label{sec:confirmed_sn}

From June 26th 2019 to February 28th 2021, we have reported 6846 new SN candidates to TNS, increasing this number by 11.8 SNe per day on average, of which 995 have been observed spectroscopically. Table~\ref{table:confirmed_supernovae} shows the number of candidates for each confirmed class, of which 971 were confirmed as SNe. Non-SNe objects reported were 5 TDEs, 5 galaxies, 4 Nova, 2 Other, 2 cataclysmic variables (CV), 2 AGNs, 2 unknowns, 1 variable star, 1 M dwarf. Even though TDEs are not SNe, follow-up of these events is still of significant interest, due to their relative scarcity (\citealt{van_velzen_seventeen_2020}). In summary, taking into consideration the conservative final candidate selection done by the team of astronomers to perform spectroscopic confirmation, our reported and confirmed candidates have around 2\% contamination by non-SNe objects.

\begin{table}[h!]
\begin{center}
\caption{Spectroscopically observed candidates discovered by ALeRCE, with a total of 971 SNe, 24 non-SN objects (5 TDE, 5 galaxies, 4 Nova, 2 Other, 2 cataclysmic variables (CV), 2 AGNs, 2 unknowns, 1 variable star, 1 M dwarf).}
\label{table:confirmed_supernovae}
\begin{tabular}{c|c}
Confirmed class   & Spectroscopically observed candidates                 \\ \hline \hline
SN Ia            &  676  \\ \hline
SN II &  148 \\  \hline
SN Ic               &  24  \\ \hline
SN Ia-91T-like    &  22  \\ \hline
SN IIn            &  21  \\  \hline
SN IIP             & 16  \\ \hline
SN Ib         &  14  \\ \hline
SN IIb  & 13 \\  \hline
SN Ic-BL & 10 \\  \hline
TDE &  5  \\ \hline
Galaxy &  5  \\ \hline
Nova  &  4  \\ \hline
SN Ia-pec  &  4  \\ \hline
SN Iax[02cx-like]  &  4  \\  \hline
SN I & 3 \\ \hline
SN Ia-91bg-like & 3 \\ \hline
SN  &  3  \\  \hline
SN Ib/c & 3 \\ \hline
SLSN-II & 3 \\ \hline
Other &  2  \\ \hline
CV &  2  \\  \hline
unknown & 2 \\ \hline
AGN & 2 \\ \hline
SN Ib-pec & 1 \\ \hline
Varstar &  1  \\  \hline
SN Ibn & 1 \\ \hline
M dwarf & 1 \\ \hline
SLSN-I & 1 \\ \hline
SN Icn & 1 \\ \hline
\end{tabular}
\end{center}
\end{table}

\begin{figure}[h]
    \centering
    \includegraphics[width=0.49\textwidth]{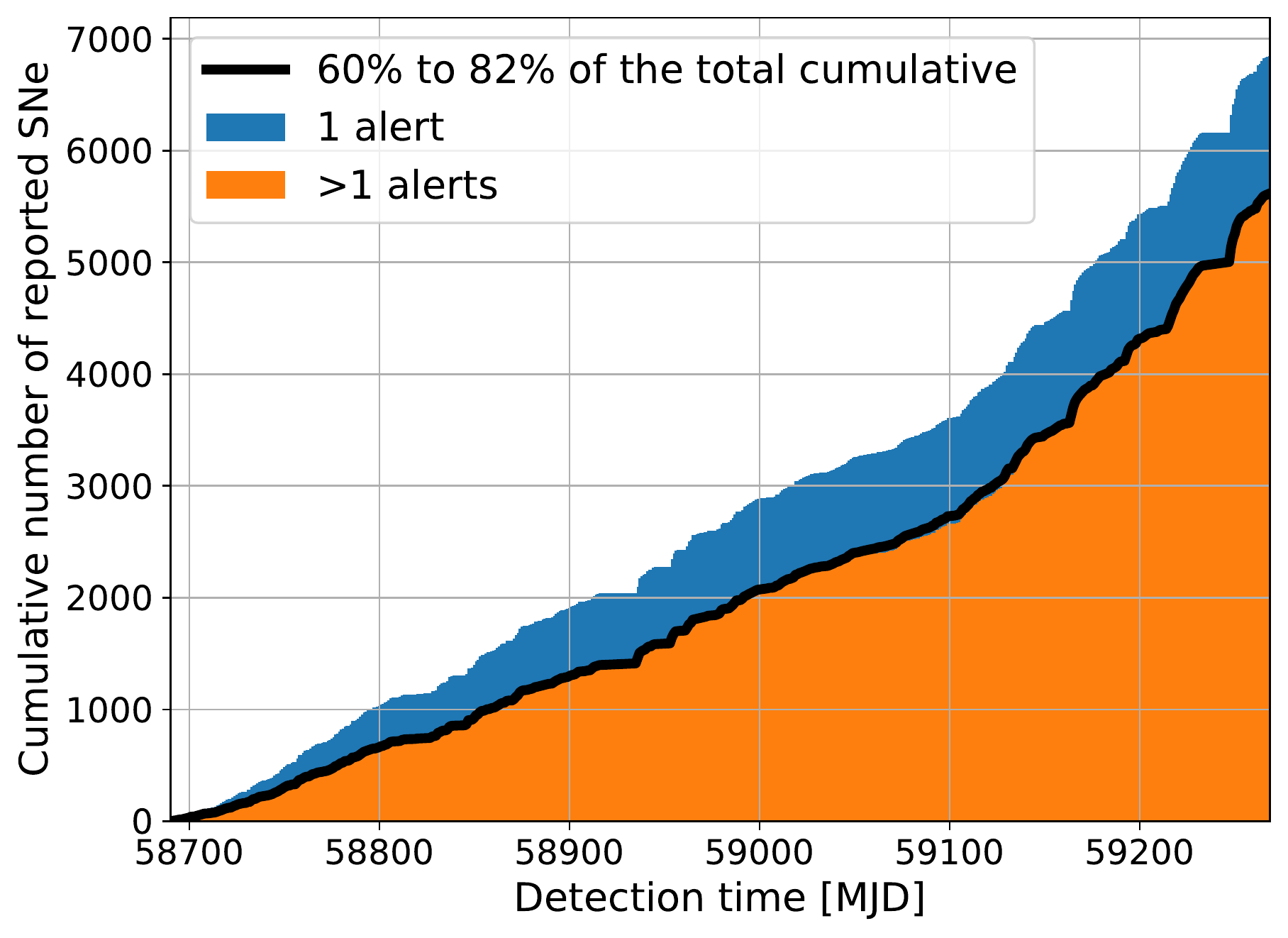}
    \caption{Cumulative number of reported SNe since we started reporting on June 26th 2019 February 28th 2021. The average rate of reporting is 11.8 candidates per day. Currently $\approx$82\% of our reported candidates are detected with multiple alerts (purity), implied they are true SNe, while $\approx$12\% have only one detection and thus less certainty. We have been increasing the purity of the reported candidates roughly linearly from $\approx$60\% to $\approx$82\%.}
    \label{fig:reporting_rate}
\end{figure}

In Figure~\ref{fig:reporting_rate}, we show the cumulative distribution of candidates reported to TNS from June 26th 2019 to February 28th 2021. The cumulative distribution is separated into two parts, namely the alerts with more than one detection to date (orange) and the alerts with a single detection to date (blue). We can consider the percentage of candidates with more than one detection to be a lower bound of real non-moving astronomical objects, since we do not have the true label for reported alerts; we define this as ``purity'', since multiple associated detections are a clear sign of a real non-moving astronomical object rather than a moving object or bogus alert. Candidates with only single detections to date could be due to several reasons: moving objects, bogus alerts, relatively short transients that were only above the detection threshold for a short period of time, and objects that are in locations which have not been visited again by the public ZTF Survey since the object detection. We have been increasing the purity approximately linearly, from $\approx$60\% to $\approx$82\% (reported candidates detected with multiple alerts to date). 

\begin{figure}[h]
    \centering
    \includegraphics[width=0.49\textwidth]{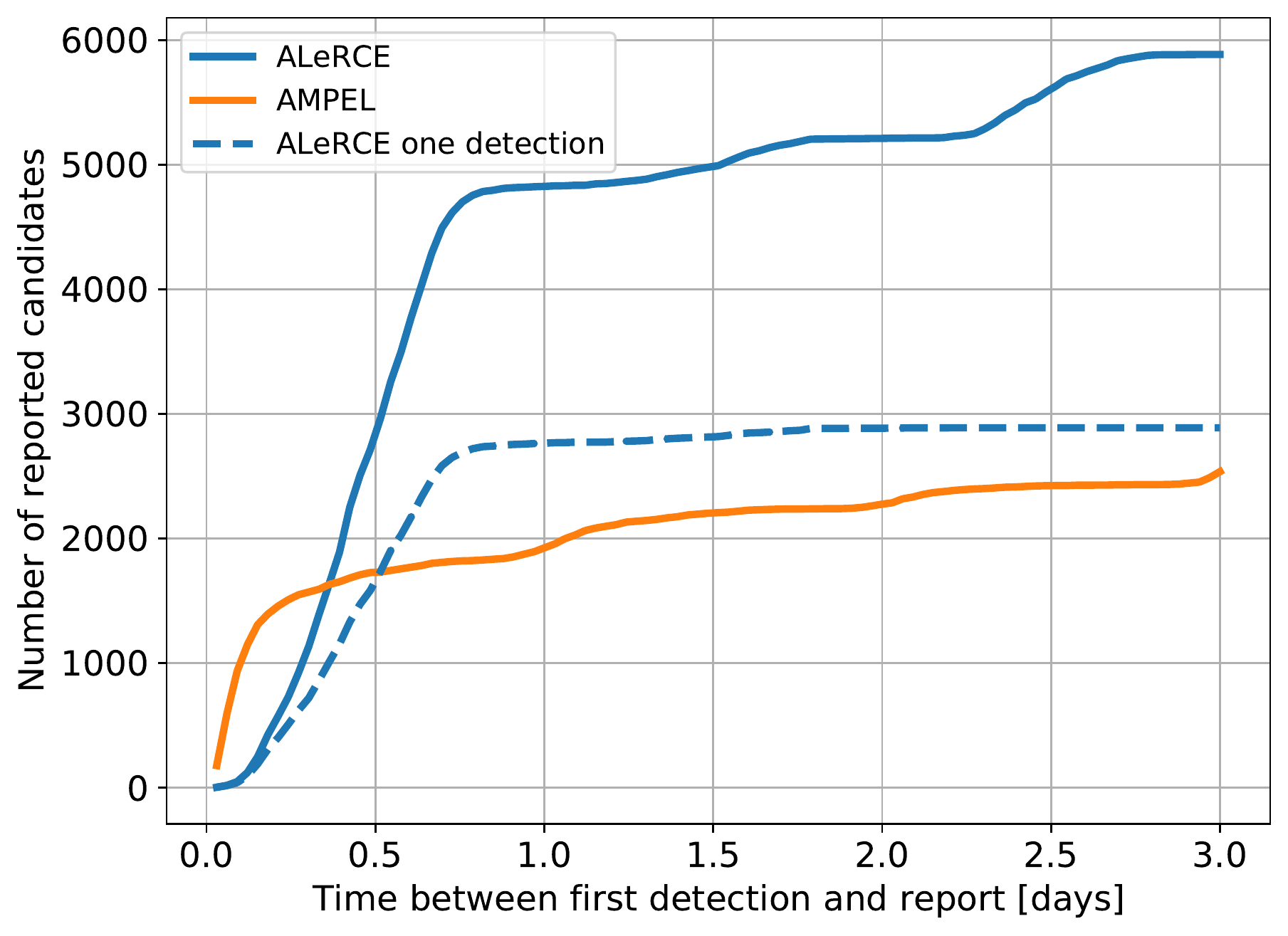}
    \caption{Cumulative distribution of time between the first detection and the reporting time from TNS, for candidates reported by ALeRCE and AMPEL. The full distributions are shown with solid lines, and the distributions of reporting time for candidates with a single detection are shown with segmented lines.}
    \label{fig:reporting_time}
\end{figure}

For comparison purposes, we gathered the objects reported by both ALeRCE and AMPEL (Alert Management, Photometry and Evaluation of Lightcurves, which is an internal ZTF classification effort; \citealt{nordin_transient_2019}) to TNS, and compared the reporting times within 3 days after the first detection. Figure~\ref{fig:reporting_time} shows the cumulative histogram of reporting times to TNS for ALeRCE and AMPEL, along with the cases where reports were done by ALeRCE before having the second detection in the public stream (one detection). Approximately 42\% of the candidates reported by ALeRCE were based on a single detection. 
An important difference between both systems is the visual inspection by experts in the reporting process to TNS. According to \cite{nordin_transient_2019}, AMPEL reports candidates automatically using their ``TNS channel'', which produces more reported candidates than our system, within 12 hours after the first detection. As described in this work, our system's final stage so far relies on human inspection, checking and reporting, which occurs within 10 to 24 hours after the first detection, without reporting transients already reported by AMPEL (only two cases were reported after AMPEL). Therefore, we report new candidates to TNS within a day after the first detection. Besides, since ALeRCE is largely reporting candidates with a single detection, 70\% (4825) of the reports were sent within one day after the first detection, compared to 25\% (1925) from AMPEL. 

\begin{figure}[h]
    \centering
    \includegraphics[width=0.49\textwidth]{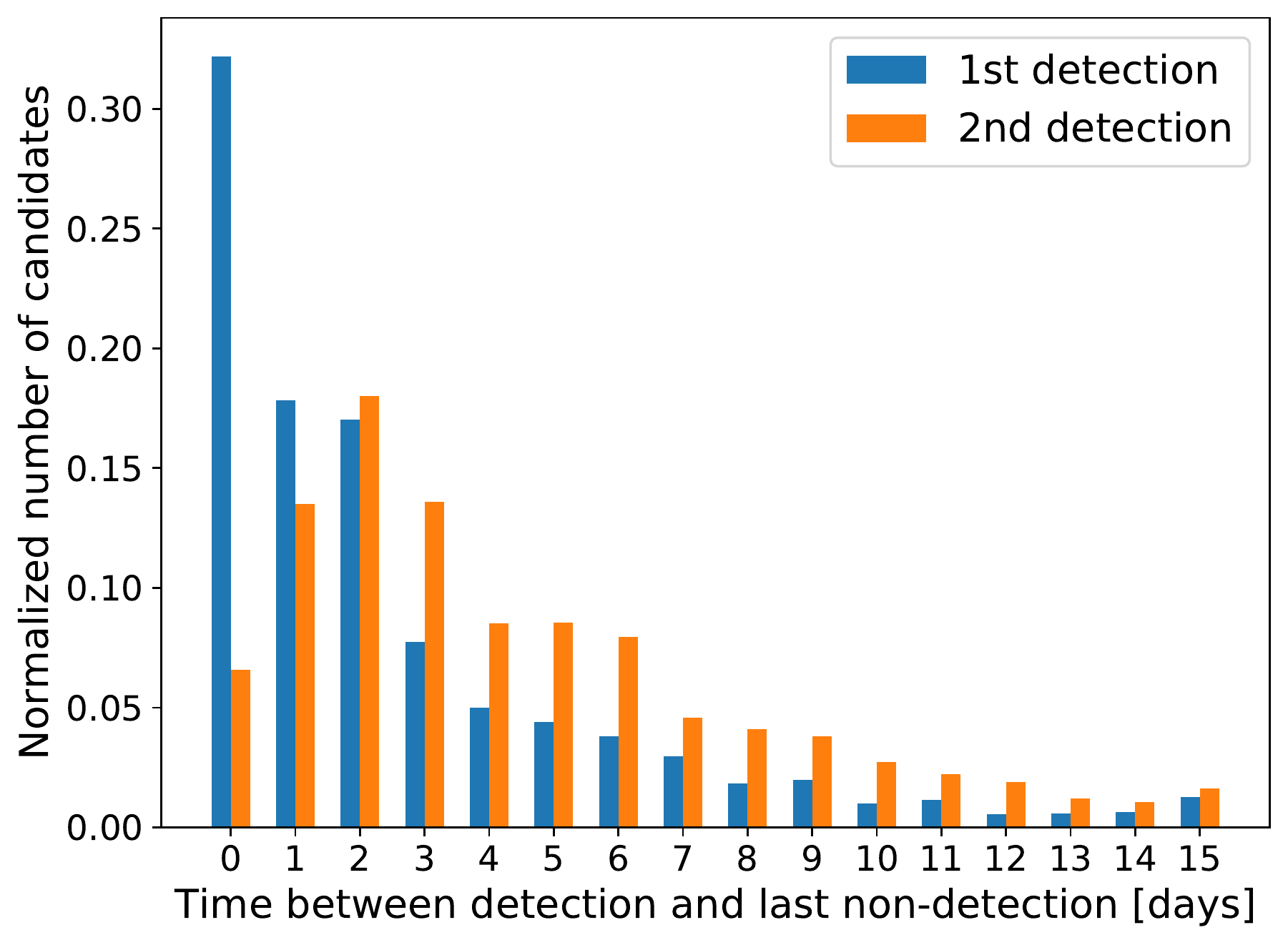}
    \caption{Histogram of time between first (second) detection and the last non-detection of the ALeRCE reported candidates.}
    \label{fig:time_between_samples}
\end{figure}

Figure~\ref{fig:time_between_samples} shows the distribution of time between first or second detection and last non-detection for candidates reported by ALeRCE. Based on the data shown in Figure~\ref{fig:time_between_samples}, the average time between the last non-detection and the first detection is 4.2 days, and 8.1 days for the second detection. Reporting candidates only after the second detection would result in a delay of 3.9 days on average, which represents a potentially critical timespan to measure the spectra at early stages of the transient event, as required in order to achieve the science goals described in Section~\ref{sec:intro}. As mentioned before, ALeRCE currently does not report candidates that were previously reported by other groups using data from ZTF, therefore our candidates reported using a single detection increase the bulk of objects available for early follow-up of transients that were not found by other groups. We will report already reported candidates in the near future, since this adds the additional information that the candidates passed our visual inspection test. In addition, the work presented is a starting point towards our goal of developing an automatic reporting systems of the most highly confident subset of SN candidates.

\section{Conclusion and Future Work} \label{sec:conclusions}

As part of the ALeRCE Broker processing pipeline, we identified characteristics of the images and metadata within the ZTF alert stream that allow us to discriminate among SN, AGN, VS, asteroids and bogus alerts using the first detection only. In order to solve this classification problem automatically and quickly identify the best SN candidates to perform follow-up, we trained a CNN. The inputs to this classifier are the science, reference and difference images, and part of the metadata of the first detection alert. In addition, our CNN architecture is invariant to rotations within the stamps, and was trained using an entropy regularized loss function. The latter is useful to improve human readability in predicted probabilities per class, in terms of certainty assigned to each sample, so an expert can gain better insights into the actual nature of the transients when inspecting SN candidates.

Among all five classes that our CNN can classify, it achieves an accuracy of $0.941 \pm 0.004$ on a balanced test set, while in the SN class reaches a true positive rate and a false positive rate of $87 \pm 1\%$ and $5 \pm 2\%$, respectively. By manually inspecting the classification of each sample, we found that the incorrectly classified objects are in concordance with our hypotheses regarding separability of classes using only the first detection images. Moreover, the CNN model successfully classify the alerts in the labeled set by using the images only, but when applied to unlabeled data we found some flaws by inspecting the spatial distribution of each predicted class, for instance, a concentration of extragalactic sources within the Galactic plane, and a higher density of asteroids far from the ecliptic. By giving the alert metadata as additional features to the classifier, we find that the spatial distributions of the events are in agreement with the expectations, according to their tentative nature. More specifically, extragalactic classes (SNe and AGNs) are found outside the Galactic plane, VS have a higher density of predicted objects within the Galactic plane, and asteroids are found around the ecliptic.

The proposed CNN classifier is deployed and its predictions are publicly available in an especially designed visualization tool for inspection of candidates with high SN class probability, called the SN Hunter (\url{https://snhunter.alerce.online/}). The predictions are also available in the ALeRCE main frontend. This tool shows relevant information about the SN candidates in order to facilitate their analysis, and we used it to report SN candidates on TNS for follow-up. We also presented a visual inspection methodology that relies on the information presented in the Supernova Hunter tool, such as the probability assigned by our classifier to the SN class, alert metadata, position in the sky with respect to the Galactic plane and the ecliptic, number of detections, etc. By manually inspecting candidates using the SN Hunter tool, from June 26th 2019 to February 28th 2021, our team has reported 6846 candidates for follow-up, out of which 995 were tested, and 971 were spectroscopically confirmed as SN. Besides, the interface allows experts to manually label bogus and SN candidates alike, which helps improve our training set.

As many as 70\% of the candidates reported to TNS by ALeRCE were reported within one day after the first detection, and 42\% of all the candidates reported by ALeRCE were done by using a single detection, where 82\% of the total alerts reported by ALeRCE have multiple detections to the date of writing this document, confirming extragalactic nature. Since ALeRCE does not report objects that had been previously reported, these results correspond to the transients that were not detected or chosen by other groups, therefore adding new early transient reports to TNS.

We are currently working on improving the training set by adding more examples from confirmed SNe, and manually adding bogus candidates to the training set. We run simple but insightful experiments to understand the contribution of each image (\textit{science, reference, bogus}) and features (metadata) to the classification task. Furthermore, we are exploring ideas for model interpretability, adding visualization tools that may help understand why the model predicts a given class for a given event. We are working on using LRP (\citealt{bach_pixel-wise_2015, montavon_layer-wise_2019}) and occlusion techniques (\citealt{zeiler_visualizing_2014}) to show what part of the input influences the decision in a specific way, so the expert can use it to choose better candidates.

Regarding the performance of the model, an additional step would be to extend the system to be able to process more than a single alert while keeping the capability of performing well with only one alert. New approaches about how to achieve this have been explored in \cite{carrasco-davis_deep_2019} and \cite{gomez_classifying_2020}, feeding a neural network sequentially with the data available so far, and improving the prediction every time a new measurement arrives. In the near future our efforts regarding alert classification, and particularly the SN detection problem, will aim towards the automatization of the entire process of classification of the data stream and reporting objects for follow-up, eliminating or bringing expert assistance to a minimum.

The methodology proposed in this work is suitable to other streams of data based on alerts, such as ATLAS (The Asteroid Terrestrial-impact Last Alert System; \citealt{tonry_atlas_2018}) and the Vera C. Rubin Observatory Legacy Survey of Space and Time (LSST; \citealt{ivezic_lsst_2019}). The latter presents a further challenge in terms of the amount of data generated, restrictions in processing time due to the data generation rate, the larger number of filters, the lack of comparison catalogs at the survey's depth, the smaller field of view per stamp (currently planned to be only 6"$\times$6") and limited contextual information, and the possibility that either the science or reference image may not be contained in the alerts. We think our work on ZTF data will be a valuable precursor for the next generation of large etendue telescopes.

\section{Acknowledgments}
The authors acknowledge support from the National Agency of Research and Development's Millennium Science Initiative through grant IC12009, awarded to the Millennium Institute of Astrophysics (RC, ER, CV, FF, PE, GP, FEB, IR, PSS, GC, SE, Ja, EC, DR, DRM, MC) and from the National Agency for Research and Development (ANID) grants: BASAL Center of Mathematical Modelling AFB-170001 (CV, FF, IR, ECN, CS, ECI) and Centro de Astrof\'isica y Tecnolog\'ias Afines AFB-170002 (FEB, PSS, MC); FONDECYT Regular \#1171678 (PE), \#1200710 (FF), \#1190818(FEB), \#1200495 (FEB), \#1171273 (MC), \#1201793(GP); FONDECYT Postdoctorado \#3200250 (PSS); Mag\'ister Nacional 2019 \#22190947 (ER). This work was funded in part by project CORFO 10CEII-9157 Inria Chile (PS).

\software{Aladin \citep{bonnarel_aladin_2000}, Apache ECharts\footnote{\url{https://echarts.apache.org}}, Apache Kafka\footnote{\url{https://kafka.apache.org/}}, Apache Spark \citep{zaharia_apache_2016}, ASTROIDE \citep{brahem_astroide_2018}, Astropy \citep{astropy_collaboration_astropy_2013}, catsHTM \citep{soumagnac_catshtm_2018}, Dask \citep{rocklin_dask_2015}, Jupyter \footnote{\url{https://jupyter.org/}}, Keras \citep{chollet_keras_2018},
Matplotlib \citep{hunter_matplotlib_2007}, NED \citep{steer_redshift-independent_2016}, Pandas \citep{mckinney_data_2010}, Prometheus\footnote{\url{https://prometheus.io/}}, Python\footnote{\url{https://www.python.org/}}, scikit-learn \citep{pedregosa_scikit-learn_2011}, Simbad-CDS \citep{wenger_simbad_2000}, Tensorflow \citep{martin_abadi_tensorflow_2015}, Vue\footnote{\url{https://vuejs.org/}}, Vuetify\footnote{\url{https://vuetifyjs.com/}}, PostgreSQL\footnote{\url{https://www.postgresql.org/}}.}

%%%%%%%%%%%%%%%%%%%%%%%%%%%%%%%%%%
%%%%%%%%%%% APPENDIX %%%%%%%%%%%%%
%%%%%%%%%%%%%%%%%%%%%%%%%%%%%%%%%%
\newpage
\appendix

\renewcommand{\thefigure}{A\arabic{figure}}
\setcounter{figure}{0}
\renewcommand{\thetable}{A\arabic{table}} \setcounter{table}{0}

\section{Exploring the relationship between features and classes}\label{sec:features_appendix}

\begin{figure*}[htbp]
    \centering
    \includegraphics[width=0.95\textwidth]{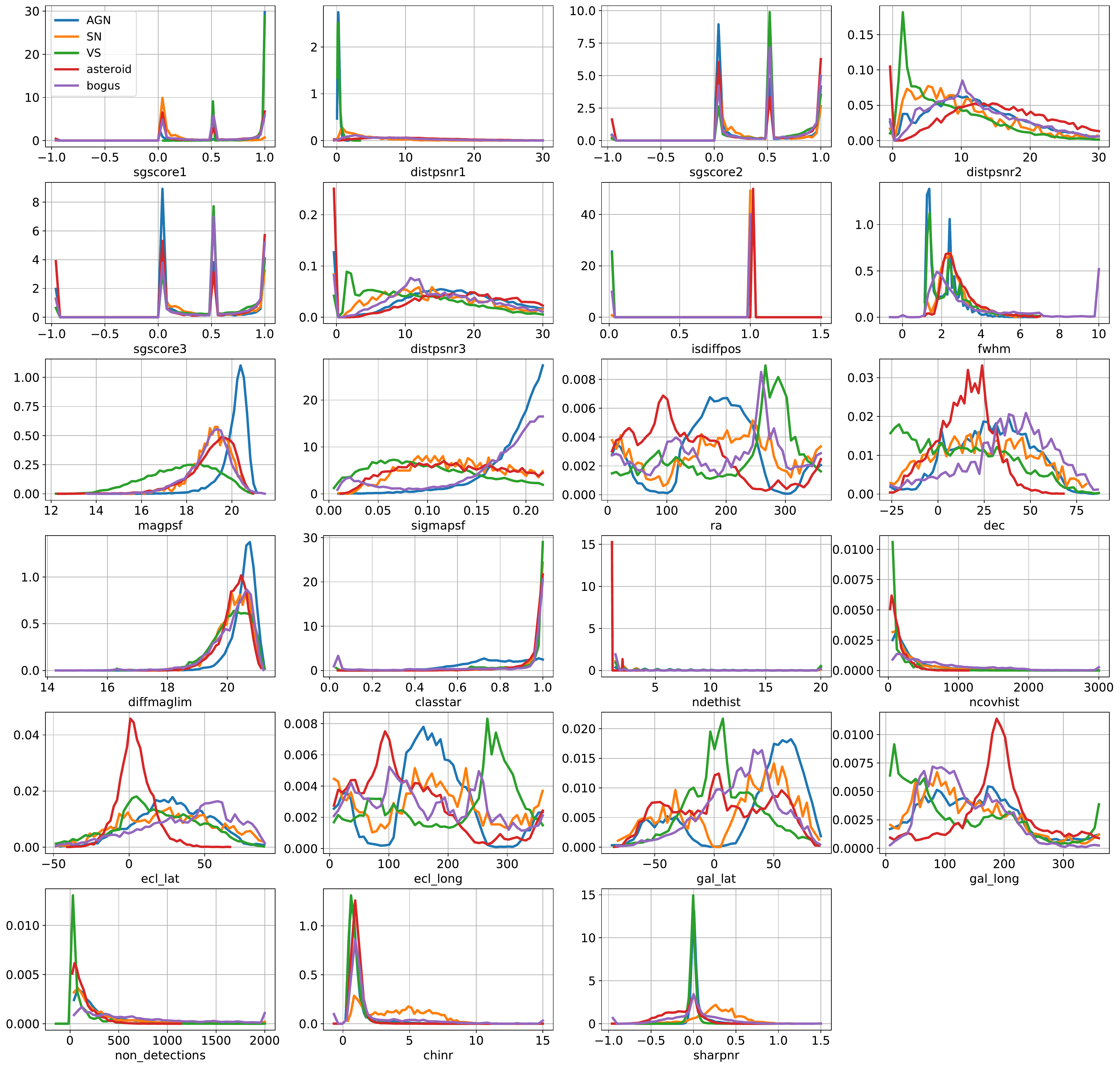}
    \caption{Feature distribution per class of the labeled dataset. Each feature was clipped to the values given in Table~\ref{table:clipping_values}.}
    \label{fig:features_grid}
\end{figure*}

\begin{table}[h!]
\begin{center}
\caption{Clipping values for each feature. ``max'' or ``min'' in the clipping range for each feature means that the maximum and minimum value is preserved for that feature respectively.}
\label{table:clipping_values}
\begin{tabular}{c|c}
Feature           & [min value, max value]                  \\ \hline \hline
\texttt{sgscore1}        & [-1, max]  \\ \hline
\texttt{distpsnr1}       & [-1, max]        \\ \hline
\texttt{sgscore2}               & [-1, max]                   \\ \hline
\texttt{distpsnr2}               & [-1, max]                   \\ \hline
\texttt{sgscore3}             & [-1, max]               \\ \hline
\texttt{distpsnr3}          & [-1, max]                      \\ \hline
\texttt{ifwhm}        & [min, 10]  \\ \hline
\texttt{ndethist}               & [min, 20]                   \\ \hline
\texttt{ncovhist}             & [min, 3000]             \\ \hline
\texttt{chinr}          & [-1, 15]                  \\ \hline
\texttt{sharpnr}             & [-1, 1.5]        \\ \hline
non detections          & [min, 2000]          \\ \hline

\end{tabular}
\end{center}
\end{table}

\section{CNN glossary and  training}\label{sec:machine_learning_appendix}

\subsection{CNN architecture}

\begin{itemize}[leftmargin=*]
\item \textbf{Fully connected layer}: Artificial neural networks (ANNs) are mathematical models that are mostly used for classification or regression. ANNs make use of basic processing units called \emph{neurons}, which receive vectors $\mathbf{x}$ of data as input, then apply a linear function to them, followed by a non-linear activation function. These neurons are grouped in \emph{layers}, which are called \emph{fully connected layers}. The output produced by a set of neurons of a specific fully connected layer is calculated as:

\begin{equation}
\mathbf{y} = \phi (\mathbf{W}\mathbf{x}+\mathbf{b}),
\label{eq:perceptron}
\end{equation}

where $\mathbf{x}\in\mathbb{R}^n$ is the input of the layer, $\mathbf{y}\in\mathbb{R}^m$ is the output of the layer, $\mathbf{W}\in\mathbb{R}^{m\times n}$ is a matrix of parameters called \emph{weights}, $\mathbf{b}\in\mathbb{R}^m$ is a vector which contains the so-called \emph{biases} of the layer, and $\phi(\cdot)$ is a non-linear activation function that follows the linear transformation of $\textbf{x}$. There are all sort of flavors of non-linear activation functions, the most commonly used are:

\begin{eqnarray}
\mathrm{sigmoid}(x) &{}={}& \frac{1}{1+e^{-x}},\\
\mathrm{tanh}(x) &{}={}& \frac{e^x-e^{-x}}{e^x+e^{-x}},\\
\mathrm{ReLU}(x) &{}={}& \max \{0,x\}.
\end{eqnarray}

To be precise, $\mathbf{W}$ and $\mathbf{b}$ are referred to as the parameters of a fully connected layer, and they are modified during training to be optimized for the task at hand.  Fully connected layers can be  sequentially stacked one after the other to integrate an ANN model. For instance, an ANN of two layers, is defined as:

\begin{eqnarray}
\mathbf{z} &{}={}& \phi (\mathbf{W}^{(1)}\mathbf{x}+\mathbf{b}^{(1)}),\\
\mathbf{y} &{}={}& \phi (\mathbf{W}^{(2)}\mathbf{z}+\mathbf{b}^{(2)}).
\end{eqnarray}

The parameters of the ANN are $\theta = (\mathbf{W}^{(1)}, \mathbf{b}^{(1)},\mathbf{W}^{(2)},\mathbf{b}^{(2)})$. The way of grouping neurons and layers in an ANN is called the \emph{architecture} of an ANN.

\item \textbf{Softmax output layer:} A commonly used activation function at the output of ANN models is the $\mathrm{sigmoid}(x)$, whose output is bounded by $(0,1)$, and can be interpreted as the probability of activation of a neuron, a property useful for binary classification. A generalization of the aforementioned function, useful for multi-class classification models, is the \emph{softmax} activation function, usually referred to as \emph{softmax output layer}, where there are $K$ neurons $x_i$, $i\in\{1,\ldots,K\}$, and it is desired to assign a probability to each one, hence, requiring that they add up to one. This is done by the softmax activation function, defined as: 

\begin{equation}
\mathrm{softmax}(x_i) = \frac{e^{x_i}}{\sum_{j=1}^K e^{x_j}},\quad i\in\{1,\ldots, K\}.
\label{eq:softmax}
\end{equation}

\item \textbf{Convolutional layer:} ANN with fully connected layers are limited to vector-like inputs, and they do not take into account the presence of correlation between adjacent features. To overcome this limitation, and preserve a degree of spatial or temporal correlation in the input of models, CNNs were proposed. The main component of CNNs are \emph{convolution layers}, which apply a filter or kernel to the input of the layer by a convolution operation. Similar to a fully connected layer, convolutional layer outputs are calculated as follow:

\begin{equation}
\mathbf{y} = \phi (\mathbf{W}*\mathbf{x}+\mathbf{b}),
\label{eq:convlayer_basic}
\end{equation}

where $\mathbf{x}$ stands for the input of the layer, $\mathbf{y}$ is the output of the layer, $\mathbf{W}$ is a set of filters to apply by convolution to the input, $\mathbf{b}$ is the vector of biases for each filter, and $\phi(\cdot)$ is the activation function. In this case the $*$ operation between $\mathbf{x}$ and $\mathbf{W}$ is a convolution. The model used in this work applies convolutions to images, so $\mathbf{x}\in\mathbb{R}^{e\times f\times g}$ and  $\mathbf{y}\in\mathbb{R}^{u\times v\times l}$ are 3d tensors, while  $\mathbf{W}\in\mathbb{R}^{d\times d\times t\times l}$ and $\mathbf{b}\in\mathbb{R}^{l}$. The calculation of every element $y_{i,j,k}$ of $\mathbf{y}$ is derived from the operation of the convolutional layer as follows:

\begin{equation}
y_{i,j,k} = \sum_{m,n,p}x_{i-m,j-n,p}W_{m,n,p,k}+b_k,
\label{eq:conv_layer}
\end{equation}

where every element $i,j,k$ of the tensor $\mathbf{y}$ is calculated by moving the filters of $\mathbf{W}$ over the tensor $\mathbf{x}$ and applying eq. \ref{eq:conv_layer}. Each time $\mathbf{W}$ moves over the first two dimensions of $\mathbf{x}$, it skips $S$ pixels, where $S$ is called stride. After applying the convolutional layer, the first two dimensions of $\mathbf{y}$ are smaller in size than the ones of $\mathbf{x}$. The spatial dimensions (first and second dimension) of the tensors $\mathbf{x}$ and  $\mathbf{y}$ relate to each other as follows:

\begin{equation}
U = \frac{E-D}{S}+1,
\label{eq:conv_dims_relation}
\end{equation}

where $U$ is the size of any of the spatial dimensions of $\mathbf{y}$, $E$ is the size of the respective spatial dimension of $\mathbf{x}$, $D$ is the respective spatial dimension of $\mathbf{W}$ and $S$ is the stride used in the convolution operation.

\item \textbf{Zero-padding:} 
It is a commonly used technique to preserve the spatial dimensions of the input  $\mathbf{x}\in\mathbb{R}^{e\times f\times g}$ at the output   $\mathbf{y}\in\mathbb{R}^{u\times v\times l}$ of a convolutional layer. Zero-padding consists on adding 0's to the edges of the spatial dimensions of $\mathbf{x}$. For a convolutional layer of stride $S=1$, and kernel size $D$, the zero-padded input to the convolutional layer must have dimensions such as $\mathbf{x}\in\mathbb{R}^{(e+\lfloor D/2 \rfloor)\times (f+\lfloor D/2 \rfloor)\times g}$, where $D/2$ is the amount of zero-padding included to achieve same spatial dimensions of $e=u \land f=v$, between the layer's input $\mathbf{x}$ and output $\mathbf{y}$.

\item \textbf{Max pooling:}  Pooling layers are used in CNNs to reduce the spatial dimensionality of their inputs. The max pooling used in the model shown in Fig. \ref{fig:early_classifier} returns the maximum
value within a window of its input  $\mathbf{x}$, in the same way as a convolutional filter, this maximum value extraction window is rolled across the spatial dimensions of the input. In the case of the architecture shown in Fig. \ref{fig:early_classifier}, the pooling window is of dimension 2$\times$2 with a stride of 2, i.e., without overlapping of the window, yielding a spatial dimensionality reduction by half each time max pooling is applied.

\item \textbf{Batch normalization layer:} It works as a trainable normalization layer that has different behavior during training and evaluation of the model. During training, batch normalization layer calculates the mean and variance of each feature, to normalize them and compute an exponential moving average of mean and variance of the training set. After training the model, for its evaluation, the whole population statistics adjusted during training are used to normalize evaluation inputs. Batch normalization not only normalizes input values to have a mean value near 0 or a variance value near 1, it also contains a linear ponderation of these inputs, that allows their scaling and shifting. This layer allows the model to emphasize or ignore specific inputs, acting as a regularizer and speeding up training. 

\item \textbf{Dropout:} It is an operation that is usually applied at the output of fully connected layers, and it is used as a regularizer of the model to avoid overfitting of layers with large number of neurons. Similar to a batch normalization layer, dropout performs different operations during training and evaluation. The dropout operation is defined by the \textbf{dropout rate} $DR \in [0,1]$, which is a parameter that, at the training phase of the model, defines the probability of setting each of its inputs to 0, and multiplying the values not set to 0 by $1/(1 - DR)$, such that the sum over all the input values remains the same. At each training step a percentage $DR$ of the outputs of a fully connected layer won't be used, reducing the effective size of that layer. On the other hand, when using the model after training, dropout is deactivated. The desired effect of dropout is to enforce the model to not depend on specific units of every layer.  
\end{itemize}

The model described in Fig. \ref{fig:early_classifier}, is based on Enhanced Deep-HiTS \cite{reyes_enhanced_2018}, a state of the art classifier for binary classification of real astronomical object and bogus samples. The architecture of this model introduced total rotational invariance, which empirically proved to enhance performance on the classification of astronomical images.

\subsection{Neural network training}
\begin{itemize}[leftmargin=*]
\item \textbf{Procedure to train a neural network:}
The objective of using a neural network $f_{\theta}$ of parameters $\theta \in \Theta$, is to approximate a function $y=f(x)$, with $x \in \mathcal{X}$. In practice there is no access to the whole data distribution $\mathcal{X}$, but to a subset of $N$ data samples of the function to approximate $\{(x^{(i)},y^{(i)})\}_{i=1}^N$, called \emph{training set}. Finding the best parameters $\theta^*$ for the neural network $f_\theta(x)$ requires solving the optimization problem:
\begin{equation}
\theta^* = \arg \min_{\theta\in\Theta} \mathcal{C}(\theta)=\arg \min_{\theta\in\Theta}
\frac{1}{N}\sum_{i=1}^N\mathcal{L}(y^{(i)},f_\theta(x^{(i)})),
\label{eq:optimization}
\end{equation}
where $\mathcal{C}$ is an error functional defined by the function $\mathcal{L}$ that is called \textbf{loss function}. The optimization depicted in eq. \ref{eq:optimization}, is achieved by the training of the model through optimization techniques based on \textbf{gradient descent}, when $\mathcal{L}$ is chosen as a differentiable function (e.g. cross-entropy). The parameters $\theta$ are iteratively adjusted by the following rule, until convergence:
\begin{equation}
\theta_{k} = \theta_{k-1}-\mu \nabla_\theta \mathcal{C}(\theta).
\label{eq:training_step}
\end{equation}
Because neural networks are composed of many consecutive layers, the direct calculation of $\nabla_\theta \mathcal{C}(\theta)$ is computationally expensive. However, they can be calculated efficiently by \textbf{back-propagation}, which is an algorithm that \emph{propagates} the error from the output of the model until it reaches the first layer of the neural network, back-propagation is based on the differentiation chain rule. 

Even when using back-propagation, for neural networks trained on large amounts of data, the calculation of the exact gradient $\nabla_\theta \mathcal{C}(\theta)$ becomes computationally expensive. As a solution to this problem, a non biased estimation of the gradient $\nabla_\theta \tilde{\mathcal{C}}(\theta)$ is used, where the gradient is calculated over a small random fraction of data, this is called a \textbf{batch} and the amount of data samples in the batch is the \textbf{batch size}. Therefore, the optimization rule for a batch $\mathcal{B}\subset \mathcal{X}$ is:
\begin{equation}
\theta_{k} = \theta_{k-1}-\mu  \frac{1}{|\mathcal{B}|}\sum_{i\in\mathcal{B}}\nabla_\theta \mathcal{L}(y^{(i)},f_\theta(x^{(i)})),
\label{eq:train_step_batch}
\end{equation}
where $\mu$ is a constant called \textbf{learning rate}, and establishes how large is the performed training step. This technique of training by batches, is a form of \emph{stochastic gradient descent}, and it guarantees convergence when $\mu$ is a well defined sequence in $k$ that satisfies $\sum_k \mu_k = \infty$ and $\sum_k \mu_k^2 < \infty$.
    
\item \textbf{Adam optimizer:} An alternative to the optimization rule of eq. \ref{eq:train_step_batch}, is Adam \cite{kingma_adam_2017}, which is an adaptive learning rate optimization algorithm that automatically adjust $\mu_k$. It uses the squared gradients to scale the learning rate and it includes the moving average of the gradients in its formulation, strategy that is known as \textbf{momentum}, and its used to avoid converge to a local minima in the optimization. The main hyperparameters of Adam are $\beta_{1}$ and $\beta_{2}$, which relate to the moving averages of the gradients and the squared gradients, respectively, and they regulate the rate at which the learning rate $\mu_k$ is adjusted.

\end{itemize}

\section{Hyperparameter Random  Search Results}
\label{sec:random_search_appendix}

For the hyperparameter random search, 133 different combinations of hyperparameters values were sampled from Table~\ref{table:hyperparam_search}, we trained 5 models for each hyperparameter set and then evaluated the test and validation set with every model. In addition, for each model the inference time over a single sample was measured. The training procedure took $\approx$3 days of continuous training on 5 NVIDIA GTX 1080Ti GPUs. From now on, every time we refer to accuracy or inference time of a model, they are the average measurement of 5 models trained with the same hyperparameters. 

\begin{figure*}[htbp]
\gridline{\fig{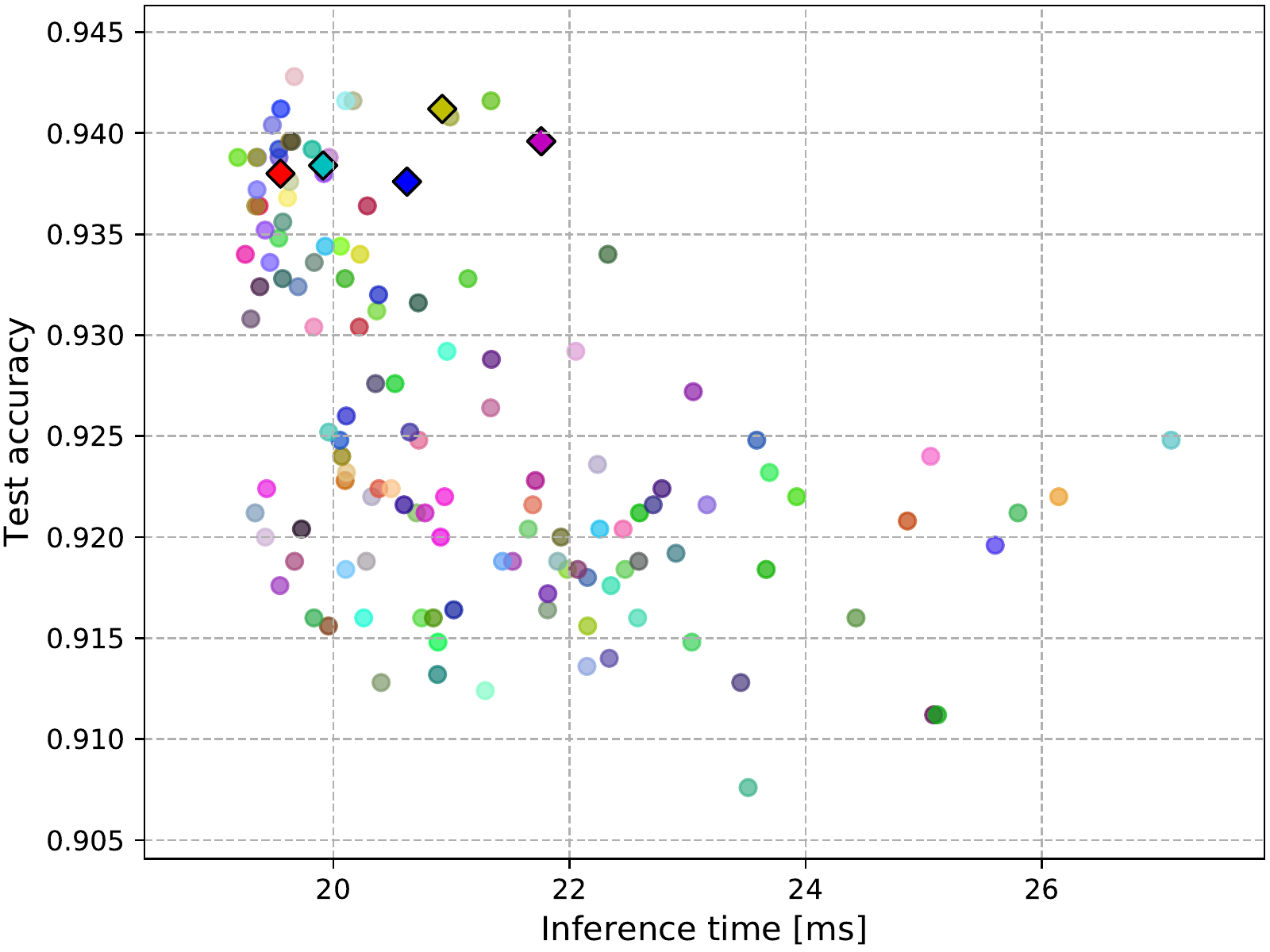}{0.44\textwidth}{(a).}
          \fig{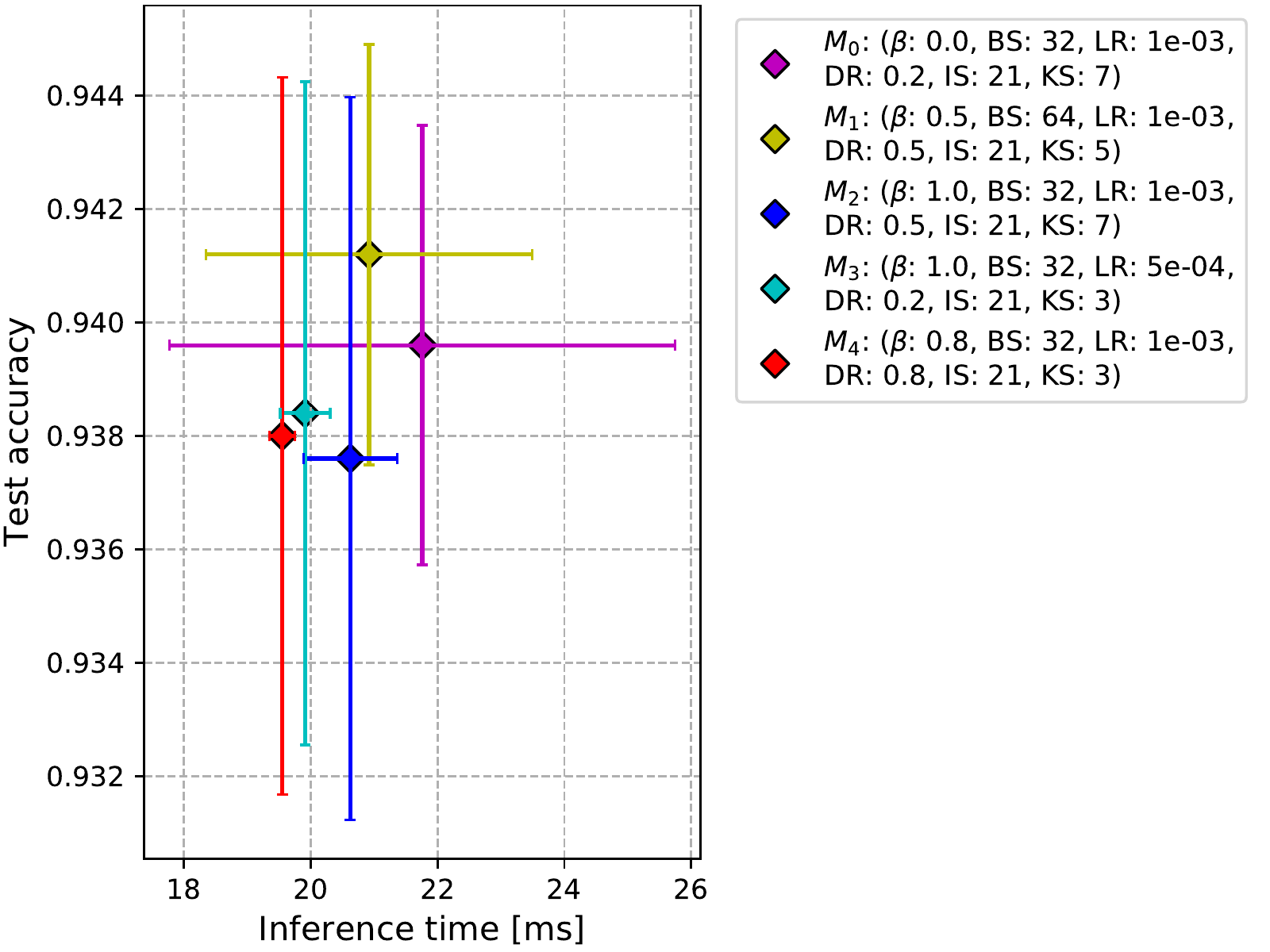}{0.44\textwidth}{(b)}}
\caption{Accuracy of 133 models from hyperparameter random search. For each model, results consider 5 trainings and respective evaluations on the test set. (a) Test accuracy versus inference time, where each dot is a model with different hyperparameters, the closer to the left top corner is a model, the better its performance. Models represented with diamonds correspond to the 5 models with best validation accuracy.  (b) Test accuracy versus inference time for the models represented as diamond in (a), each model has its error bars corresponding to one standard deviation and every model is denoted as $M_i$, where $i$ corresponds to the ranking of its validation accuracy performance among all the 133 models of (a). \label{fig:random_search}}
\end{figure*}

The selection of the best model from the hyperparameter random search was done by looking at the performance in the validation set. We took the 5 models with highest validation accuracy and performed a Welch’s hypothesis test between the model with highest and lowest validation accuracy among the top-5 models, obtaining a p-value of 0.594, which means that the accuracy differences among the models are not statistically significant. Another important factor when processing the volume of data encountered in astronomy, is the inference time of the used models, we measure inference time for the top-5 models and a Welch’s hypothesis test between the fastest and the slowest model got a p-value of 0.330, i.e. no statistically meaningful difference. Finally, the best model among the top-5 is chosen as the one with $\beta=0.5$, because it shows the most interpretable range of prediction probabilities, according to our team of astronomers. The model chosen as the best has a validation accuracy of 0.950$\pm$0.003, test accuracy of 0.941$\pm$0.004 and inference time of 20.5$\pm$2.6 [ms]. The performance of the top-5 models over the test set is shown both in Figure~\ref{fig:random_search} and Table~\ref{table:hyperparam_search_results}. Every time we refer to the top-5 models we mean the aforementioned models.

Figure~\ref{fig:random_search} shows plots of test accuracy versus the inference time, Figure~\ref{fig:random_search}a shows the results of all the 133 models from the random hyperparameter search: In Figure~\ref{fig:random_search}a, the nearest to the top left corner, the better the model, the diamond shapes in this figure correspond to the models with top-5 validation accuracy. Figure~\ref{fig:random_search}b shows in detail the performance of these top-5, where each model has its one standard deviation error bars. These models are named $M_i$, where $i$ corresponds to the position of its validation accuracy w.r.t. to all the 133 models, the lower its validation accuracy, the higher is $i$. Table~\ref{table:hyperparam_search_results} shows validation, test accuracy and inference time for the models with top-5 validation accuracy of Figure~\ref{fig:random_search}b, where $M_1$ is selected as the best model, which is used for experiments shown in previous sections. In Table~\ref{table:hyperparam_search_results}, metrics of the best model $M_1$ are underlined, whereas bold metrics correspond to the highest of their respective column. Coincidentally, model $M_1$ chosen as the best, has the higher test accuracy among the top-5 models, and when compared to the model $M_2$ with worst test accuracy, the Welch’s hypothesis test p-value is 0.364, meaning that the difference is not statically meaningful.

Figure~\ref{fig:random_search}b shows that test accuracy and inference time error bars of each of the top-5 models contain each other.

\begin{table}[h!]
\begin{center}
\caption{Top-5 models with highest validation accuracy from the hyperparameter random search, ranked from $M_0$ to $M_4$. There are no statistical difference between the accuracy and inference time of the displayed models. $M_1$ is chosen as the best model because it has $\beta=0.5$, which shows the most interpretable range of prediction probabilities, according to astronomers. Metrics of the best model $M_1$ are underlined, whereas bold metrics are the best of their respective column.}
\label{table:hyperparam_search_results}
%\scriptsize
\setlength{\tabcolsep}{4pt}
\begin{tabular}{c|c|c|c|c}

Model Name  & Model's Hyperparameters & Validation Accuracy & Test Accuracy & Inference Time {[}ms{]}\\
 \hline \hline
$M_0$ & $\beta$: 0, BS: 32, LR: 1e-03, DR: 0.2, IS: 21, KS: 7 & \textbf{0.950$\pm$0.003} &  0.940$\pm$0.004 &  21.8$\pm$4.0 
\\ \hline
 $M_1$ & $\beta$: 0.5, BS: 64, LR: 1e-03, DR: 0.5, IS: 21, KS: 5 &  \underline{0.950$\pm$0.005} &  \underline{\textbf{0.941$\pm$0.004}} &  \underline{20.9$\pm$2.6} \\ \hline
 $M_2$ & $\beta$: 1.0, BS: 32, LR: 1e-03, DR: 0.5, IS: 21, KS: 7 &  0.949$\pm$0.002 &  0.938$\pm$0.006 &  20.6$\pm$0.7
 \\ 
\hline
 $M_3$ & $\beta$: 1.0, BS: 32, LR: 5e-04, DR: 0.2, IS: 21, KS: 3 &  0.948$\pm$0.003 & 0.938$\pm$0.006 & 19.9$\pm$0.4 
 \\ \hline
 $M_4$ & $\beta$: 0.8, BS: 32, LR: 1e-03, DR: 0.8, IS: 21, KS: 3 &  0.949$\pm$0.003 &  0.938$\pm$0.006 &  \textbf{19.6$\pm$0.2} \\ 
  \hline \hline
\multicolumn{2}{c|}{Welch's t-test p-value $M_0$ v/s $M_4$ | $M_1$ v/s $M_4$ | $M_0$ v/s $M_4$}  & 0.594 & 0.364 & 0.330 \\
\hline
\end{tabular}

\end{center}
\end{table}

\section{Additional Results}\label{sec:additional_results}

\begin{figure}[htbp]
    \centering
    \includegraphics[width=0.47\textwidth]{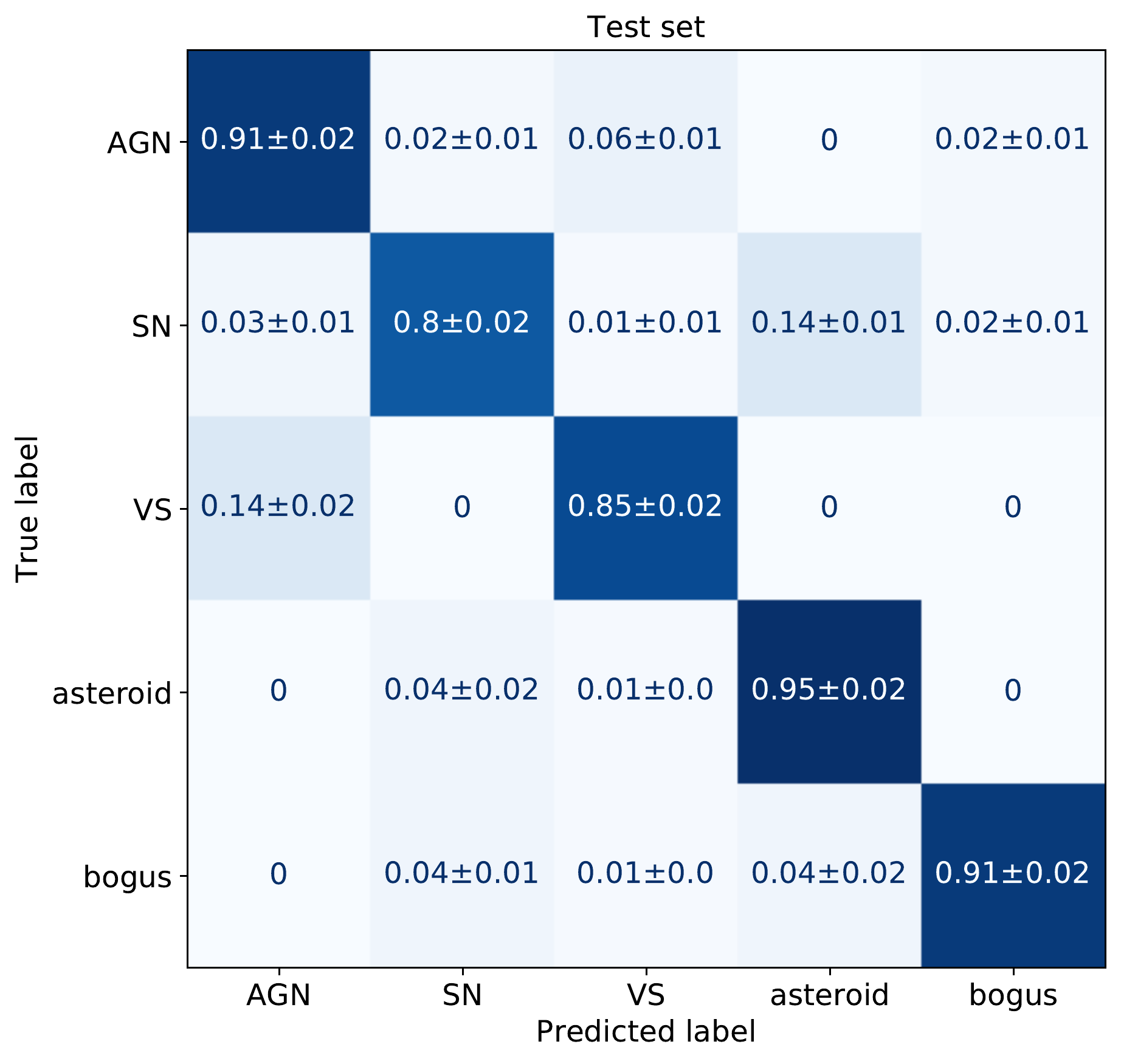}
    \caption{Confusion matrix for the test set when using the stamp classifier only on the three images, without alert metadata features.}
    \label{fig:cm_test_no_feat}
\end{figure}

\begin{figure*}[htbp]
    \centering
    \includegraphics[width=1\textwidth]{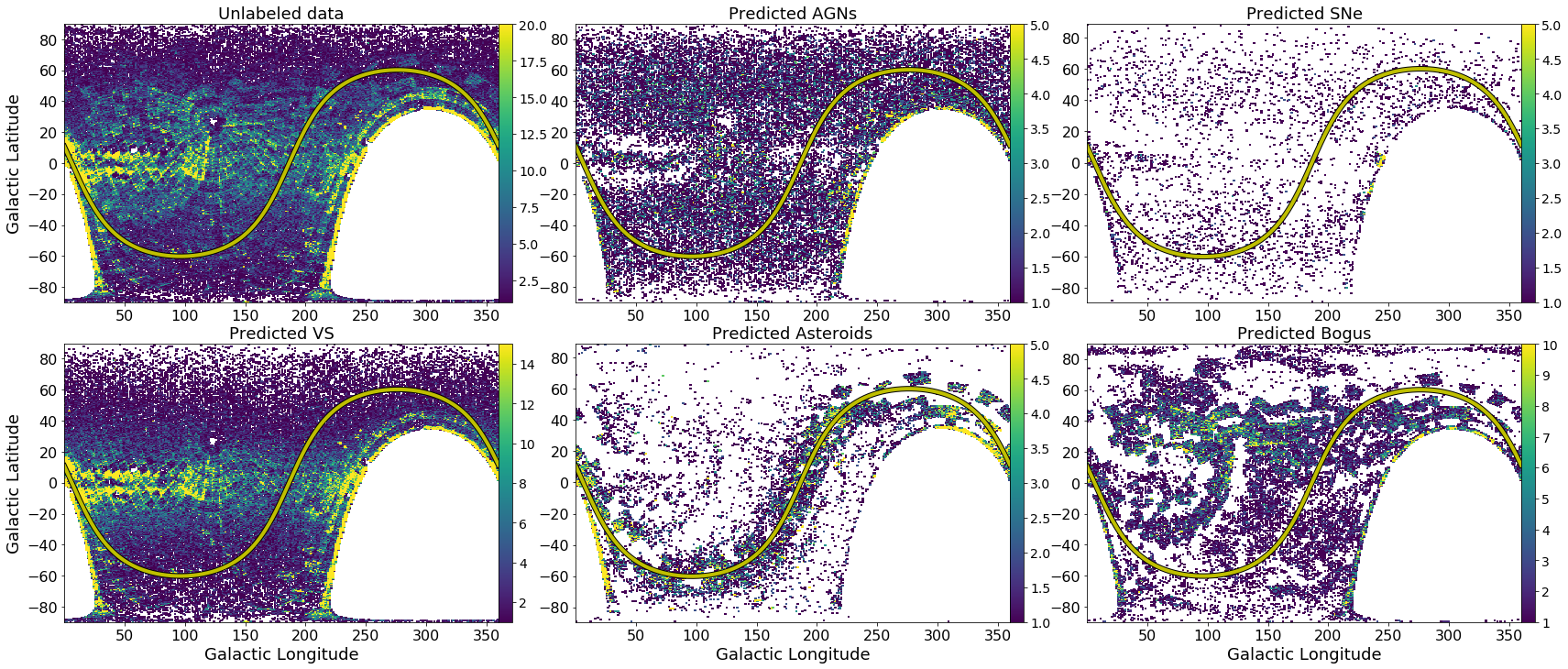}
    \caption{Space distribution for the unlabeled data, and distribution of predictions per class using the stamp classifier only on the three images, without alert metadata features. The ecliptic is shown with a yellow line with black edges. Extragalactic classes (SNe and AGNs) still show a lower density at low galactic latitudes, but are found with higher density, compared to the predictions of the stamp classifier with features, shown in Figure~\ref{fig:space_distribution}. Also, the predicted asteroids by the classifier without using alert metadata features have higher density far from the ecliptic compared to the classifier that uses features.}
    \label{fig:space_distr_no_feat}
\end{figure*}

\begin{table}[h!]
\centering
\caption{Classes assigned to the unlabeled set described in Section \ref{sec:results}, by the light curve classifier classifier (LC) from \citealt{sanchezsaez_alert_2020} and the stamp classifier (SC). The stamp classifier finds 78\% of the SN classified by the light curve classifier, 85\% for AGNs and 96\% for VS. The false positive in the SN class are 9\% of AGNs, 6\% of VS, 4\% of asteroids and 3\% of bogus. The false positives of AGNs are 4\% of SN and 1\% of VS, and false positives of VS is only 3\% with AGNs \label{table:lc_comparisonn}}
%\scriptsize
\setlength{\tabcolsep}{4pt}
\scriptsize

\begin{tabular}{l|r|r|r|r|r|r|r|r|r|r|r|r|r|r|r}
\diagbox[innerleftsep=-28pt, width=1.3cm, innerrightsep=0pt]{SC}{LC} &  SNIa &  SNIbc &  SNII &  SLSN &   AGN &  Blazar &   QSO &  CV/Nova &   YSO &  DSCT &    RRL &  Ceph &    LPV &     EB &  Periodic-Other \\
\hline \hline
SN       &   355 &    124 &   246 &   257 &   657 &      87 &    13 &      241 &    76 &     1 &     22 &     4 &     58 &      7 &              18 \\ \hline
AGN      &     5 &      2 &    27 &    83 &  4057 &    1310 &  9553 &      592 &   309 &   227 &   1586 &    67 &     61 &   1133 &            1738 \\ \hline
VS       &    10 &      7 &    22 &    38 &   393 &     623 &   691 &     2545 &  5023 &  5098 &  29635 &  9657 &  36478 &  56587 &           16539 \\ \hline
Asteroid &    19 &      8 &     6 &    17 &     1 &       4 &     1 &       38 &    10 &     0 &      7 &     1 &    107 &      8 &               1 \\ \hline
Bogus    &     7 &      1 &     9 &    19 &    47 &      16 &    13 &       84 &    36 &     1 &      8 &     4 &     23 &     14 &              22 \\
\hline \hline
\textbf{Recall} & 90\% & 87\%   & 79\% & 62\% & 79\%   & 64\%    & 93\%  & 73\%    & 92\% & 96\% & 95\% & 99\% & 99\% & 98\% & 90\%           \\ 
\end{tabular}

%\diagbox[innerwidth=1.1cm]{SC}{LCC} 

%\begin{tabular}{c|c|c|c|c|c|c|c|c|c|c|c|c|c|c|c}

% Stamp classifier       & \multicolumn{4}{c|}{SN}     & \multicolumn{3}{c|}{AGN} & \multicolumn{8}{c}{VS}                                            \\ \hline \hline
%Light-curve class & SNIa & SNIIbc & SNII & SLSN & AGN    & Blazar  & QSO   & CV/Nova & YSO  & DSCT & RRL  & Ceph & LPV  & EB   & Per-Other \\ \hline
%Recall                       & 90\% & 87\%   & 79\% & 62\% & 79\%   & 64\%    & 93\%  & 73\%    & 92\% & 96\% & 95\% & 99\% & 99\% & 98\% & 90\%           \\ 
%\end{tabular}
\end{table}

\clearpage

\newpage

\section{Bogus Analysis}
\label{sec:bogus_analysis}

We elaborated an analysis of both the bogus samples from ZTF's alert stream and the ones present in our training set.

With our model we estimate the proportion of bogus present in ZTF's alert stream, proportion that is unfeasible to accurately calculate via direct visualization, due to the large number of alerts generated each night. We run through the stamp classifier 176,376 alerts chosen at random from ZTF's alert stream on 10 nights, each of these nights are also chosen at random between 08-18-2020 and 03-03-2021. Our model classifies 34,438 alerts as bogus ($\sim$20\% of the alerts). 

As the stamp classifier used by the SN Hunter only processes first detection alerts, to estimate the amount of bogus alerts processed in a night, from the previous 176,376 alerts we only take the 51,481 alerts that have 1 detection. Our model identifies 31,001 samples as bogus ($\sim$60\% of the total 51,481 alerts with 1 detection).

Based on the manual observations of alerts classified as bogus by our experts, we were able to identify 9 different types of bogus that most commonly appear in ZTF’s stream:

\begin{itemize}
\item \textbf{Region near saturated star}: Bright sources systematically found around very bright targets. When stars are too bright, they can saturate the camera of the telescope and produce artifacts that appear as light flux variations and trigger alerts.

\item \textbf{Bad difference}: Produced by a misalignment of the bright source in the template and science images, which translates into artifacts in the difference image. They appear in the images as dark/light dipoles around every source in the field.

\item \textbf{Near bad difference}: A bad difference can produce artifacts that affect wide portions of an image and trigger alerts far away from the misaligned source.

\item \textbf{Ghost}: Residuals from saturated observations, they usually look like large extended circles with diffraction spikes.

\item \textbf{Extended dark region}: Extended dark regions in the science image.

\item \textbf{Satellite}: We consider alerts triggered by passing human-made satellites as bogus. They trigger alerts that often show multiple point-like or extended sources on the same image in a line (due to rotation and reflection of the satellite), and appear convolved with the PSF.

\item \textbf{Bad pixel columns}: Science images captured by regions of the telescope’s camera where whole columns of pixels go bad. They look like a clear change in background (up or down) over a small portion of columns in the field of the images.

\item \textbf{Bad pixel}: Science images captured by regions of the telescope’s camera where a single pixel is bad.

\item \textbf{Cosmic ray}: High energy particles that interact with a few pixels of the telescope’s camera, they generate alerts that look smaller than or have shapes which are distinctly different from a point source or moving source convolved with PSF.
\end{itemize}

\begin{figure}[t]
    \centering
    \includegraphics[width=0.5\textwidth]{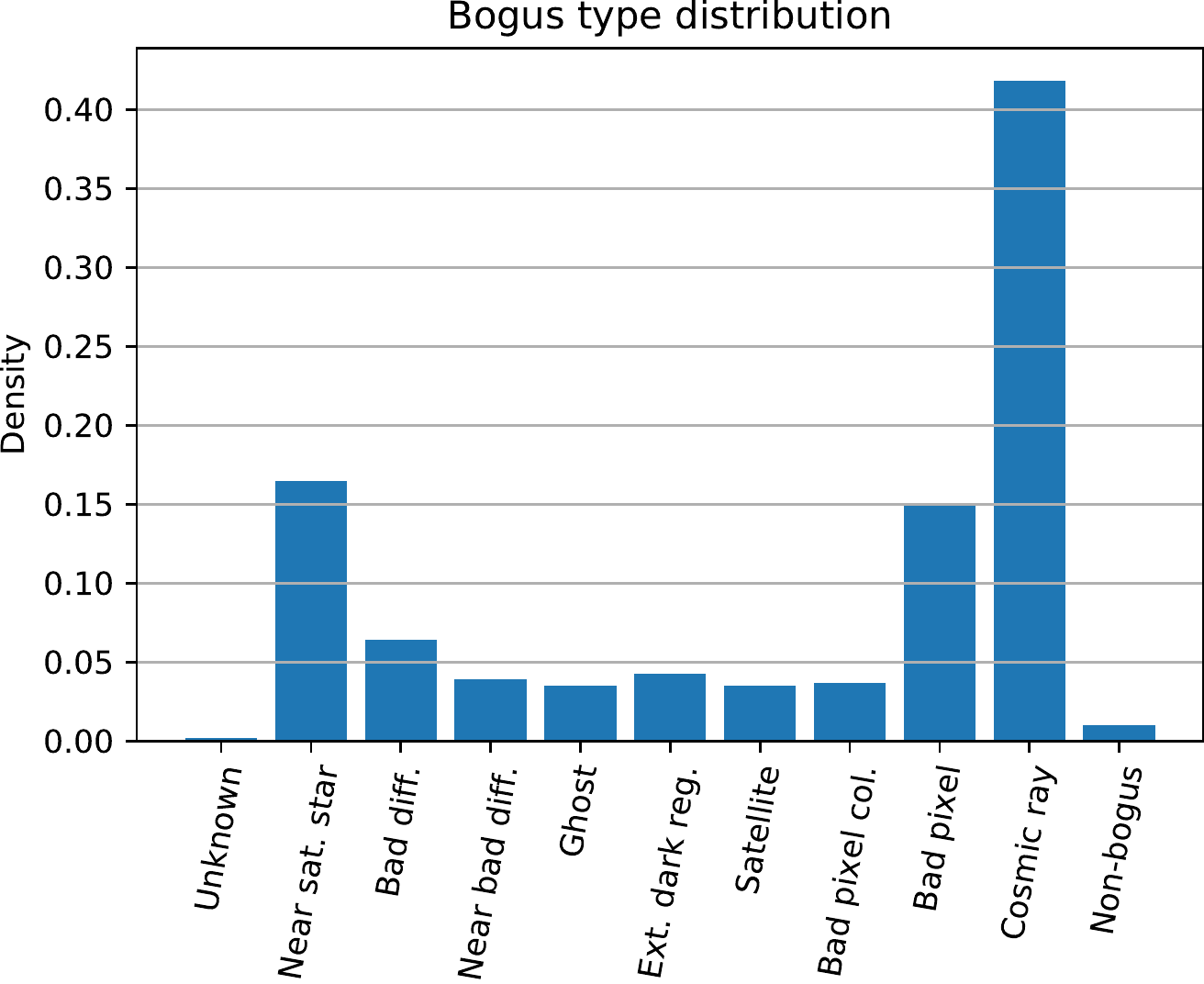}
    \caption{Normalized distribution of different types of bogus present in 1,000 bogus samples from our training set. Bogus types were visually analyzed by an astronomer.}
    \label{fig:bogus_type_distribution}
\end{figure}

We elaborated a recognition of the previous types of bogus present in our training set. A subset of 1000 bogus samples randomly taken from the training set was analyzed by an astronomer, were all the 9 different types of bogus described above were identified. A distribution of the 1000 identified bogus can be seen in Figure~\ref{fig:bogus_type_distribution}. Although we didn’t characterize types of bogus for all the bogus in the training set (because of time restrictions involved in manually analyzing $\sim$10000 bogus events), as the 1000 bogus samples from Figure~\ref{fig:bogus_type_distribution} were randomly sampled, it is an approximation of the true distribution of bogus types in the whole training set.

Figure~\ref{fig:bogus_type_distribution} shows that the most common type of bogus in our training set are cosmic rays, this can be explained by how most of the bogus samples were obtained; bogus missclassified as SN by the stamp classifier, and cosmic rays is the most common type of bogus missclassified as SN. Figure~\ref{fig:bogus_type_distribution} also shows that 10 samples were identified as non-bogus (alerts triggered by a real astronomical source), which can be interpreted as the fact that our manual bogus labeling process has an estimated error/contamination of 1\% percent. One of the bogus analyzed doesn't match any known type, so it is assigned the \emph{unknown} tag in Figure~\ref{fig:bogus_type_distribution}.

The previous analysis was performed over bogus of our training set. We propose as future work to perform a similar study over alerts from the ZTF's stream which are classified as bogus by our stamp classifier. It is worth mentioning that the distribution of types of bogus shown in Figure~\ref{fig:bogus_type_distribution} is not representative of the actual distribution from ZTF's alert stream, since it is biased by our labeling process.

As bogus in the training set were manually labeled in 2 steps, we analyzed how the distribution of bogus vary from one step to the other. Step 1 is composed of 1980 bogus examples reported by ZTF (based on human inspection). Many bogus coming from step 1 are characterized for containing many NaN patches, being near saturated sources or subtraction misalignments; bogus types that are easily identifiable by eye. These bogus were used to train an early version of the stamp classifier and detect SNe. Step 2 bogus include 8783 samples and they are alerts that were confused with a SN by the early version of the stamp classifier and astronomers visually confirmed as bogus. As we increased the amount of labeled bogus, we iteratively improved the stamp classifier and kept visually identifying bogus that were misclassified as SNe. Because of the way we obtained our bogus samples, we expect them not to be representative or have the same distribution of all the bogus in ZTF’s alert stream. To avoid this type of biases, in the future we plan to label by hand bogus alerts that are confused with the rest of the classes (VS, AGN and asteroids).

\begin{figure}[t]
    \centering
    \includegraphics[width=0.8\textwidth]{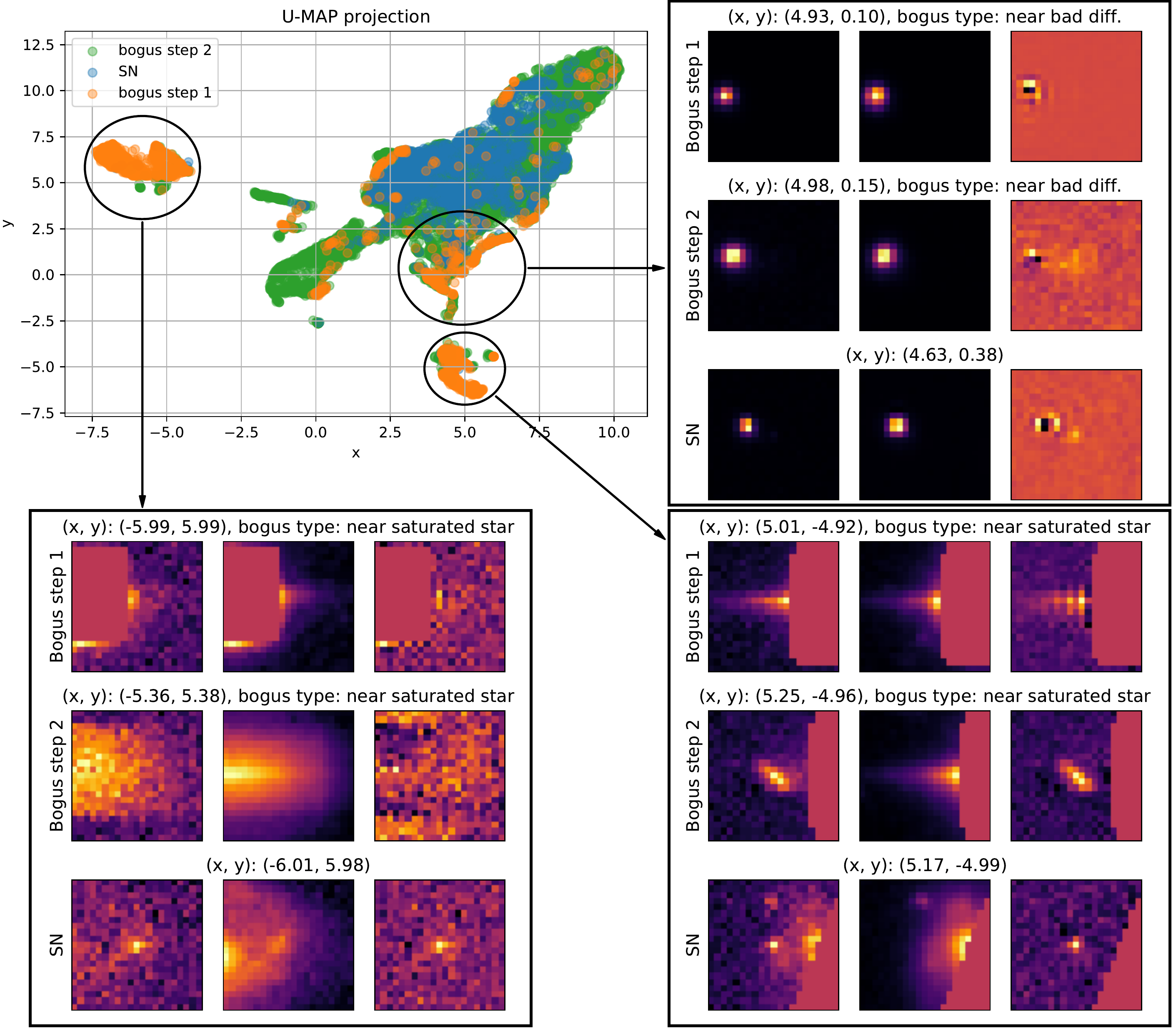}
    \caption{U-MAP projection of image triplets of bogus labeled in step 1 and step 2, alongside SNe alerts. Three main clusters can be identified in the projection, for each of them, images of samples from each class are shown, their projection coordinates are displayed to denote that they are close in the U-MAP projection space, and close by samples tend to look alike. Bogus types for each sample are assigned by an astronomer. Each triplet shows images of science, reference and difference, in that order. The analyzed regions are enclosed by circles labeled from 1 to 3.}
    \label{fig:umap_projection_bogus}
\end{figure}

The effect of using bogus coming form these 2 steps can be seen in Figure~\ref{fig:umap_projection_bogus}, where we used U-MAP (\citealt{mcinnes_umap_2020}) to project alert image triplets in a 2D space, alerts with similar images should appear as neighbor points in the projection. Alongside each cluster of alerts we plotted images of bogus and SNe samples, to visualize how alike are SNe and step 2 bogus alerts. Figure~\ref{fig:umap_projection_bogus} shows 3 main clusters, 2 small cluster at the right and bottom of the U-MAP projection, composed mainly by step 1 bogus, and a big cluster mainly composed of SNe alerts and step 2 bogus, with a few step 1 bogus in it, the ones that most resemble SNe. The big cluster shows that SNe samples overlap with step 2 bogus, where the later fill spaces between and around SNe. The behavior of step 2 bogus w.r.t. SNe is no surprise to us, given the way step 2 bogus were generated (from misclassified SNe by old versions of the stamp classifier) they are expected to resemble or look like SNe.

By visual inspection of the clusters, we verified that close-by samples in the U-MAP projection do look alike and have similar geometric structure. For the 3 main clusters of Figure~\ref{fig:umap_projection_bogus}, we analyzed the regions enclosed by dark circles (labeled from 1 to 3 in the figure) and visualized the samples of SNe and bogus of step 1 and 2:

\begin{itemize}
    \item Circle 1 encloses the small cluster at the left, which is dominated by samples with NaN patches or bright sources at the left of the images. The displayed samples correspond to a near saturated star bogus with a NaN patch at its left, for step 1 bogus. The step 2 bogus sample is also of type near saturated star with the bright source at its left. The displayed SN sample has a bright source at the left of the template image (middle image).
    
     \item Circle 2 encloses a region of the biggest cluster. This region is the one with most step 1 bogus, and it is mainly composed of samples with a bad difference at the left or top of the images. The displayed samples of type step 1 and step 2 bogus are near bad differences located at the left of the images, while the SN sample also has a bad difference at its left.
    
    \item Circle 3 encloses the small cluster at the bottom of Figure~\ref{fig:umap_projection_bogus}, which is mainly composed of samples with NaN patches or bright sources at the right of the images. The displayed step 1 and step 2 bogus samples correspond to near saturated star bogus with NaN patch at their right, while the SN sample has a bright source and a NaN patch at its right.
\end{itemize}

For all the regions analyzed, clusterization was dominated by geometric compositions within images, also bogus clusters can be related to the different bogus types previously described, where bogus within a region tend to be of the same type. In the future it would be useful to use a visualization technique invariant to geometric orientation of samples, to avoid the case where clusters with bogus samples of the same type but different geometric orientation are positioned far away in the projection.

We do not show all 5 classes together in the U-MAP projection because trying to project so many classes of high-dimensional data in a 2D embedding is a problem that is too hard to solve for U-MAP and not enough insightful clusters arise, only blobs of highly overlapped samples can be seen.

%MODIFY ANSWER ACCORDING TO CITATION OF TEXT (STATE THAT WHOLE SECTIONS DONT GO WITH BOLD)

%MENTION EVERYTHING IN PAPER TEXT
\clearpage

\begin{figure*}[htbp]
    \centering
    \includegraphics[width=1\textwidth]{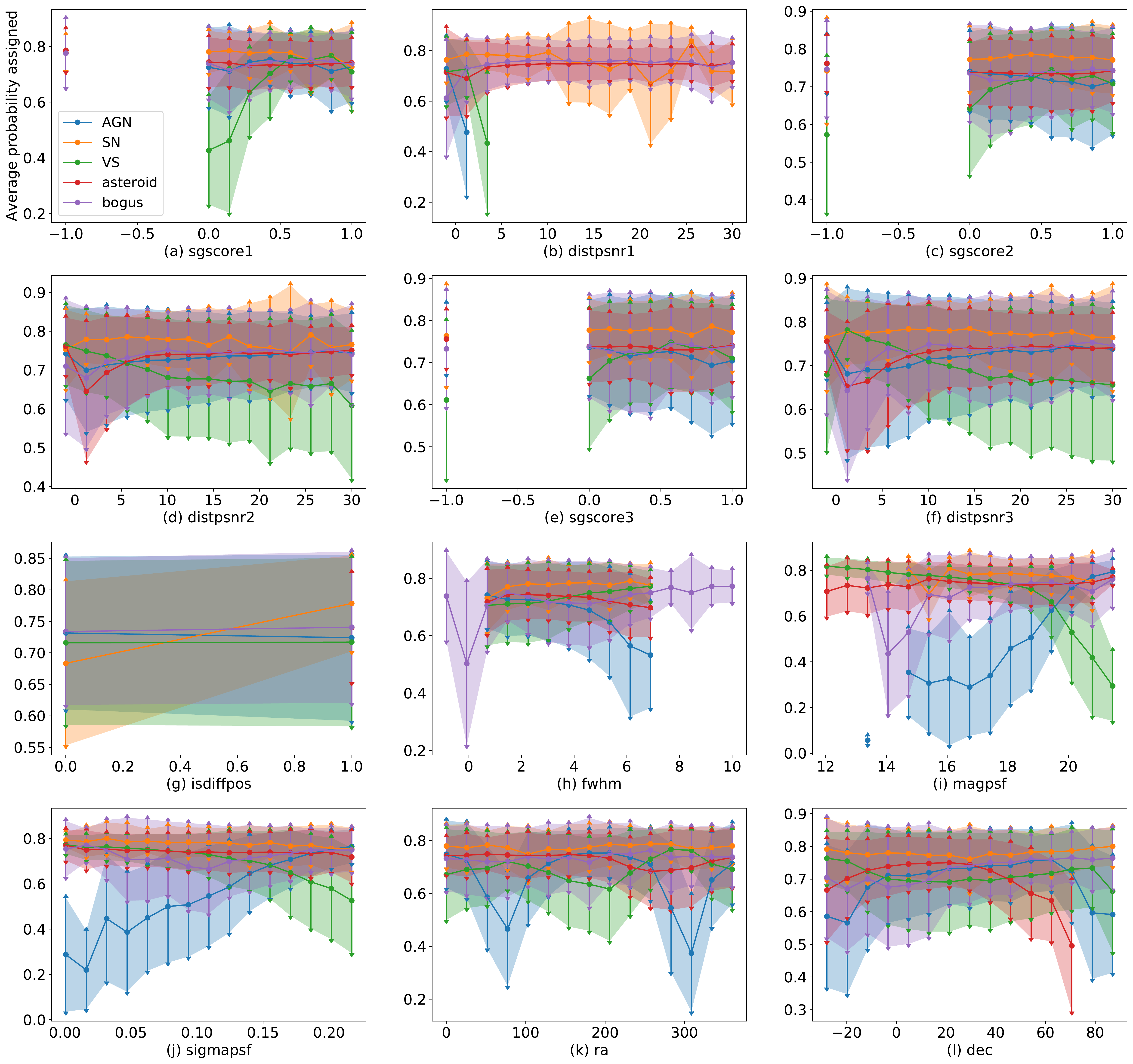}
    \caption{Average model probabilities assigned to the correct class in the training set vs feature values. Each plot from (a) to (l) contains the probabilities for a specific feature.}
    \label{fig:model_probabilities_part1}
\end{figure*}

\begin{figure*}[htbp]
    \centering
    \includegraphics[width=1\textwidth]{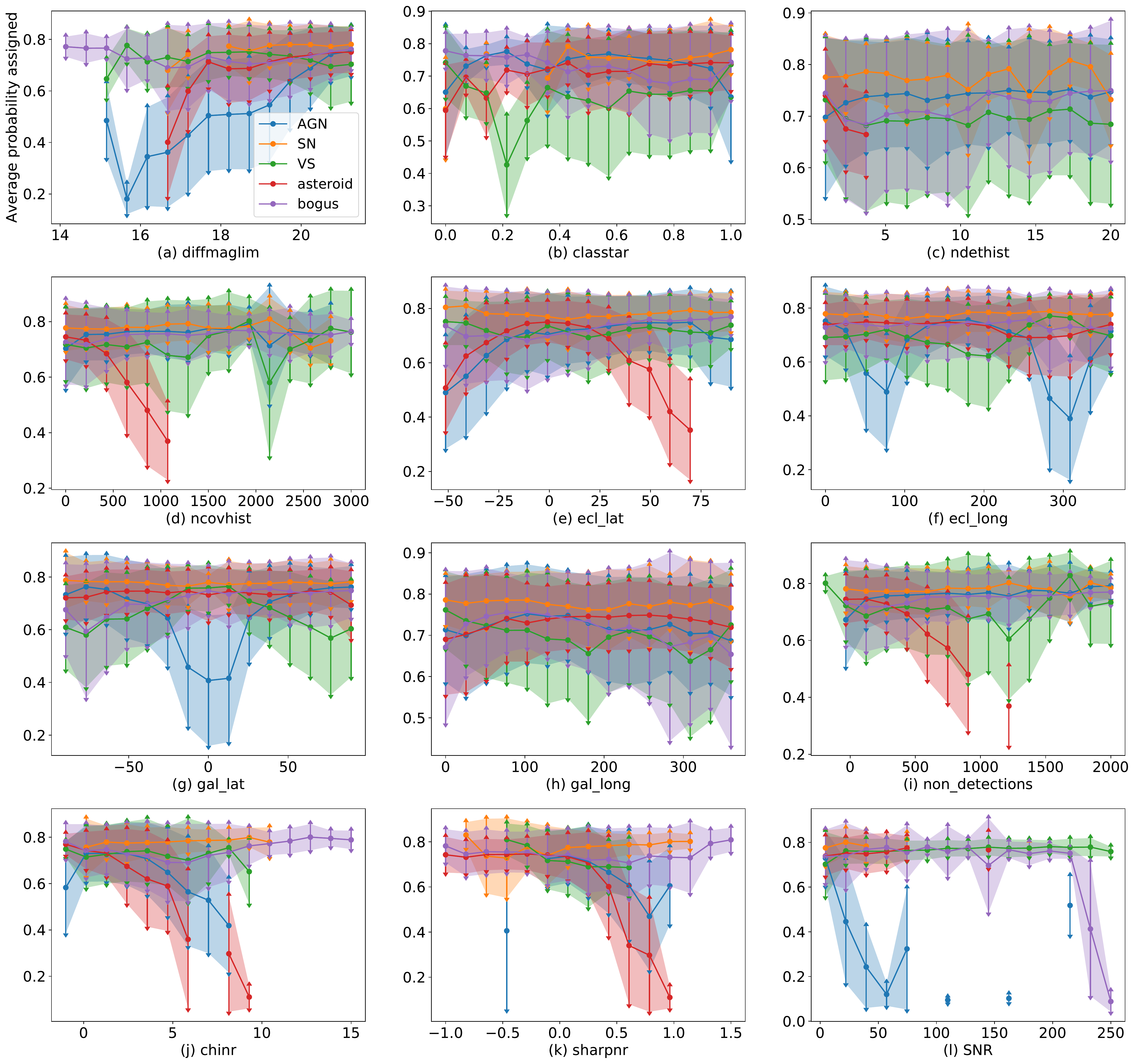}
    \caption{Average model probabilities assigned to the correct class in the training set vs feature values. Each plot from (a) to (l) contains the probabilities for a specific feature. Note that the SNR was not added as a feature to the classifier, this is shown for explanatory purposes mentioned in Section \ref{sec:results}.}
    \label{fig:model_probabilities_part2}
\end{figure*}

\clearpage

\bibliography{stamp_classifier}{}
\bibliographystyle{aasjournal}

%\appendix

%\section{Appendix information}



\end{document}